\documentclass[preprint]{aastex}

\usepackage{graphicx}
\usepackage{txfonts}
\usepackage{natbib}
\usepackage[caption=false]{subfig}
\usepackage{float}
\usepackage{bm}

\newcommand{\noprint}[1]{}
\newcommand{\figsetstart}{{\bf Fig. Set} }
\newcommand{\figsetend}{}
\newcommand{\figsetgrpstart}{}
\newcommand{\figsetgrpend}{}
\newcommand{\figsetnum}[1]{{\bf #1.}}
\newcommand{\figsettitle}[1]{ {\bf #1} }
\newcommand{\figsetgrpnum}[1]{\noprint{#1}}
\newcommand{\figsetgrptitle}[1]{\noprint{#1}}
\newcommand{\figsetplot}[1]{\noprint{#1}}
\newcommand{\figsetgrpnote}[1]{\noprint{#1}}
 
\newcommand{\herschel}{\emph{Herschel}}
\newcommand{\cmc}{cm$^{-3}$}
\newcommand{\cms}{cm$^{-2}$}

\newcommand{\tmb}{$T_{\rm mb}$}
\newcommand{\tpeak}{$T_{\rm peak}$}

\newcommand{\de}{CH$_{3}$OCH$_{3}$}
\newcommand{\ssize}{$\theta_{s}$}
\newcommand{\bsize}{$\theta_{b}$}
\newcommand{\trot}{$T_{\rm rot}$}
\newcommand{\tkin}{$T_{\rm kin}$}
\newcommand{\ntot}{$N_{\rm tot}$}
\newcommand{\vlsr}{v$_{\rm lsr}$}
\newcommand{\dv}{$\Delta$v}

\newcommand{\nhh}{$N_{\rm H_{2}}$}
\newcommand{\nht}{$n_{\rm H_{2}}$}
\newcommand{\td}{$D_{\rm tel}$}
\newcommand{\kapd}{$\kappa_{1.3\rm mm}$}
\newcommand{\betad}{$\beta$}
\newcommand{\eup}{$E_{\rm up}$}
\newcommand{\hh}{H$_{2}$}

\newcommand{\taud}{$\tau_{d}$}
\newcommand{\soff}{$\Delta \theta$}
\newcommand{\sn}{~$\times$~10}
\newcommand{\nf}{$N_{\rm L}$}
\newcommand{\intflux}{$\int{T_{mb}}d\rm v$}
\newcommand{\chisq}{$\chi^{2}$}
\newcommand{\redchisq}{$\chi_{\rm red}^{2}$}
\newcommand{\tvib}{$T_{\rm vib}$}
\newcommand{\um}{$\mu$m}
\newcommand{\sseff}{$\theta_{\rm eff}$}
\newcommand{\nchan}{$N_{\rm chan}$}
\newcommand{\sigtot}{$\sigma_{tot}$}
\newcommand{\nmols}{39} 
\newcommand{\niso}{79} 
\newcommand{\nxclass}{83}
\newcommand{\nlines}{13,000}
\newcommand{\nuchan}{8650}
\newcommand{\nulines}{1730}
\newcommand{\ulfrac}{12\%}
\newcommand{\durl}{http://herschel.esac.esa.int/UserProvidedDataProducts.shtml}
\newcommand{\hotcores}{hot core (s)}
\newcommand{\hts}{H$_{2}$S}

\shorttitle{analysis of the HIFI spectral survey toward Orion KL I}
\shortauthors{Crockett et al.}

\begin{document}
\title{\herschel\ observations of EXtra-Ordinary Sources: ANALYSIS OF THE HIFI 1.2~THz WIDE SPECTRAL SURVEY TOWARD ORION KL I. METHODS\footnote{Herschel is an ESA space observatory with science instruments provided by European-led Principal Investigator consortia and with important participation from NASA.}}

\author{Nathan R. Crockett\altaffilmark{1, a}, Edwin A. Bergin\altaffilmark{1}, Justin L. Neill\altaffilmark{1}, C\'{e}cile Favre\altaffilmark{1}, 
Peter Schilke\altaffilmark{2},
Dariusz C. Lis\altaffilmark{3}, 
Tom A. Bell\altaffilmark{4}, 
Geoffrey Blake\altaffilmark{5,11},
Jos\'{e} Cernicharo\altaffilmark{4},
Martin Emprechtinger\altaffilmark{3},
Gisela B. Esplugues\altaffilmark{4},
Harshal Gupta\altaffilmark{6},
Maria Kleshcheva\altaffilmark{5},
Steven Lord\altaffilmark{7},
Nuria Marcelino\altaffilmark{8},
Brett A. McGuire\altaffilmark{11},
John Pearson\altaffilmark{6},
Thomas G. Phillips\altaffilmark{3},
Rene Plume\altaffilmark{9},
Floris van der Tak\altaffilmark{10, 12},
Bel\'{e}n Tercero\altaffilmark{4}, and 
Shanshan Yu\altaffilmark{6}
}

\altaffiltext{1}{Department of Astronomy, University of Michigan, 500 Church Street, Ann Arbor, MI 48109, USA}
\altaffiltext{2}{Physikalisches Institut, Universit\"at zu K\"oln, Z\"ulpicher Str. 77, 50937 K\"oln, Germany}
\altaffiltext{3}{California Institute of Technology, Cahill Center for Astronomy and Astrophysics 301-17, Pasadena, CA 91125 USA}
\altaffiltext{4}{Centro de Astrobiolog\'ia (CSIC/INTA), Laboratiorio de Astrof\'isica Molecular, Ctra. de Torrej\'on a Ajalvir, km 4 28850, Torrej\'on de Ardoz, Madrid, Spain}
\altaffiltext{5}{California Institute of Technology, Division of Geological and Planetary Sciences, MS 150-21, Pasadena, CA 91125, USA}
\altaffiltext{6}{Jet Propulsion Laboratory, California Institute of Technology, 4800 Oak Grove Drive, Pasadena, CA 91109, USA}
\altaffiltext{7}{Infrared Processing and Analysis Center, California Institute of Technology, MS 100-22, Pasadena, CA 91125}
\altaffiltext{8}{National Radio Astronomy Observatory, 520 Edgemont Road, Charlottesville, VA 22903, USA}
\altaffiltext{9}{Department of Physics and Astronomy, University of Calgary, 2500 University Drive NW, Calgary, AB T2N 1N4, Canada}
\altaffiltext{10}{SRON Netherlands Institute for Space Research, PO Box 800, 9700 AV, Groningen, The Netherlands}
\altaffiltext{11}{Division of Chemistry and Chemical Engineering, California Institute of Technology Pasadena, CA 91125, USA}
\altaffiltext{12}{Kapteyn Astronomical Institute, University of Groningen, The Netherlands}
\altaffiltext{a}{Current address: California Institute of Technology, Division of Geological and Planetary Sciences, MS 150-21, Pasadena, CA 91125, USA}

\begin{abstract}
We present a comprehensive analysis of a broad band spectral line survey of the Orion Kleinmann-Low nebula (Orion KL), one of the most chemically rich regions in the Galaxy, using the HIFI instrument on board the \herschel\ Space Observatory. This survey spans a frequency range from 480 to 1907~GHz at a resolution of 1.1~MHz. These observations thus encompass the largest spectral coverage ever obtained toward this high-mass star-forming region in the sub-mm with high spectral resolution, and include frequencies $>$~1~THz where the Earth's atmosphere prevents observations from the ground. In all, we detect emission from \nmols\ molecules (\niso\ isotopologues). Combining this dataset with ground based mm spectroscopy obtained with the IRAM 30~m telescope, we model the molecular emission from the mm to the far-IR using the XCLASS program which assumes local thermodynamic equilibrium (LTE). Several molecules are also modeled with the MADEX non-LTE code. Because of the wide frequency coverage, our models are constrained by transitions over an unprecedented range in excitation energy. A reduced \chisq\ analysis indicates that models for most species reproduce the observed emission well. In particular, most complex organics are well fit by LTE implying gas densities are high ($>$~10$^6$~\cmc) and excitation temperatures and column densities are well constrained. Molecular abundances are computed using \hh\ column densities also derived from the HIFI survey. The distribution of rotation temperatures, \trot, for molecules detected toward the hot core is significantly wider than the compact ridge, plateau, and extended ridge \trot\ distributions, indicating the hot core has the most complex thermal structure.
\end{abstract}

\section{Introduction}
\label{s-intro}

The origin of chemical complexity in the interstellar medium (ISM) is still not well understood. Approximately 175 molecules, not counting isotopologues, have been detected in the ISM \citep{menten11}. The majority of complex molecules are thought to originate on grain surfaces although it is possible that gas-phase processes play a significant but, as of yet, unknown role \citep{herbst09}. One of the best ways to probe the chemistry that is occurring within the ISM is via unbiased spectral line surveys of star-forming regions in the mm and sub-mm, where molecular line emission is strong. High mass star-forming regions are among the most prolific emitters of complex organic molecules, which are produced primarily by energetic protostars that heat the surrounding material liberating molecules from dust grains and driving chemical reactions that cannot occur at lower temperatures \citep[see e.g.][and references therein]{herbst09, garrod06, garrod08}. Unbiased spectral line surveys offer a unique avenue to explore the full chemical inventory and active molecular pathways in the dense ISM.  Molecular rotational emissions are, for the most part, concentrated at mm and sub-mm wavelengths but the interference of atmospheric absorption has left broad wavelength regimes unexplored which has hampered our ability to obtain a complete view of the molecular content in these regions.
 
In this study, we present a comprehensive full band analysis of the HIFI 1.2~THz wide spectral survey toward the Orion Kleinmann-Low nebula (Orion KL), one of the archetypal massive star-forming regions in our Galaxy. Specifically, we model the emission in order to obtain reliable molecular abundances. Because of its close distance \citep[420~pc;][]{menten07, hirota07} and high luminosity \citep[$\sim$~10$^{5}$~L$_{\sun}$;][]{wynn84}, Orion KL has been exhaustively studied not only in the (sub-)mm but throughout the electromagnetic spectrum \citep{genzel89, odell01}. As such, numerous high spectral resolution single dish line surveys have been carried out toward Orion KL in the mm \citep{johansson84, sutton85, blake87, turner89, greaves91, ziurys93, lee01, lee02, goddi09, tercero10, tercero11}, sub-mm \citep{jewell89, schilke97, schilke01, white03, comito05, olofsson07, persson07}, and far-IR \citep{lerate06}, though the far-IR survey was obtained at a much lower spectral resolution ($\nu/\Delta \nu$~$<$~10$^{4}$) than the (sub-)mm surveys ($\nu/\Delta \nu$~$>$~10$^{5}$). These studies show that Orion KL is one of the most chemically rich sources in the Milky Way and that the molecular line emission originates from several spatial/velocity components representing a diverse set of environments within Orion KL. Although not spatially resolved by single dish observations, these components can be differentiated using high resolution spectroscopy because they have significantly different line widths and central velocities.  Furthermore, interferometric observations have mapped the spatial distributions of these components using different molecular tracers revealing a complex morphology \citep[see e.g.][]{blake96, wright96, beuther05, beuther06, friedel08, wang10, goddi11, favre11, peng12, brouillet13}. A more detailed description of these components is given in Sec.~\ref{s-orion}.

The observations presented in this study were obtained as part of the \emph{Herschel Observations of EXtra Ordinary Sources} (HEXOS) guaranteed time key program and span a frequency range from 480 to 1907~GHz, providing extraordinary frequency coverage in the sub-mm and far-IR. This dataset alone provides a factor of 2.5 larger frequency coverage than all previous high resolution (sub-)mm spectral surveys combined. As a result, we are able to robustly constrain the emission of both complex organics, using hundreds to thousands of lines, and lighter species with more widely spaced transitions over an unprecedented range in excitation energy. Molecular abundances derived in this study thus span the entire range of ISM chemistry, from simple molecules to complex organics. The HIFI spectrum at $\nu$~$\gtrsim$~1~THz, furthermore, represents the first high spectral resolution observation of Orion KL in this spectral region, which is not accessible from the ground, giving access to transitions of light hydrides such as H$_{2}$O and H$_{2}$S. We emphasize that these data were obtained with the same instrument and near uniform efficiency meaning that relative line intensities across the entire band are tremendously reliable. 

This paper is organized in the following way. In Sec.~\ref{s-obs}, we present the observations and outline our data reduction procedure. Our modeling methodology using two different computer codes is described in Sec.~\ref{s-model}. Our results are presented in Sec.~\ref{s-res}. This includes line statistics and reduced \chisq\ calculations for our models (Sec.~\ref{s-lstats}), our derived molecular abundances (Sec.~\ref{s-abund}), calculated vibration temperatures for HCN and CH$_{3}$CN (Sec.~\ref{s-vib}), and unidentified (U) line statistics (Sec.~\ref{s-uline}). Descriptions of individual molecular fits are given in Sec.~\ref{s-indiv}. We give a discussion of our results in Sec.~\ref{s-disc}. Finally, our conclusions are summarized in Sec.~\ref{s-con}.

\section{Observations and Data Reduction}
\label{s-obs}

\subsection{The HIFI Survey}
\label{s-hifi}

The data presented in this work were obtained using the HIFI instrument \citep{degraauw10} on board the \herschel\ Space Observatory \citep{pilbratt10}. The full HIFI spectral survey toward Orion KL is composed of 18 observations, each of which is an independent spectral scan obtained using the wide band spectrometer (WBS) covering the entire frequency range of the band in which the observation was taken. All available HIFI bands (1a -- 7b) are represented in this dataset meaning total frequency coverage between 480 and 1900~GHz with two gaps at 1280 -- 1430~GHz and 1540 -- 1570~GHz. The WBS has a spectral resolution of 1.1~MHz (corresponding to 0.2 -- 0.7~km/s across the HIFI scan) and provides separate observations for horizontal (H) and vertical (V) polarizations. The scans were taken such that any given frequency was covered by 6 subsequent LO settings for bands 1 -- 5, or 4 LO settings for bands 6 -- 7. Additional details concerning HIFI spectral surveys can be found in \citet{bergin10}. 

The telescope was pointed toward $\alpha_{J2000} = 5^h35^m14.3^s$, $\delta_{J2000} = -5^{\circ}22'33.7''$ in bands 1 -- 5, where the beam size, \bsize, was large enough ($\sim$~44 -- 17\arcsec) to include emission from all spatial/velocity components. For bands 6 and 7, however, the beam size was small enough ($\sim$~15 -- 11\arcsec) that individual pointings toward the hot core ($\alpha_{J2000} = 5^h35^m14.5^s$, $\delta_{J2000} = -5^{\circ}22'30.9''$) and compact ridge ($\alpha_{J2000} = 5^h35^m14.1^s$, $\delta_{J2000} = -5^{\circ}22'36.5''$) were obtained. We assume the nominal \herschel\ pointing uncertainty of 2\arcsec\ \citep{pilbratt10}. All data were taken using dual beam switch (DBS) mode with the reference beam 3\arcmin\ east or west of the target position. 

The method we used to reduce the data is described in 
\citet{crockett14}. This procedure begins with standard HIPE \citep{ott10} pipeline processing (version 5.0, build 1648) to produce calibrated ``Level 2" double sideband (DSB) spectra at individual LO settings. Spurious spectral features (``spurs") and baselines were also removed from each scan before they were deconvolved into a single sideband (SSB) spectrum \citep{comito02}. The finished product of this procedure is a deconvolved, H/V polarization averaged, SSB spectrum for each band with the continuum emission removed. We note that even though the continuum emission was removed from the DSB data prior to deconvolution, baseline offsets as large as $\pm$~0.1~K in the SSB spectra are present. Table~\ref{t-obs} lists the date, operational day (OD), observation ID (OBSID), frequency coverage, and RMS on an antenna temperature intensity scale for each observation. We note that the RMS level can vary by as much as a factor of $\sim$~2 across a given band. Values reported in Table~\ref{t-obs} are, therefore, merely the most representative RMS estimates. 

\begin{figure}
\epsscale{1.0}
\plotone{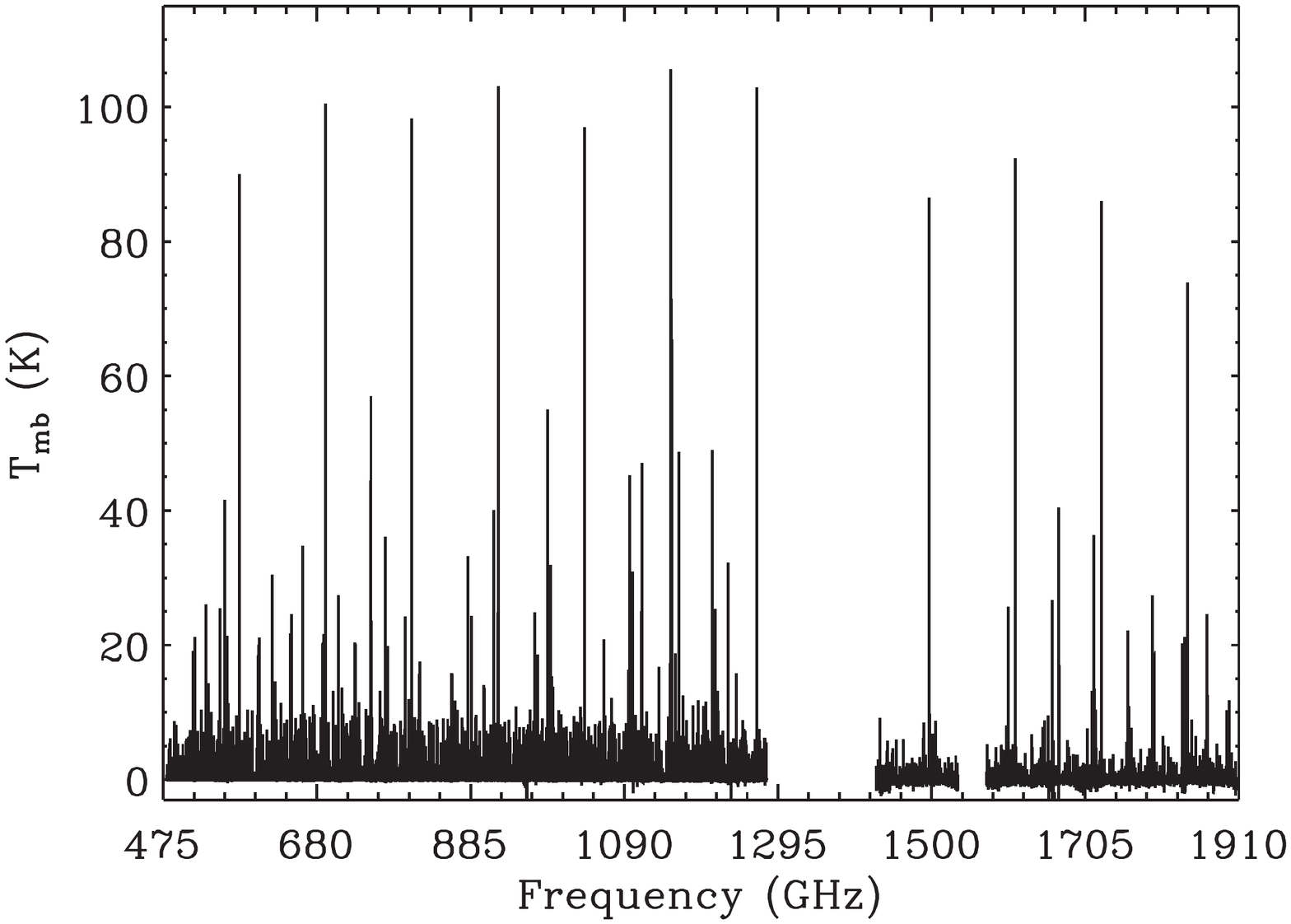}
\caption{The Orion KL HIFI spectral survey is plotted after baseline subtraction. For bands 6 and 7, the hot core pointing is plotted. The data are resampled uniformly to a velocity resolution of $\sim$~10~km/s to improve the appearance at this scale. \label{p-fball}}
\end{figure}

\begin{figure}
\epsscale{1.0}
\plotone{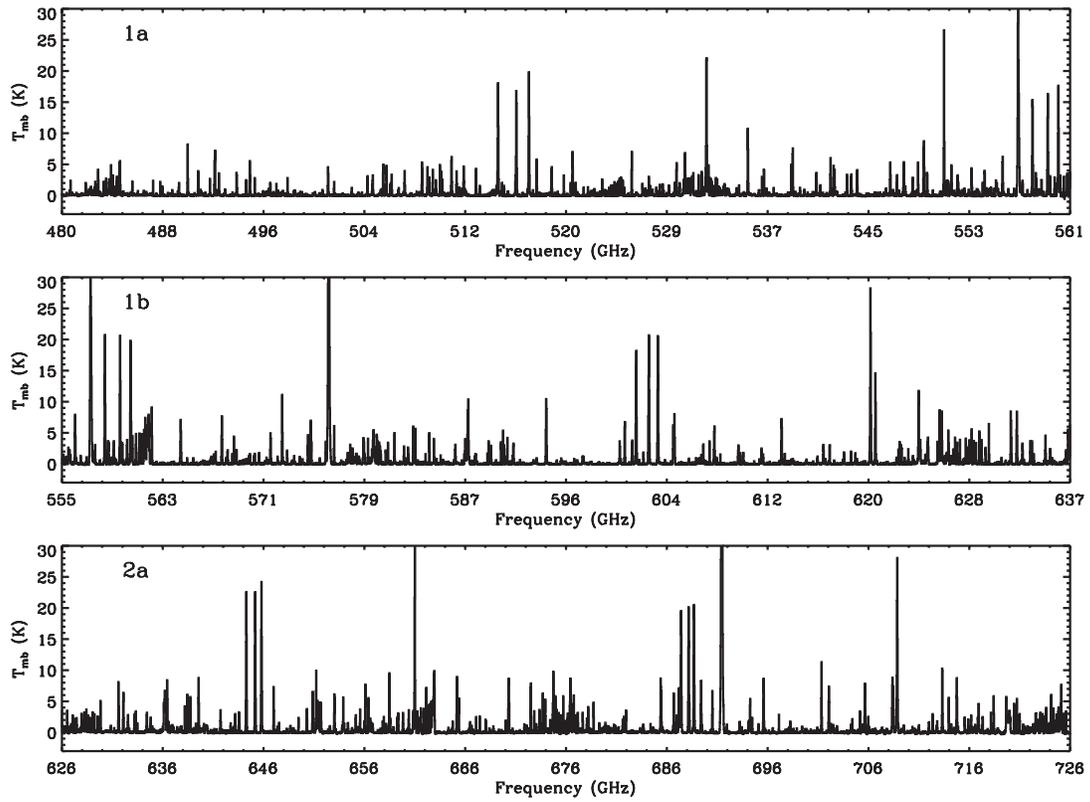}
\caption{Bands 1a, 1b, and 2a of the HIFI spectral survey toward Orion KL. The data are resampled to a velocity resolution of $\sim$~10~km/s to improve the appearance at this scale. Each band is labeled in the upper left hand corner of each panel. \label{p-fullband1}}
\end{figure}

\begin{figure}
\epsscale{1.0}
\plotone{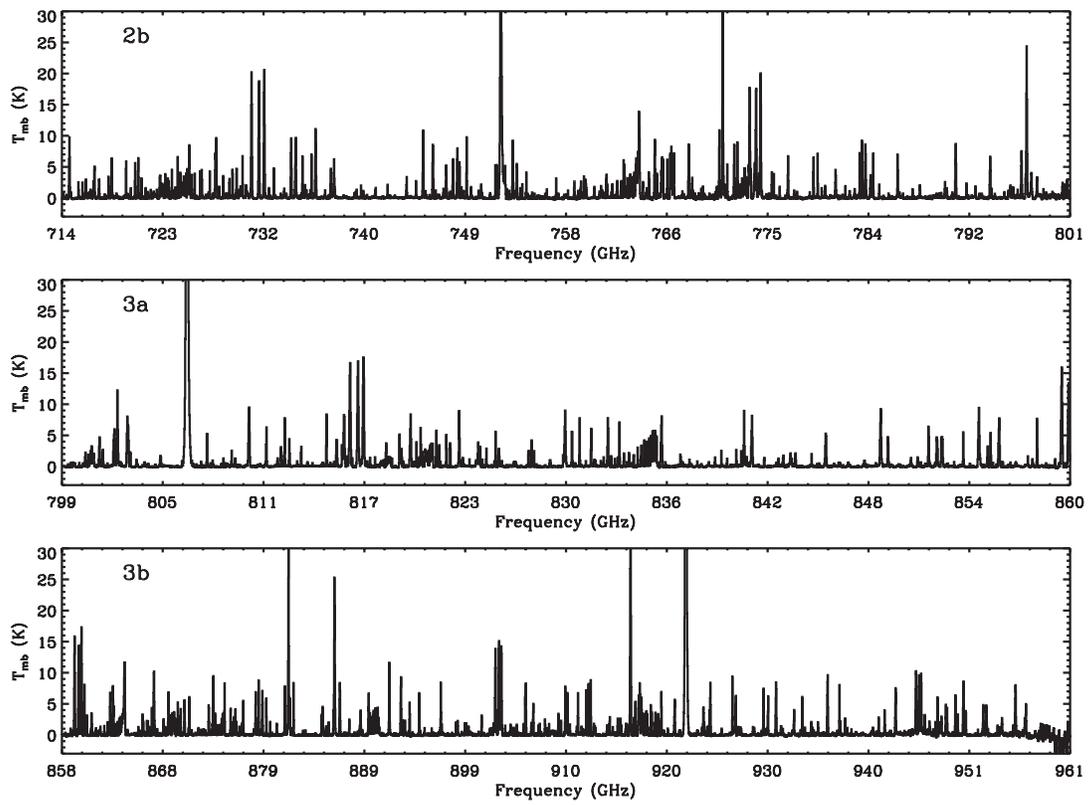}
\caption{Same as Fig.~\ref{p-fullband1} for bands 2b, 3a, and 3b. \label{p-fullband2}}
\end{figure}

\begin{figure}
\epsscale{1.0}
\plotone{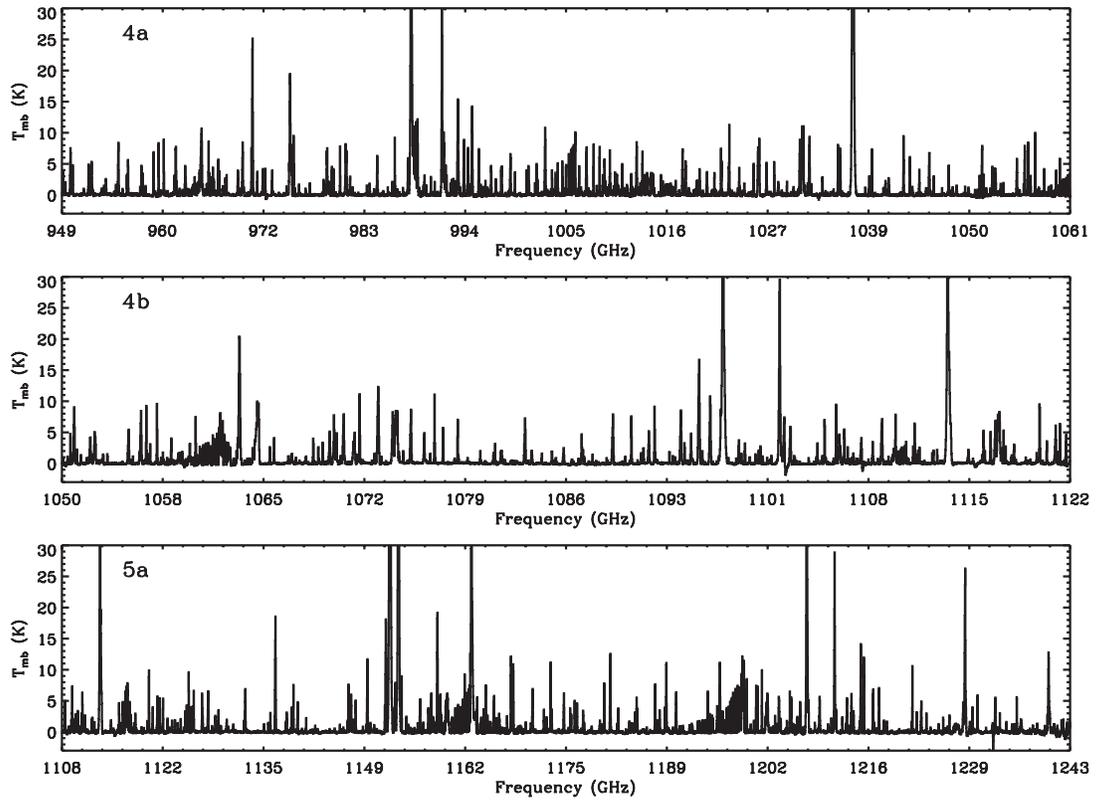}
\caption{Same as Fig.~\ref{p-fullband1} for bands 4a, 4b, and 5a.  \label{p-fullband3}}
\end{figure}

\begin{figure}
\epsscale{1.0}
\plotone{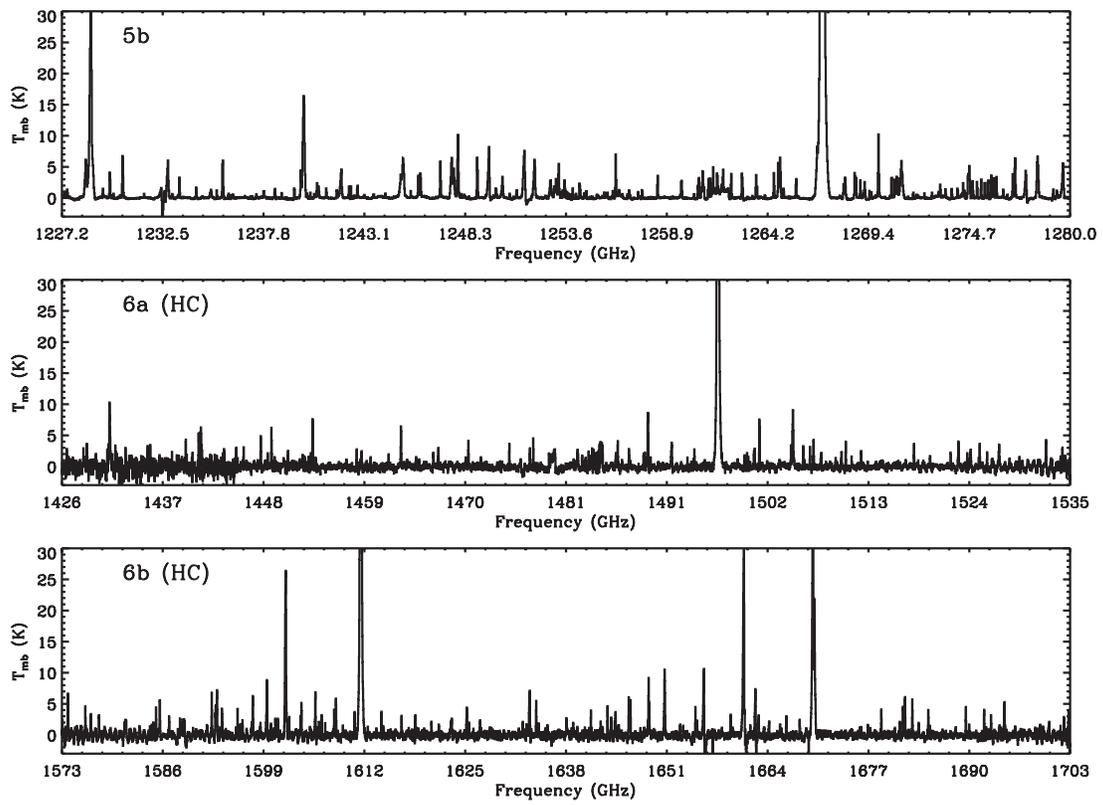}
\caption{Same as Fig.~\ref{p-fullband1} for bands 5b, 6a, and 6b. The hot core pointing is plotted for band 6. \label{p-fullband4}}
\end{figure}

\begin{figure}
\epsscale{1.0}
\plotone{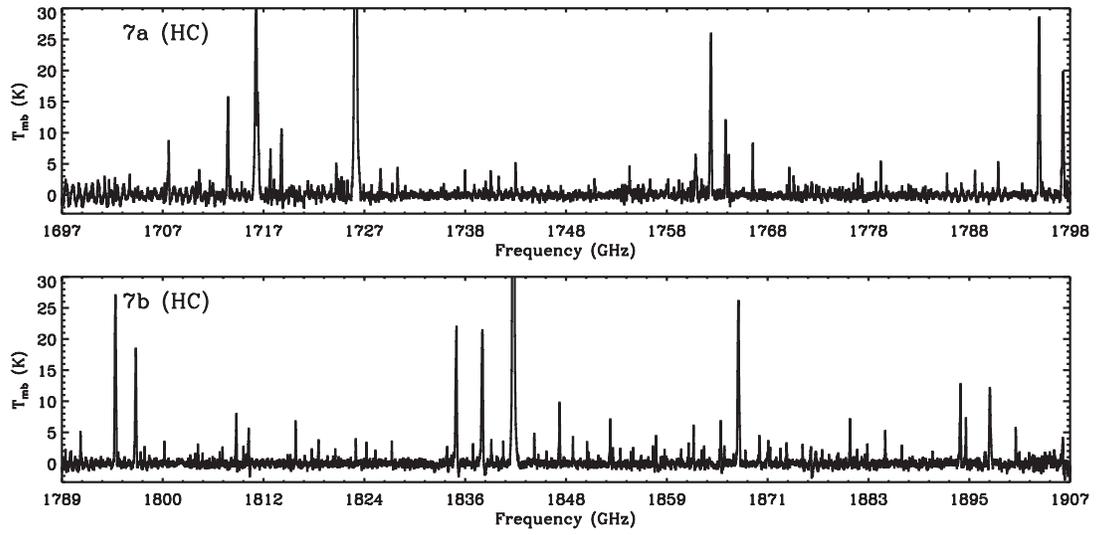}
\caption{Same as Fig.~\ref{p-fullband1} for bands 7a and 7b. Only the hot core pointing is plotted. \label{p-fullband5}}
\end{figure}

All scans were corrected for aperture efficiency using Eqs. 1 and 2 from \citet{roelfsema12}. In bands 1 --  5, we applied the aperture efficiency, which is linked to a point source, because the \herschel\ beam is large relative to the size of the hot core and compact ridge ($\sim$10\arcsec). For bands 6 and 7, on the other hand, we applied the main beam efficiency because the beam size is comparable to the size of the hot core and compact ridge and the main beam efficiency is coupled to an extended source. We, however, refer to all line intensities as main beam temperatures, \tmb, in this work for the sake of simplicity. Fig.~\ref{p-fball} plots the entire HIFI spectral survey toward Orion KL. From the figure, the high sub-mm line density, characteristic of Orion KL, is readily apparent. Figs~\ref{p-fullband1} -- \ref{p-fullband5} plot each band individually so that more details of the spectrum can be seen. In particular, these figures show a marked decrease in the observed line density as a function of frequency, which is mainly due to a drop off in the number of emissive transitions from complex organics \citep{crockett10}. 

In addition to these data products, we also provide SSB spectra with the continuum present. These data were produced by deconvolving DSB spectra after spur removal but before baseline subtraction. The resulting SSB spectra thus contained the continuum but had higher noise levels due to baseline offsets between scans. We next fit a second order polynomial to the continuum in each band.  We then added these polynomial fits to the baseline subtracted SSB scans, thus yielding a spectrum which includes the continuum but does not contain the extra noise brought about by baseline offsets. All reduced data products are available online at \durl\ in ASCII, CLASS, and HIPE readable FITS formats. Model fits of individual molecular species (Sec.~\ref{s-model}) are also available there.

\subsection{The IRAM Survey}

In order to constrain the molecular emission at mm wavelengths, we include a spectral survey obtained with the IRAM 30~m telescope in our analysis. This dataset is described in \citet{tercero10} and covers frequency ranges 200 -- 280~GHz, 130 -- 180~GHz, and 80 -- 116~GHz, corresponding to spectral windows at 1.3, 2, and 3~mm, respectively, at a spectral resolution of 1.25~MHz corresponding to 1.3 -- 4.7~km/s across the IRAM survey. These observations were pointed toward IRc2 at $\alpha_{2000.0}$~=~$5^{h}45^{m}14^{s}.5$ and $\delta_{2000}$~=~$-5\degr22\arcmin30.0\arcsec$. Because the beam size varies between 29\arcsec\ and 9\arcsec, these observations are most strongly coupled to the hot core especially at high frequencies where the beam size is smallest. As a result, our models often over predict emission toward the compact ridge relative to the data in the 1.3~mm band.

\subsection{The ALMA Survey}
\label{s-obs_alma}

We take advantage of the publicly available ALMA band 6 line survey to investigate the spatial distribution of a subset of molecules detected toward the hot core. This survey was observed as part of ALMA's science verification (SV) phase and covers a frequency range of 214 -- 247~GHz at a spectral resolution of 0.488~MHz ($\sim$~0.6~km/s at 231~GHz). The observations were obtained with an array of 16 antennas on 20 January 2012. All antennas had a diameter of 12~m and projected baselines had lengths between 13 and 202~k$\lambda$. The phase center was pointed at coordinates $\alpha_{J2000} = 5^{h}35^{m}14^{s}.35$ and $\delta_{J2000} = -05^{\circ}22'35''$. At 231~GHz, ALMA's primary beam size (field of view) is $\sim$~27\arcsec, similar to \herschel\ at HIFI frequencies. Callisto and the quasar J0607-085 were used as the absolute flux and phase calibrators, respectively. The Common Astronomy Software Applications package, CASA, was used to produce the maps presented in this work using the same methodology as \citet{neill13a}. We also employ the publicly available continuum map of Orion~KL, derived from 30 line free channels near 230.9~GHz. Both the line survey and continuum map can be downloaded from the ALMA SV website at https://almascience.nrao.edu/alma-data/science-verification.

\section{Modeling Methodology}
\label{s-model}

We modeled the emission of each detected molecule, including isotopologues, one at a time. Summing all of the individual fits, thus, yielded the total molecular emission, i.e. the ``full band model".  This procedure was carried out by multiple individuals simultaneously, each person modeling several molecules and incorporating the best fit results for other species as determined by other participants, thereby allowing for blended lines to be more quickly identified. The molecular emission was fit using two programs: XCLASS\footnote{http://www.astro.uni-koeln.de/projects/schilke/XCLASS} and MADEX \citep{cernicharo12}. We note that two species, CO and H$_{2}$O (main isotopologues only), were too optically thick to model with either program as described below. Consequently, we fit Gaussian profiles to these species and include those fits in the full band model.

We use both the HIFI and IRAM surveys to constrain our molecular fits. By combining these datasets, we are essentially modeling the entire spectrum of a given molecule including low energy states, i.e. ground state transitions or close to it, up to energy levels where emission is no longer detected. This is true even for lighter molecules, with widely spaced transitions. Because of the extremely large number of observed lines in the HIFI survey ($\sim$~\nlines; see Sec.~\ref{s-lstats}), we do not compile line lists for detected molecules. Rather, we provide a model spectrum for each detected species, from which line intensities for individual transitions can be obtained.

\subsection{Description of Orion KL Components}
\label{s-orion}

\begin{figure}
\epsscale{0.93}
\plotone{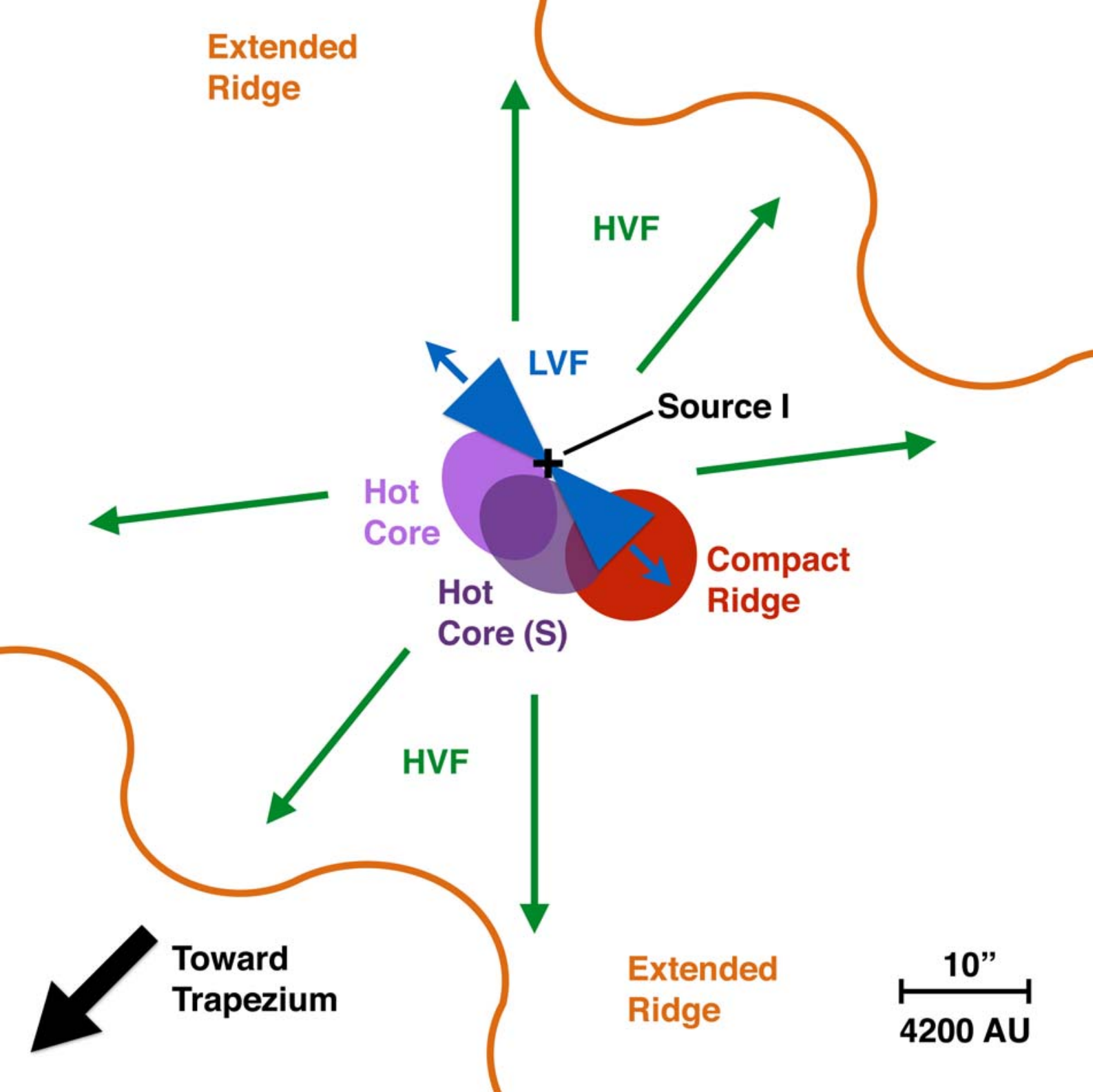} 
\caption{A cartoon illustrating the morphology of Orion~KL as seen in the plane of the sky. The spatial/velocity components, which are labelled in the figure, are represented by different colors. The location of radio source I is indicated with a black cross, and the direction of the outflowing gas associated with the LVF and HVF are represented by blue and green arrows, respectively. For orientation, the location of the Trapezium, which lies outside the spatial extent of the cartoon, is indicated with a black arrow. An approximate size scale in arc seconds and AU, assuming a distance of 420~pc, is also given in the lower right hand corner. \label{p-cartoon}}
\end{figure}

\begin{figure}
\epsscale{1.0}
\plotone{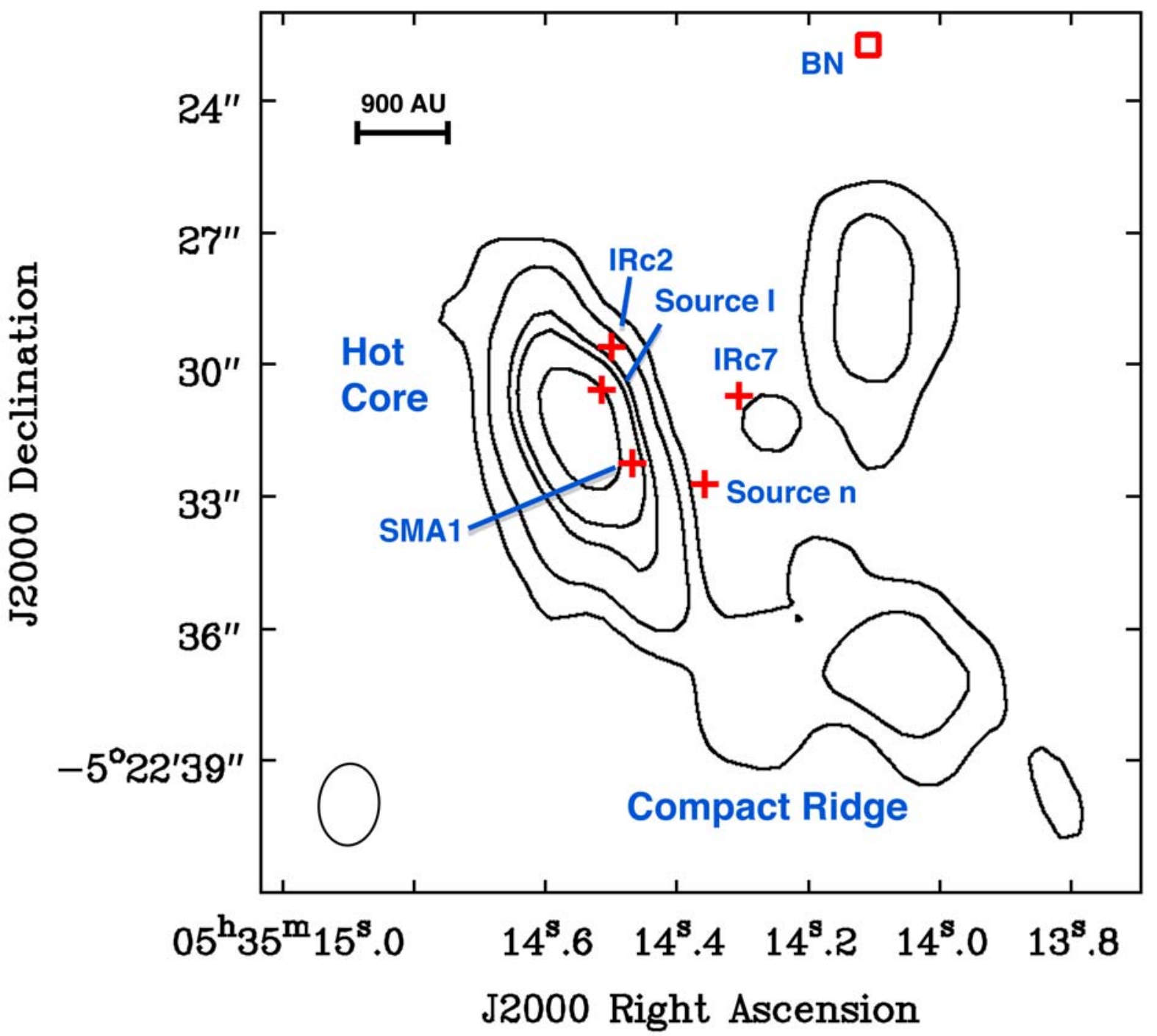}
\caption{The continuum map of Orion~KL at 230.9~GHz  taken from the ALMA SV survey. The continuum clumps associated with the hot core and compact ridge are labelled for clarity. Locations of various sources, which are labeled in the plot, are indicated by red crosses (see the text in Sec.~\ref{s-orion} for details concerning these sources). The position of BN is also represented with a red square. The contour levels correspond to (0.1, 0.2, 0.4, 0.5, 0.75) $\times$ 1.334 Jy beam$^{-1}$. A size scale in AU, assuming a distance of 420~pc, is given in the upper left corner and the size of the synthesized beam is indicted by an oval in the lower left corner. 
\label{p-closeup}}
\end{figure}

Molecular line emission toward Orion KL originates from several distinct spatial/velocity components. These components are typically labeled as the ``hot core", ``compact ridge", ``plateau", and ``extended ridge" \citep{blake87}. Using HIFI's high spectral resolution, we are able to differentiate these components even though they all lie within the \herschel\ beam because they have line profiles with significantly different velocities relative to the Local Standard of Rest, \vlsr, and full widths at half maximum, \dv\ \citep{blake87}. A cartoon illustrating the spatial distribution of these components in the plane of the sky is given in Fig.~\ref{p-cartoon} with the different spatial/velocity components labelled. Additionally, we present the ALMA SV continuum map at 230.9~GHz in Fig.~\ref{p-closeup} which shows more detailed structure toward Orion KL on smaller spatial scales. The continuum clumps associated with the hot core and compact ridge are labelled in Fig.~\ref{p-closeup} for clarity. We briefly describe each component in the following. When spatial scales (in AU) are reported, we assume an Orion~KL distance of 420~pc \citep{menten07, hirota07}.

The hot core, so named because it is a hot (\tkin~$\gtrsim$~150~K) and dense ($\gtrsim$10$^{7}$~\cmc) clump of gas which may be harboring one or more massive protostars, is characterized by lines with \vlsr~$\approx$~4 -- 6~km/s and \dv~$\approx$~7--12~km/s. This region, originally detected via inversion lines of ammonia \citep[NH$_{3}$;][]{ho79}, is in general rich in nitrogen bearing molecules. Interferometric observations of methyl cyanide (CH$_{3}$CN) and NH$_{3}$ reveal an intricate, clumpy structure on size scales $\lesssim$~1--2\arcsec\ ($\lesssim$~800~AU) and a non-uniform temperature distribution with measured rotation temperatures varying between $\sim$150 -- 600~K \citep{wang10, goddi11}. The ultimate heating source, however, remains unclear. Both \citet{zapata11} and \citet{goddi11} conclude that the Orion KL hot core is most likely externally heated, though they disagree on the source, while \citet{devicente02} argues for internal heating by an embedded massive protostar. 

The compact ridge is a group of dense clumps of quiescent gas, which are likely externally heated \citep{wang11, favre11}, located approximately 12\arcsec\ ($\sim$~5000~AU) south-west of the hot core (Figs.~\ref{p-cartoon} and \ref{p-closeup}). This region is characterized by high densities ($\sim$~10$^{6}$~\cmc) and temperatures of $\sim$~80--150~K \citep{blake87}. Line profiles emitted from the compact ridge have narrow line widths, \dv~$\approx$~3 -- 6~km/s and line centers at \vlsr~$\approx$~7 -- 9~km/s. Compared to the hot core, the compact ridge is much richer in complex oxygen bearing organics \citep{blake87, friedel08, beuther05}. 

The plateau, which is characterized by wide line widths (\dv~$\gtrsim$~20~km/s) at \vlsr~$\approx$~7 -- 11~km/s, includes at least two outflows often referred to as the low velocity flow (LVF) and high velocity flow (HVF).  The LVF is oriented along a NE -- SW axis \citep{genzel89, blake96, stolovy98, greenhill98, nissen07, plambeck09, goddi09b}, and is thought to be driven by radio source I, an embedded massive protostar with no sub-mm or IR counterpart \citep{menten95, plambeck09}. The HVF, on the other hand, is more spatially extended ($>$~30\arcsec) than the LVF and is oriented along a NW--SE axis, perpendicular to the LVF \citep{allen93, chernin96, schultz99, odell01, doi02, nissen12}. Fig.~\ref{p-cartoon} illustrates the relative orientation of these two outflows. The most compact part of the LVF as traced by interferometric observations of SiO \citep{plambeck09} is indicated by a blue hour-glass, though we note the full spatial extent of the LVF is somewhat larger ($\sim$~30\arcsec, see Sec.~\ref{s-xclass}). There have been several suggestions as to the ultimate power source behind the HVF. Vibrationally excited transitions of methanol (CH$_{3}$OH), cyanoacetylene (HC$_{3}$N), and sulfur dioxide (SO$_{2}$) have been detected toward the sub-mm source SMA1 \citep{beuther04}, possibly indicating the presence of an embedded protostar, which \citet{beuther08} suggest may be driving the HVF. \citet{plambeck09}, on the other hand, argue that the HVF is merely a continuation of the LVF. Yet another possibility is that the HVF is powered by the dynamical decay of a multi star system possibly involving radio source I, IR source n, and BN \citep{rodriguez05, gomez05, gomez08, zapata09, bally11, nissen12}. Both source n and BN are themselves strong IR continuum emitters which likely harbor embedded self-luminous sources \citep{becklin67, lonsdale82, menten95, gezari98, debuizer12}. For orientation, we indicate the positions of radio source~I, SMA1, IR source~n and BN in Fig.~\ref{p-closeup}. We also indicate the locations of two ``infrared clumps", IRc2 and IRc7, which are adjacent to the hot core \citep{rieke73, gezari98}.

The extended ridge represents the most widespread, quiescent gas toward Orion KL ( Fig.~\ref{p-cartoon}). Measured rotation temperatures are typically $\lesssim$~60~K and the line profiles are narrow, \dv~$\approx$~2 -- 4~km/s, with line centers at \vlsr~$\approx$~8 -- 10~km/s \citep{blake87}. The extended ridge is rich in unsaturated, carbon rich species indicating the dominance of exothermic ion-molecule reactions that do not require activation energies \citep{herbst73, watson73, smith92, ungerechts97}. 

\subsection{Hot Core South}
\label{s-hcs}

In the course of modeling the data, we noticed that several molecules contained a spectral component with \dv~$\sim$~5 -- 10~km/s and \vlsr~$\sim$~6.5 -- 8~km/s, in between line parameters typically associated with the hot core and compact ridge. The presence of this type of component was noted previously by 
\citet{neill13a} who present a detailed analysis of water and HDO emission in the Orion KL HIFI survey. Specifically, HD$^{18}$O lines detected in the HIFI scan have an average \vlsr~=~6.7~km/s and \dv~=~5.4~km/s, consistent with this ``in between" component. Using HDO interferometric maps obtained from the ALMA-SV line survey of Orion KL (Sec.~\ref{s-obs_alma}), the 
\citet{neill13a} study showed that this emission likely originates from a high water column density clump approximately 1\arcsec\ south of the hot core sub-mm continuum peak. 

\begin{figure}
\epsscale{1.0}
\plotone{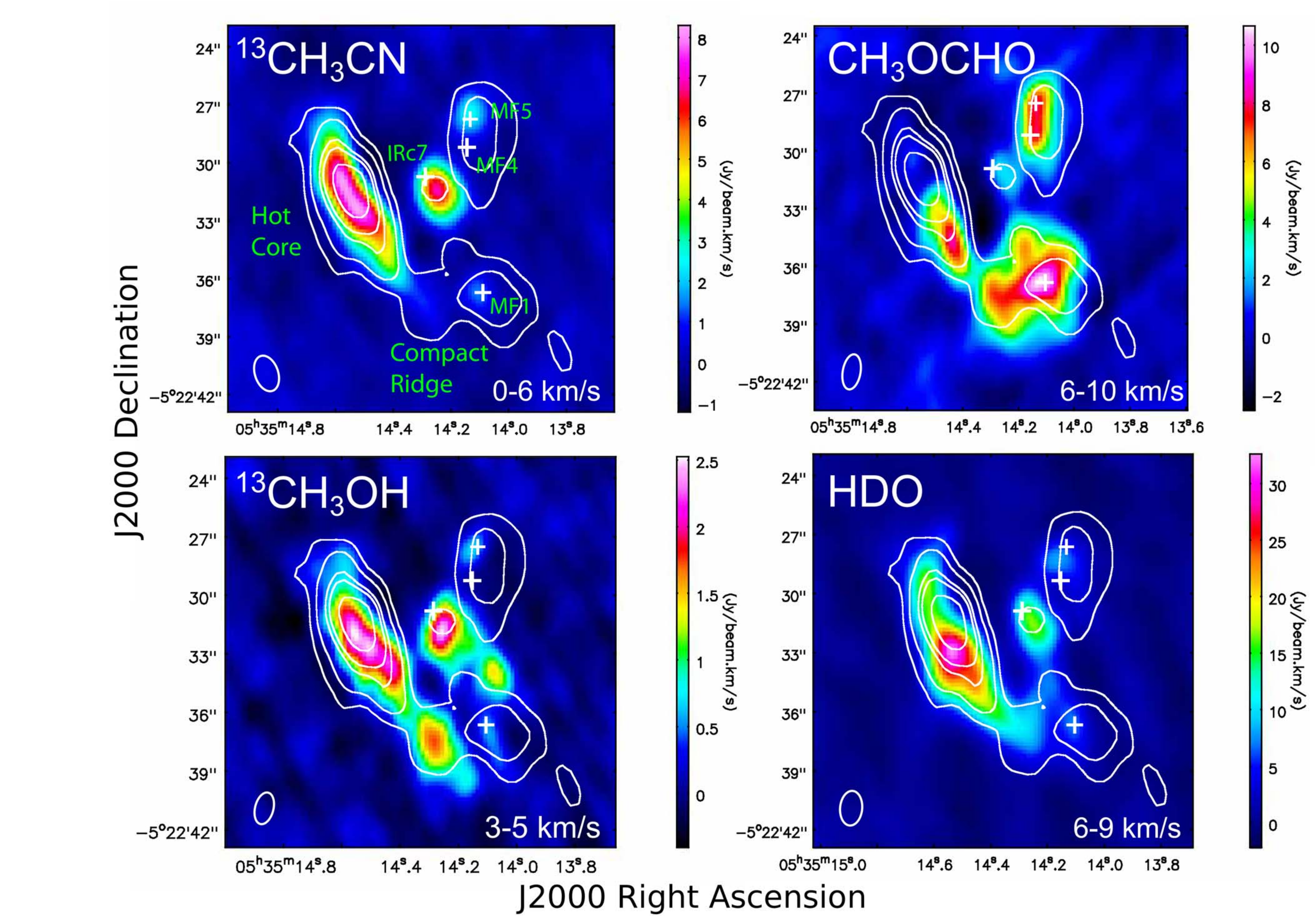}
\caption{Integrated intensity maps obtained from lines in the ALMA-SV dataset plotted as a color scale. The velocity range over which the intensity is integrated is given in each panel. The transitions plotted are $^{13}$CH$_{3}$CN 13$_{2}$ -- 12$_{2}$ (upper left), $^{13}$CH$_{3}$OH-E 5$_{2}$ -- 4$_{2}$ (lower left), HDO 3$_{1,2}$ -- 2$_{2,1}$ (lower right), and CH$_{3}$OCHO 19$_{6,13}$ -- 18$_{6,12}$ (upper right). We only show integrated emission in the range 3 -- 5~km/s for $^{13}$CH$_{3}$OH because we wanted to avoid contamination from the compact ridge. The continuum at 230.9~GHz is overlaid in each panel as white contours. The contour levels correspond to (0.1, 0.2, 0.4, 0.5, 0.75) $\times$ 1.334 Jy beam$^{-1}$. The synthesized beam size is indicated by an oval in the lower left corner of each panel. \label{p-alma_maps}}
\end{figure}

We employ the ALMA-SV dataset here to map, in addition to HDO, $^{13}$CH$_{3}$OH, another species which contains an ``in between" component in the HIFI scan (\vlsr~=~7.5~km/s, \dv~=~6.5~km/s), $^{13}$CH$_{3}$CN, a hot core tracer with typical line parameters for that region (\vlsr~$\approx$~5.5~km/s, dv~$\approx$~8~km/s), and methyl formate (CH$_{3}$OCHO), a prominent compact ridge tracer also with typical line parameters (\vlsr~$\approx$~8~km/s, \dv~$\approx$~3~km/s). Fig.~\ref{p-alma_maps} contains four panels each plotting an integrated intensity map (color scale) of a transition from one of these molecules. The continuum at 230.9~GHz is overlaid as white contours in each panel. White crosses indicate the locations of IRc7, and methyl formate peaks MF1, MF4, and MF5 in the notation of \citet{favre11}. The sub-mm clump associated with the hot core is also labeled. We used the same data product presented in 
\citet{neill13a} to make Fig.~\ref{p-alma_maps}.

\begin{figure}
\epsscale{1.0}
\plotone{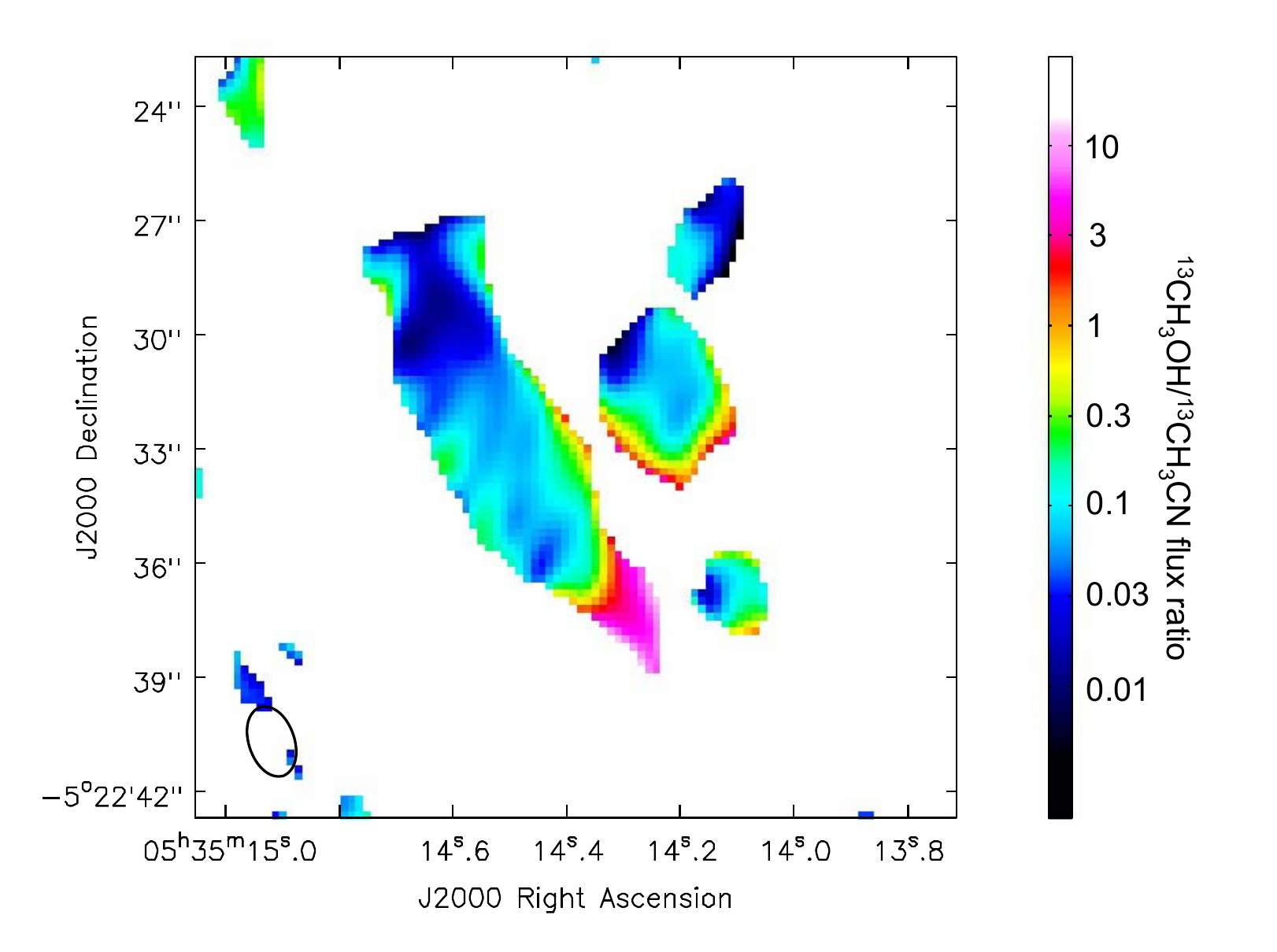}
\caption{The integrated intensity ratio of the $^{13}$CH$_{3}$OH-E map to $^{13}$CH$_{3}$CN given in Fig.~\ref{p-alma_maps} showing the increase in methanol emission relative to methyl cyanide from north to south. The synthesized beam size is indicated by an oval in the lower left corner. \label{p-hc_lratio}}
\end{figure}

From Fig.~\ref{p-alma_maps}, we see that $^{13}$CH$_{3}$CN traces the hot core continuum closely while HDO, as first pointed out by 
\citet{neill13a}, traces a clump $\sim$~1\arcsec\ south of the continuum peak. The $^{13}$CH$_{3}$OH map in Fig.~\ref{p-alma_maps} is integrated in the velocity range 3 -- 5~km/s to avoid emission from the compact ridge, which methanol also traces. As such, Fig.~\ref{p-alma_maps} shows that $^{13}$CH$_{3}$OH emission from the ``in between" component does not originate from the compact ridge as traced by methyl formate. Rather, this emission is strongest just south of where $^{13}$CH$_{3}$CN peaks. Because the difference is more subtle than with HDO, Fig.~\ref{p-hc_lratio} plots the $^{13}$CH$_{3}$OH/$^{13}$CH$_{3}$CN integrated intensity ratio, which shows a clear gradient in $^{13}$CH$_{3}$OH emission relative to $^{13}$CH$_{3}$CN from north to south. Given that HDO and $^{13}$CH$_{3}$OH both trace regions south of the $^{13}$CH$_{3}$CN peak, we assume other molecules displaying emission from this ``in between" component originate from a similar region. We thus label this component ``hot core south" or \hotcores, which we represent schematically in the cartoon presented in Fig.~\ref{p-cartoon}.

\subsection{XCLASS modeling}
\label{s-xclass}

All molecular species are modeled using XCLASS. This program uses both the CDMS \citep[][http://www.cdms.de]{muller01, muller05} and JPL \citep[][http://spec.jpl.nasa.gov]{pickett98} databases to produce model spectra assuming local thermodynamic equilibrium (LTE). Input parameters are the telescope diameter, \td, source size, \ssize, rotation temperature, \trot, total column density, \ntot, line velocity relative to the Local Standard of Rest, \vlsr, and line full width at half maximum, \dv. In order to account for dust extinction, the dust optical depth, \taud, is parameterized by a power law,
\begin{equation}
\tau_{d} = 2 m_{H} (\chi_{\rm dust})  N_{H_{2}} \kappa_{1.3 \rm mm} \left(\frac{\nu}{230\ GHz} \right)^{\beta},
\label{e-taud}
\end{equation}
where \nhh\ is the \hh\ column density, \kapd\ is the dust opacity at 1.3mm (230~GHz), \betad\ is the spectral index, $m_{\rm H}$ is the mass of a hydrogen atom, and $\chi_{\rm dust}$ is dust to gas mass ratio. As outlined below, we hold \nhh\ and \ssize\ fixed for a given spatial/velocity component, but note that there are several exceptions in which we varied \ssize\ to obtain better agreement between the models and data. These instances are explained in Sec.~\ref{s-indiv}. We also set \td~=~3.5~m and 30~m when comparing our models to the HIFI and IRAM surveys, respectively. We varied \trot, \ntot, \vlsr, and \dv\ as free parameters. Additional information regarding XCLASS, e.g. specific equations used in the code, can be found in \citet{comito05} and \citet{zernickel12}. 

We assume that all molecules emitting from the same spatial/velocity component have the same source size, which is a simplifying assumption. The aim of this study, however, is not a detailed analysis of any single molecule. It is a holistic analysis of the entire spectrum. This is therefore a reasonable approximation and is in line with previous spectral survey papers of Orion KL \citep[see e.g.][]{tercero10, tercero11}. Adopted source sizes for each spatial/velocity component are given in Table~\ref{t-standard}. We estimate \ssize\ for the hot core and compact ridge using interferometric observations from \citet{beuther08} and \citet{favre11}, respectively. The plateau source size was obtained from \herschel/HIFI water maps taken as part of the HEXOS program (Melnick et al. 2014, in preparation). Finally, we assume a source size of 180\arcsec\ for the extended ridge to reflect the fact that the extended ridge completely fills the \herschel\ beam at all frequencies. For several molecules, we were forced to use \ssize\ values that differed from those given in Table~\ref{t-standard}. These deviations are explained in the descriptions of individual molecular fits presented in Sec.~\ref{s-indiv}. Estimates of \nhh\ for each spatial/velocity component are also given in Table~\ref{t-standard}. We obtained these values from \citet{plume12}, who use C$^{18}$O lines within the Orion~KL HIFI scan to derive total C$^{18}$O column densities which they convert to \nhh\ estimates by assuming a CO abundance of 1.0\sn$^{-4}$ and $^{16}$O~/~$^{18}$O~=~500. We modified the \hh\ column densities for the compact ridge and plateau because the Plume study assumed source sizes for these components that are different from what we adopt here. Consequently, we recalculated the C$^{18}$O upper state columns assuming the source sizes used in this study and applied the same correction factors reported by \citet{plume12}.

\begin{figure}
\epsscale{0.78}
\plotone{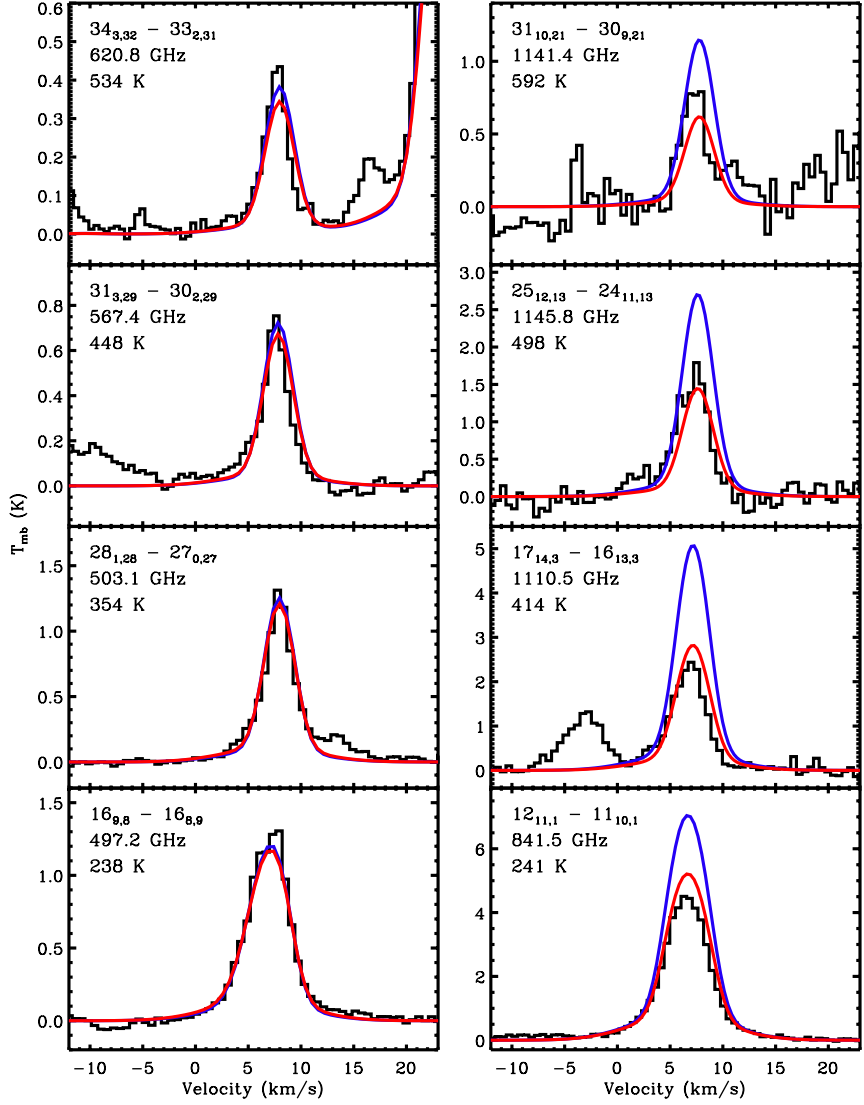}
\caption{A sample of 8 \de\ lines from the HIFI scan. The data are plotted in black and the quantum numbers, rest frequency, and \eup\ for each transition are labeled from top to bottom in each panel. We use the quantum numbers from the EA spin isomer, but note that transitions from AA, EE, AE, and EA-\de\ are blended together at HIFI frequencies. The red and blue lines represent XCLASS models which assume \nhh~=~2.5\sn$^{24}$~\cms\ and 3.9\sn$^{23}$~\cms\ and set \ntot~=~6.5\sn$^{16}$~\cms\ and 5.9\sn$^{16}$~\cms, respectively. Both models set \trot~=~110~K and \ssize~=~10\arcsec. Dimethyl ether is also detected toward the hot core, but this component is not emissive for the transitions plotted here. \label{p-dtau_de}}
\end{figure}

\begin{figure}
\epsscale{0.78}
\plotone{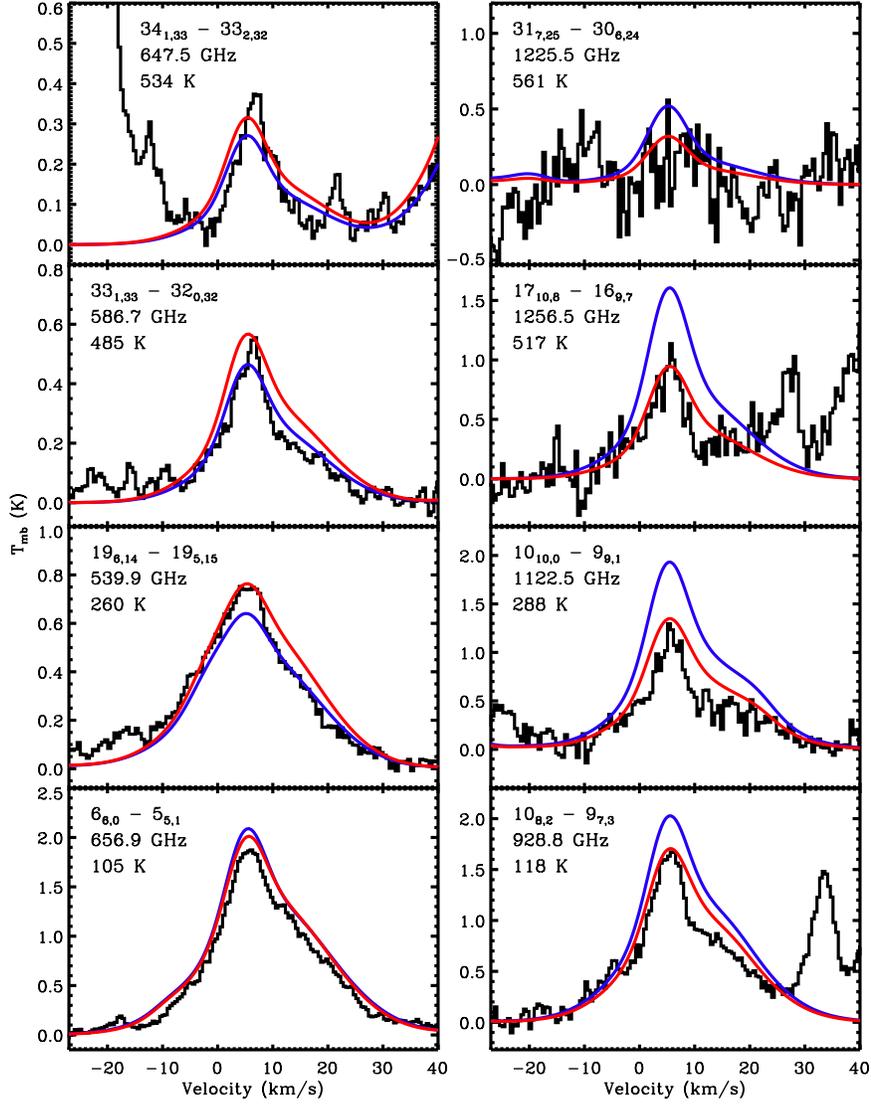}
\caption{A sample of 8 $^{34}$SO$_{2}$ lines from the HIFI scan. The data are plotted in black and the quantum numbers, rest frequency, and \eup\ for each transition are labeled from top to bottom in each panel. The red and blue lines represent XCLASS models which assume \nhh~=~2.5\sn$^{24}$~\cms\ and 3.9\sn$^{23}$~\cms, respectively. Each model has a hot core and plateau component. The model plotted in red sets \ntot~=~5.0\sn$^{15}$ and 4.7\sn$^{15}$~\cms, while the model plotted in blue sets \ntot~=~3.5\sn$^{15}$ and 3.0\sn$^{15}$~\cms\ for the hot core and plateau components, respectively. Both models set \trot~=~240~K and 150~K and \ssize~=~10\arcsec and 30\arcsec\ for the hot core and plateau components, respectively. There are additional cooler sub-components present in these models to fit the emission at lower excitation for both the hot core and plateau. These sub-components, however, are not significantly emissive for the plotted transitions. \label{p-dtau_so2}}
\end{figure}

For the dust extinction power law, we assume \kapd~=~0.42~cm$^{2}$~g$^{-1}$, corresponding to the midpoint between bare grains and grains with thin ice mantles \citep{ossenkopf94}. We also set \betad~=~2 and $\chi_{\rm dust}$~=~0.01. In the course of modeling the data, we found that the \nhh\ values given Table~\ref{t-standard} tended to underestimate the extinction necessary to reproduce the emission in the highest frequency bands where the dust optical depth is highest. In other words, molecules fit well at frequencies below $\sim$~1~THz always tended to be over predicted at higher frequencies. As a result, we use a higher \nhh\ estimate to compute the dust optical depth. We adopt a value of \nhh~=~2.5\sn$^{24}$~\cms\ for the hot core, compact ridge, and plateau. Because the extended ridge represents lower density gas that is not as heavily embedded, we retain the \nhh\ value given in Table~\ref{t-standard} for this spatial/velocity component. Fig.~\ref{p-dtau_de} plots 8 transitions of dimethyl ether (\de), a prominent compact ridge tracer. The panels are organized so that both columns span a range in upper state energy, \eup, from $\sim$~200 to 600~K with \eup\ increasing from bottom to top. Lines in the left and right columns occur at frequencies below and above 800~GHz, respectively. Transitions in the left column are therefore less affected by dust extinction than the right with both columns covering similar ranges in \eup. The red line corresponds to an XCLASS fit which sets \nhh~=~2.5\sn$^{24}$~\cms, while the blue line represents a similar model that assumes the \hh\ column density given in Table~\ref{t-standard}. Dimethyl ether column densities for the higher and lower extinction fits are 6.5\sn$^{16}$~\cms\ and 5.9\sn$^{16}$~\cms, respectively. Both models set \trot~=~110~K and \ssize~=~10\arcsec. From the plot, we see that the model with greater dust extinction fits the data better than the lower extinction model over all frequencies. Because similar ranges in \eup\ are covered at low and high frequencies, it is not possible to improve the fit at lower extinction by changing either \trot\ or \ssize. We observed the same trend for molecules detected toward the hot core and plateau. Fig.~\ref{p-dtau_so2} plots a sample of 8 transitions of $^{34}$SO$_{2}$, a molecule with strong hot core and plateau components, organized in the same way as Fig.~\ref{p-dtau_de}, with red and blue lines representing an analogous set of XCLASS models. We again see better agreement for the higher extinction model. The \nhh\ we adopt to compute \taud\ is between 6 and 9 times larger than the \hh\ column densities derived toward the hot core, compact ridge, and plateau using C$^{18}$O line emission, but is commensurate with other \nhh\ estimates derived from mm and sub-mm observations which report \nhh~$\gtrsim$~10$^{24}$~\cms\ \citep{favre11, mundy86, genzel89}. This difference could be resolved by assuming a higher dust opacity, i.e. increasing \kapd\ by the same factor \nhh\ is reduced (see Eq.~\ref{e-taud}). However, because this adjustment would result in the same \taud, we use the higher \nhh\ and conclude a dust optical depth which obeys the relation,
\begin{equation}
\tau_{d} = 3.5 \times 10^{-2} \left(\frac{\nu}{230\ GHz} \right)^{2}
\end{equation}
produces the required dust extinction to accurately fit the data across the entire HIFI band.

We fit an XCLASS model to the emission of each molecule by first selecting a sample of transitions with varying line strengths that covered the entire range in excitation energy over which emission was detected. For simple species, with relatively few lines, this was straightforward. For more complex organics, however, we had many lines, thousands in some cases, from which to choose. Care was taken to select lines that were not blended with any other species. This was done by overlaying the full band model and observed spectrum while we selected transitions on which to base our fit. Because the full band model is the sum of all molecular fits, it evolved as the individual molecular fits changed. Once a sample of lines was selected, each transition was plotted simultaneously in a different panel with the panels arranged so that the upper state energy increased from the lower left panel to the upper right. The emission was then fit by varying the free parameters (i.e. \trot, \ntot, \vlsr, and \dv) by hand until good agreement, as assessed by eye, was achieved between the data and models at all excitation energies. At this point, we computed a reduced \chisq\ metric (described in the Appendix) for each model across the entire HIFI band in order to determine how well the models reproduced the data relative to one another, and to identify those fits which could be improved. Model fits were then revised iteratively. 

We found that automated fitting algorithms had difficulty reaching reasonable solutions, especially when multiple spatial/velocity components and/or temperature gradients were required. Weaker species (\tpeak~$\lesssim$~1~K) that had observed line intensities close to the noise, combined with the prevalence of line blends presented additional difficulties in assessing the goodness of fit with these algorithms. As a result, we derived most models by hand. However, in some instances, when a sample of strong unblended transitions was available, we employed the MAGIX program \citep{moller13}, which optimizes the output of other numerical codes (XCLASS in this case), to automate the fitting process using a Levenberg-Marquardt algorithm. MAGIX utilizes a subset of observed transitions supplied by the user to assess the goodness of fit for a given molecule. Consequently, we carried out by hand alterations to these fits once reduced \chisq\ calculations were performed over the entire HIFI band as described above.

When we observed more than one isotopologue for a given molecule, effort was made to produce models which used consistent values for \trot, \dv, and \vlsr\ for all isotopic species. We, however, sometimes made small adjustments to these parameters to get the optimum fit. The major difference between the models is thus the column density, the ratio of which should be equal to the isotopic abundance ratio. Among rarer isotopologues, isotopic ratios inferred from our models are commensurate with those derived previously toward Orion KL \citep{tercero10, blake87} indicating optically thin emission. Our models, however, also indicate that many of the most abundant isotopologues are optically thick, meaning our models likely underestimate \ntot. Furthermore, we often had to fit very optically thick isotopologues with higher rotation temperatures than their more optically thin counterparts in order to reproduce the observed line intensities over all energies. Emission from these species therefore is too optically thick from which to derive reliable \trot\ and \ntot\ values using XCLASS. As a result, these models serve mainly as templates for the molecular emission. Species that fall into this category are marked with an ``X" in Table~\ref{t-thick}. This table is broken down by spatial/velocity component because a particular molecule may not be optically thick in all of its components. 

While modeling the hot core and plateau with XCLASS, we found that, for some molecules, a single temperature fit failed to reproduce the observed emission, which suggests the presence of temperature gradients in these components. This was most apparent when trying to simultaneously fit both the IRAM and HIFI data. Because the IRAM data, in general, probed lower energy transitions compared to HIFI, the IRAM spectra sometimes required additional cooler sub-components in order for a single model to fit both datasets well. We, therefore, included additional sub-components when necessary to simulate temperature gradients. For the hot core, the sub-component responsible for most of the emission in the HIFI scan always had a source size of 10\arcsec. Hotter or cooler sub-components were then added such that the source size increased or decreased by successive factors of two. Temperature gradients were organized such that \trot\ increased as \ssize\ decreased (i.e. the more compact emission is hotter), corresponding to an internally heated clump. We chose this convention based on more detailed non-LTE models presented by \citet{neill13a} and \citet{crockett14} which fit the H2O/HDO and H2S emission, respectively, within the HIFI scan. In order to reproduce the observed emission of these species, their models require enhanced near and far-IR radiation fields relative to what is observed, suggesting the presence of a self-luminous source or sources within the hot core. Moreover, ethyl cyanide (C$_{2}$H$_{5}$CN) transitions observed within the ALMA SV survey show that more highly excited C$_{2}$H$_{5}$CN lines originate from a more compact region than lower lying transitions (Favre et al. 2014, in preparation). For the plateau, we kept a 30\arcsec\ source size for all sub-components to simulate the fact that the plateau fills most of the \herschel\ beam at HIFI frequencies. We did not need temperature gradients to fit the compact ridge or extended ridge in our XCLASS models. 

\begin{figure}
\figurenum{13.13}
\epsscale{0.90}
\plotone{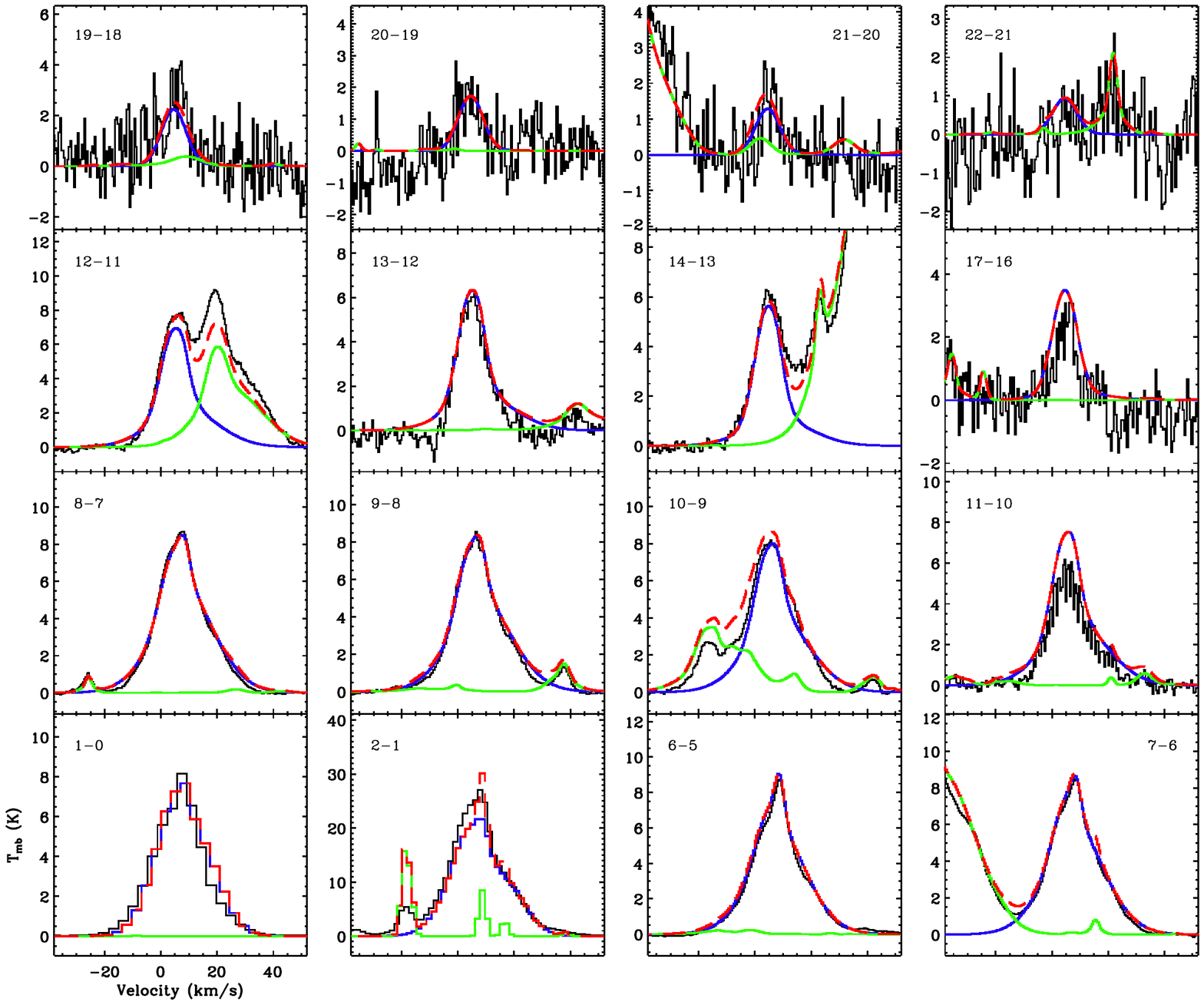}
\caption{Observed transitions in the HIFI and IRAM surveys (black) with the XCLASS model for H$^{13}$CN overlaid as a blue solid line. Modeled emission from all other species (solid green) along with the total fit (dashed red) are also overlaid. Quantum numbers labeling each transition are given in each panel and excitation energy increases from the lower left panel to the upper right. See Sec.~5 for a description of the quantum numbers. \label{p-panels1}}
\end{figure}

\begin{figure}
\figurenum{13.33}
\epsscale{0.90}
\plotone{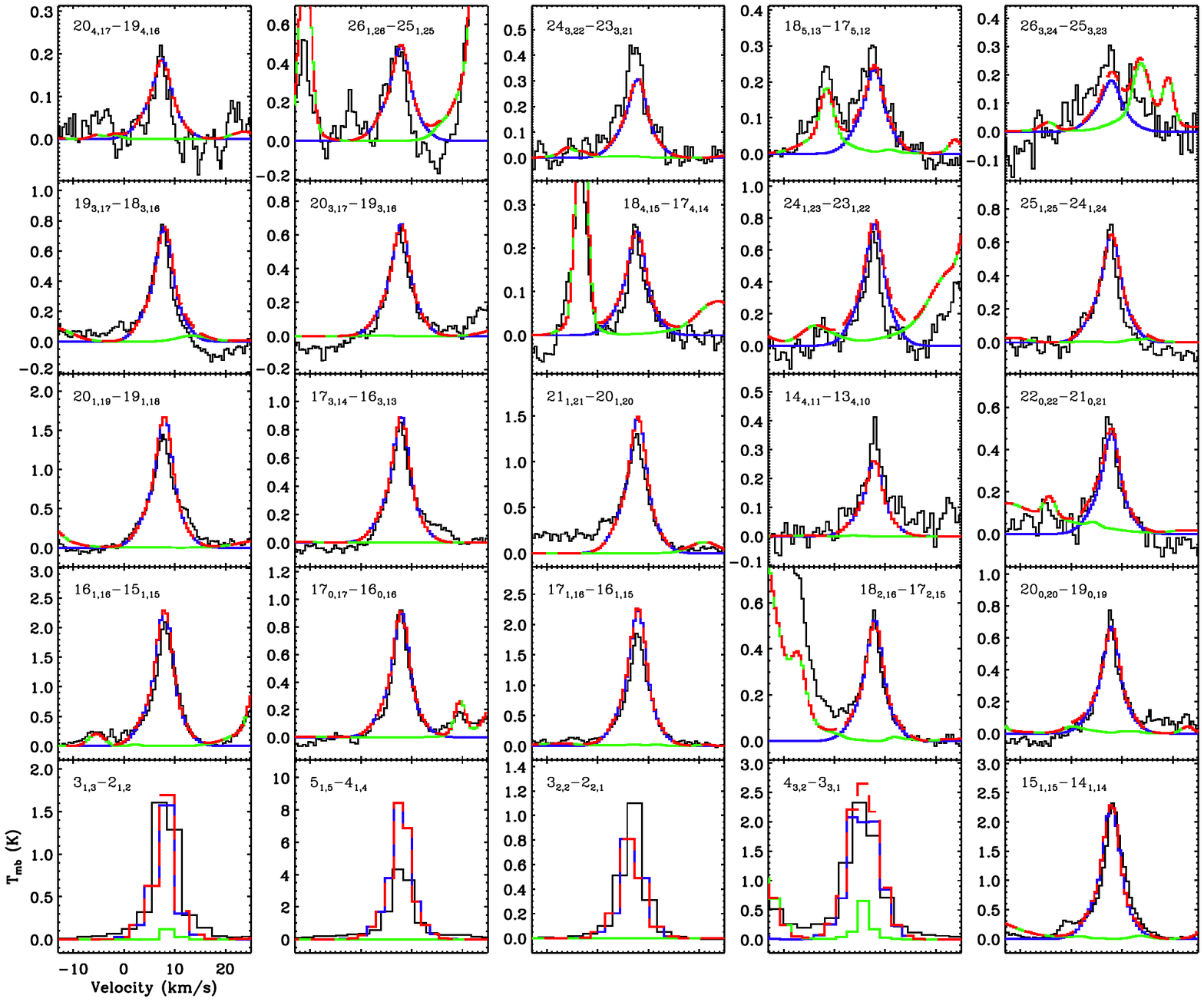}
\caption{Observed transitions in the HIFI and IRAM surveys (black) with the XCLASS model for H$_{2}$CS overlaid as a blue solid line. Modeled emission from all other species (solid green) along with the total fit (dashed red) are also overlaid. Quantum numbers labeling each transition are given in each panel and excitation energy increases from the lower left panel to the upper right. See Sec.~5 for a description of the quantum numbers. \label{p-panels2}}
\end{figure}

\begin{figure}
\figurenum{13.66}
\epsscale{0.90}
\plotone{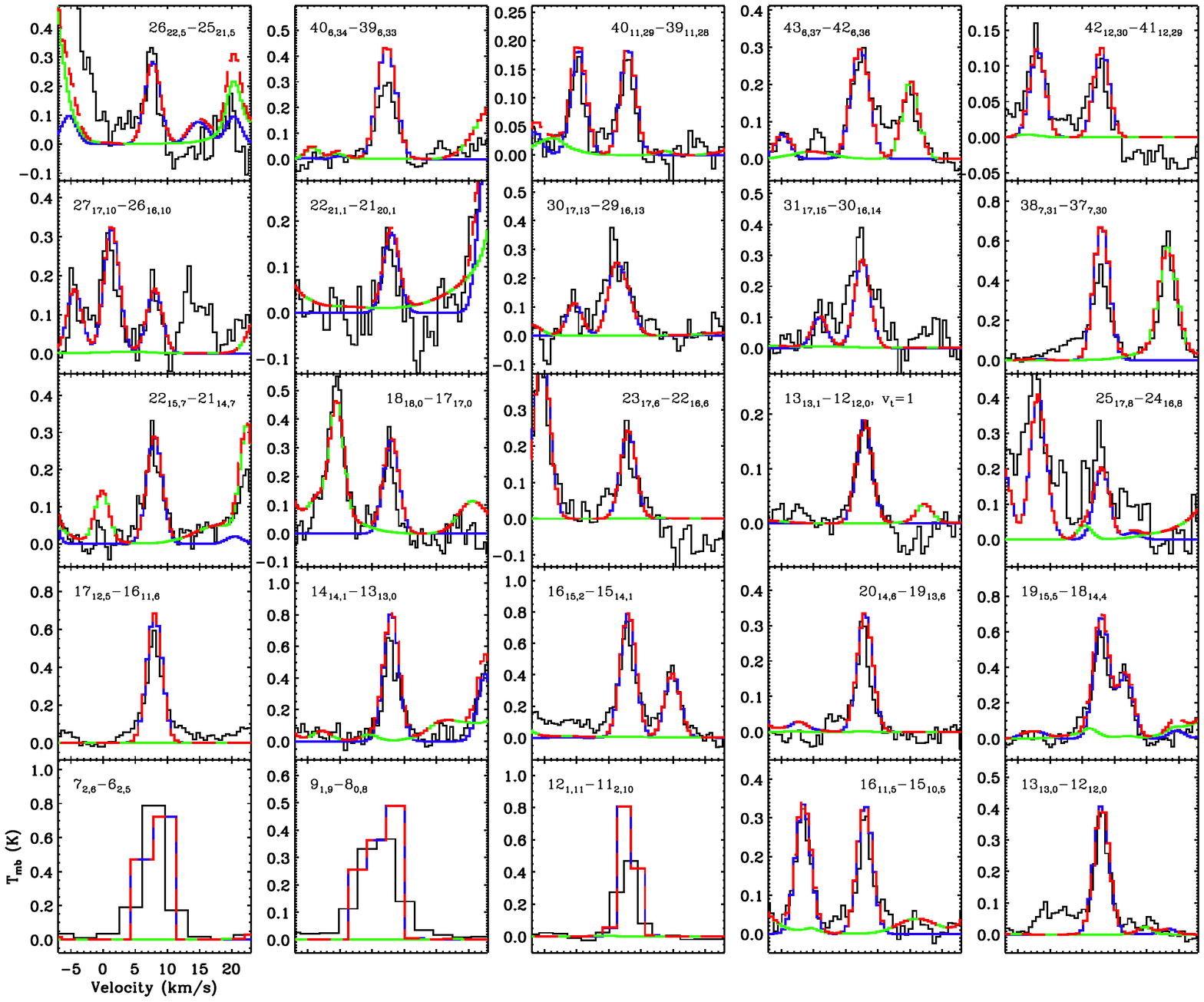}
\caption{Observed transitions in the HIFI and IRAM surveys (black) with the XCLASS model for CH$_{3}$OCHO overlaid as a blue solid line. Modeled emission from all other species (solid green) along with the total fit (dashed red) are also overlaid. Quantum numbers labeling each transition are given in each panel and excitation energy increases from the lower left panel to the upper right. See Sec.~5 for a description of the quantum numbers. \label{p-panels3}}
\end{figure}

Our final XCLASS models are plotted in Figure Set 13 available in the online edition. In this set, each molecule is represented by a figure in which a sample of transitions is plotted covering the entire range in excitation energy over which that species is detected. At least one transition is plotted in each panel, and the panels are organized so that \eup\ increases from the lower left panel to the upper right. The lowest energy panels often show transitions observed in the IRAM scan. The quantum numbers corresponding to the transition at the center of each panel are labeled. The solid blue line represents the XCLASS model for the molecule being considered, the solid green line corresponds to the model emission from all other molecules, and the dashed red line is the model for all detected species (the sum of the former two curves). Figs.~\ref{p-panels1}, \ref{p-panels2}, and \ref{p-panels3} show examples from Figure Set 13, which plot the molecular fits for H$^{13}$CN, a polyatomic linear rotor, H$_{2}$CS, an asymmetric rotor, and CH$_{3}$OCHO, a complex organic, respectively. These figures illustrate the diversity in observed line profiles not only between molecules which trace different spatial/velocity components but also from the same species at different excitation energies. The latter arises because the hot core, compact ridge, plateau, and extended ridge are emissive over different ranges in excitation energy. 

XCLASS model parameters for molecular fits from which we obtain robust \trot\ and \ntot\ information are given in Table~\ref{t-xclass}. We do not include models marked in Table~\ref{t-thick} because they do not provide any physical information. We estimate the uncertainty in our derived \trot\ and \ntot\ values to be approximately 10\% and 25\%, respectively. Our estimated error in \vlsr\ is $\pm$~1~km/s for the hot core, compact ridge, and extended ridge, and $\pm$~2~km/s for the plateau. We also estimate \dv\ errors of $\pm$~0.5~km/s, 1.5~km/s, and 5.0~km/s, for the compact/extended ridge, hot core, and plateau, respectively. These uncertainty calculations are described in the Appendix.

\subsection{MADEX modeling}
\label{s-madex}

A subset of the molecules detected in the HIFI scan were also modeled using the MADEX code. When collisional excitation rates are available, this non-LTE program solves the equations of statistical equilibrium assuming the large velocity gradient (LVG) approximation based on the the formalism of \citet{goldreich74}. MADEX computes transition frequencies and line strengths for most molecules directly from an internal database of rotation constants and dipole moments. For a small fraction ($\sim$6\%) of molecules, however, frequencies and line strengths are taken directly from the JPL and CDMS catalogs. If collision rates do not exist for a given molecular species, model spectra can also be computed assuming LTE. Just as with XCLASS, model input parameters include: \td, \ssize, \ntot, \vlsr, and \dv. Because MADEX is a non-LTE code, the user must also set the kinetic temperature, \tkin, and the \hh\ volume density, \nht. We take \ssize, \tkin, \nht, \ntot, \vlsr, and \dv\ to be free parameters. However, as described in more detail below, we attempt to use consistent \tkin\ and \nht\ values for a given spatial/velocity component, especially when modeling a gradient, and apply \ssize\ values that are similar to those used with XCLASS. We also set \td~=~3.5~m and 30~m when modeling the HIFI and IRAM surveys, respectively, consistent with XCLASS. Additionally, the user can also specify a spatial offset from the center pointing position, \soff, to account for differences in the telescope response. For the hot core and compact ridge we adopt a \soff\ value of 3\arcsec\ and do not apply any spatial offsets for the plateau or extended ridge. These values were obtained from a 2D line survey of Orion KL taken with the IRAM 30~m telescope (Marcelino et al. 2014, in preparation). 

Our modeling approach is similar to previous studies which use MADEX to model molecular emission within the IRAM survey 
\citep[Marcelino et al. 2014, in preparation;][]{tercero10, tercero11, esplugues13a, esplugues13b}. The molecules we model using MADEX are: CH$_{3}$CN, HCN, HNC, HCO$^{+}$, SO, SO$_{2}$, and their isotopologues. These molecules are chosen because they have existing collision rates for states probed by HIFI. This group also includes a complex organic as well as simpler 2 and 3 atom molecules. In addition, these species are detected toward almost every spatial/velocity component. (We do not detect CH$_{3}$CN and SO toward the extended ridge and compact ridge, respectively). Temperature and density gradients surely exist within these components \citep[see e.g.][]{wang11, wang10}. As a result, we model all but the extended ridge and, in some cases, the compact ridge with multiple sub-components which vary both \tkin\ and \nht\ to simulate such gradients. Utilizing MADEX in this way, thus, allows us to compare the column densities, and ultimately molecular abundances, derived from our XCLASS LTE models to more advanced non-LTE calculations, which include both temperature and density gradients.  In particular, we are able to determine if including density gradients significantly affects the determination of molecular abundances toward the different spatial velocity/components. Where possible, we have used the same values for \dv, \vlsr, \ssize, \tkin, and \nht\ for the sub-components, allowing only the column density to vary. In order to obtain better agreement between the data and model, however, slight adjustments were needed in some cases. The need for these adjustments likely indicates the sensitivity of these molecules to the complex underlying physical structure of Orion KL. Just as with XCLASS, these parameters were varied by hand.

Temperature and density gradients are organized in the following way. For the hot core, sub-components have source sizes between 10\arcsec\ and 5\arcsec\ with \nht\ and \tkin\ decreasing with increasing source size, thus simulating an internally heated cloud which is densest closest to the central heating source. For the compact ridge, sub-component source sizes vary between 10\arcsec\ and 20\arcsec\ with temperature and density gradients organized such that \nht\ decreases and \tkin\ increases with increasing source size. This setup corresponds to an externally heated dense clump, roughly approximating the structure of the compact ridge \citep{wang11, favre11}. Sub-components for the plateau have source sizes in the range 10\arcsec\ -- 30\arcsec\ with temperature and density gradients organized in the same way as the hot core. The assumption here being that the denser parts of the outflow subtend a smaller area in the \herschel\ beam. We note that the same source size is used when fitting a temperature gradient to SO and SO$_{2}$ in the plateau, which are both fit with lower (\vlsr~=~6.0~km/s) and higher (\vlsr~=~11.0~km/s) velocity components. Source sizes used in the MADEX models are therefore commensurate with those used in the XCLASS fits. 

\begin{figure}
\figurenum{14.1}
\epsscale{0.90}
\plotone{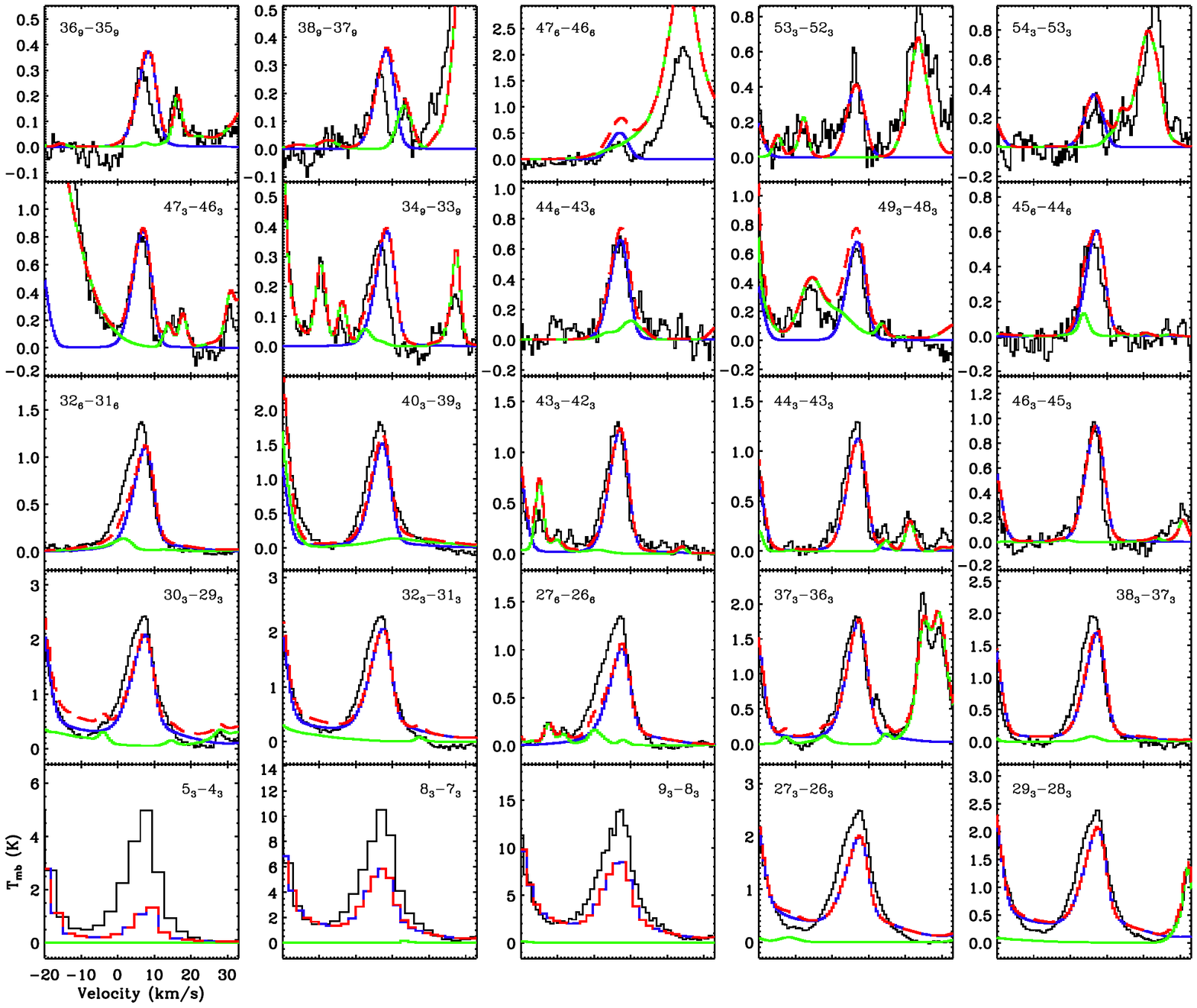}
\caption{Observed transitions in the HIFI and IRAM surveys (black) with the MADEX model for CH$_{3}$CN-A overlaid as a blue solid line. Modeled emission from all other species (solid green) along with the total fit (dashed red) are also overlaid. Quantum numbers labeling each transition are given in each panel and excitation energy increases from the lower left panel to the upper right. See Sec.~5 for a description of the quantum numbers. \label{p-lvg1}}
\end{figure}

\begin{figure}
\figurenum{14.14}
\epsscale{0.90}
\plotone{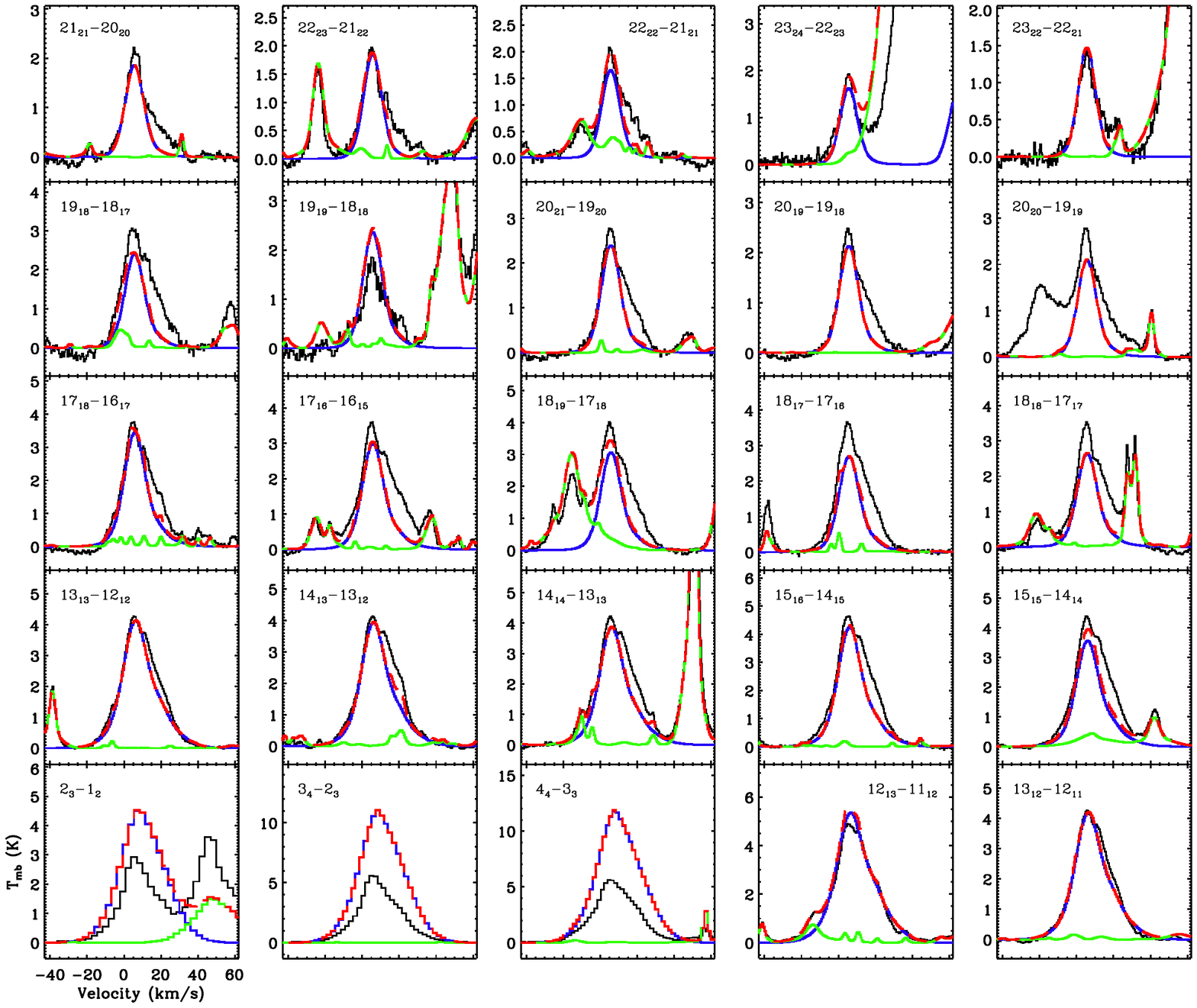}
\caption{Observed transitions in the HIFI and IRAM surveys (black) with the MADEX model for $^{34}$SO overlaid as a blue solid line. Modeled emission from all other species (solid green) along with the total fit (dashed red) are also overlaid. Quantum numbers labeling each transition are given in each panel and excitation energy increases from the lower left panel to the upper right. See Sec.~5 for a description of the quantum numbers. \label{p-lvg2}}
\end{figure}

Our MADEX models are plotted in Figure Set 14 available in the online edition. In this set, we follow the same conventions as Figure Set 13, plotting the same sample of transitions for the MADEX and XCLASS fits. Figs.~\ref{p-lvg1} and \ref{p-lvg2} show samples from Figure Set 14, which plot our models for CH$_{3}$CN-A, a complex organic, and $^{34}$SO, a linear rotor with electronic angular momentum, respectively. Model parameters for those species fit with MADEX are given in Table~\ref{t-madex}. Just as in Table~\ref{t-xclass}, we only list model parameters for the optically thin isotopologues from which reliable column densities can be derived. We estimate the uncertainty in our derived column densities to be approximately 25\%, and adopt the same errors for \vlsr\ and \dv\ as our XCLASS models. The uncertainty in these values are described in more detail in the Appendix. Although \tkin\ and \nht\ are, in principle, free parameters in our MADEX models, we do not report uncertainties for these quantities because, during the modeling process, these values were, for the most part, held fixed for a given spatial/velocity component. That is, we a priori assumed the temperature/density structure of a spatial/velocity component, (via one or more temperature/density sub-components), and only changed these parameters when varying \ntot\ did not significantly improve the fit.

\section{Results}
\label{s-res}

\subsection{The Full Band Model and Line Statistics}
\label{s-lstats}

\begin{figure}
\figurenum{15}
\includegraphics[angle=90, scale=0.63]{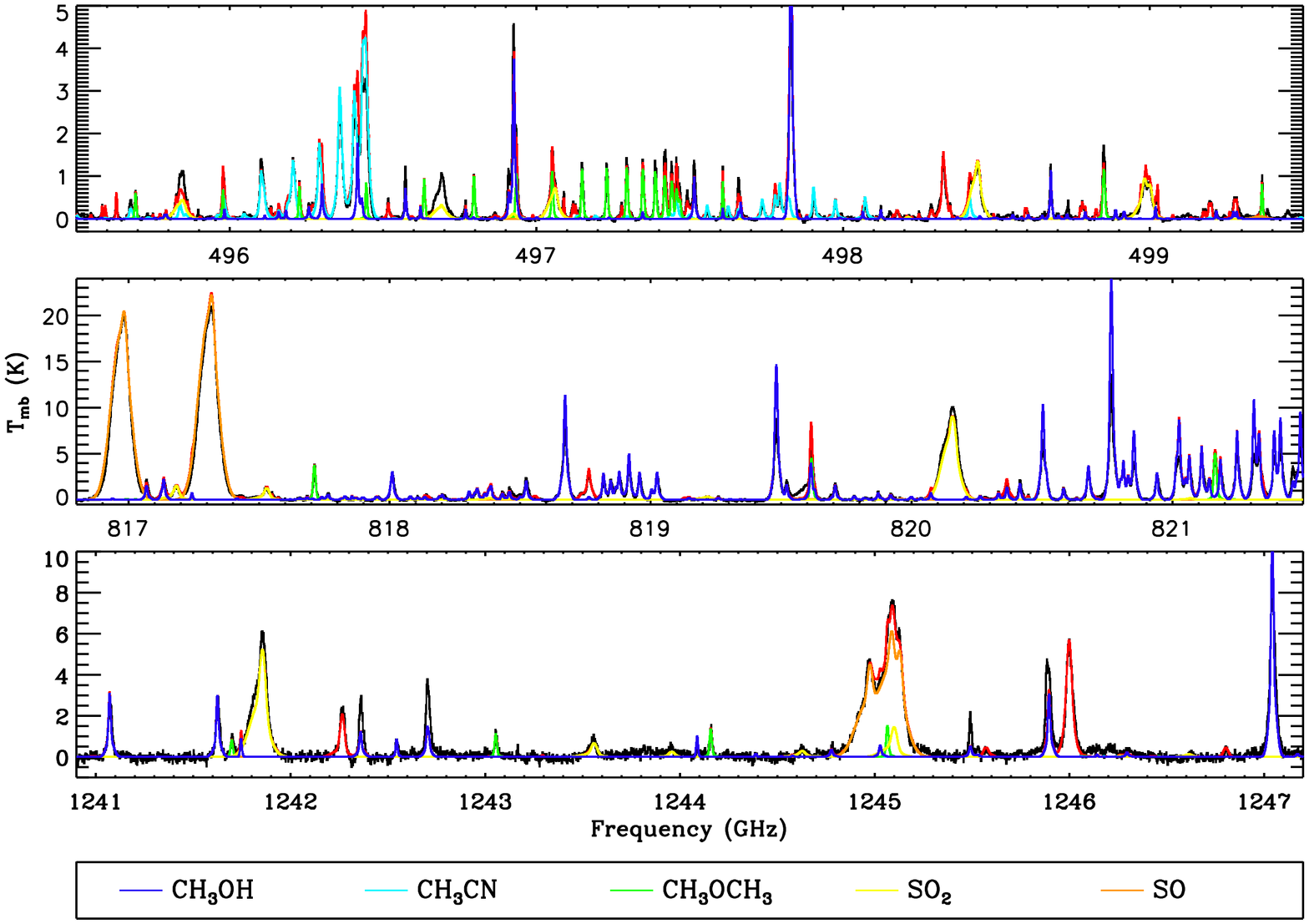}
\caption{Three selected spectral regions from the HIFI survey. The data are plotted in black and the full band model is overlaid in red. Other colors correspond to individual molecular fits which are labeled in a legend at the bottom of the plot. The overlaid individual fits include emission from all detected isotopologues and vibrationally excited states. \label{p-sstrong}}
\end{figure}

\begin{figure}
\figurenum{16}
\epsscale{1.0}
\includegraphics[angle=90, scale=0.63]{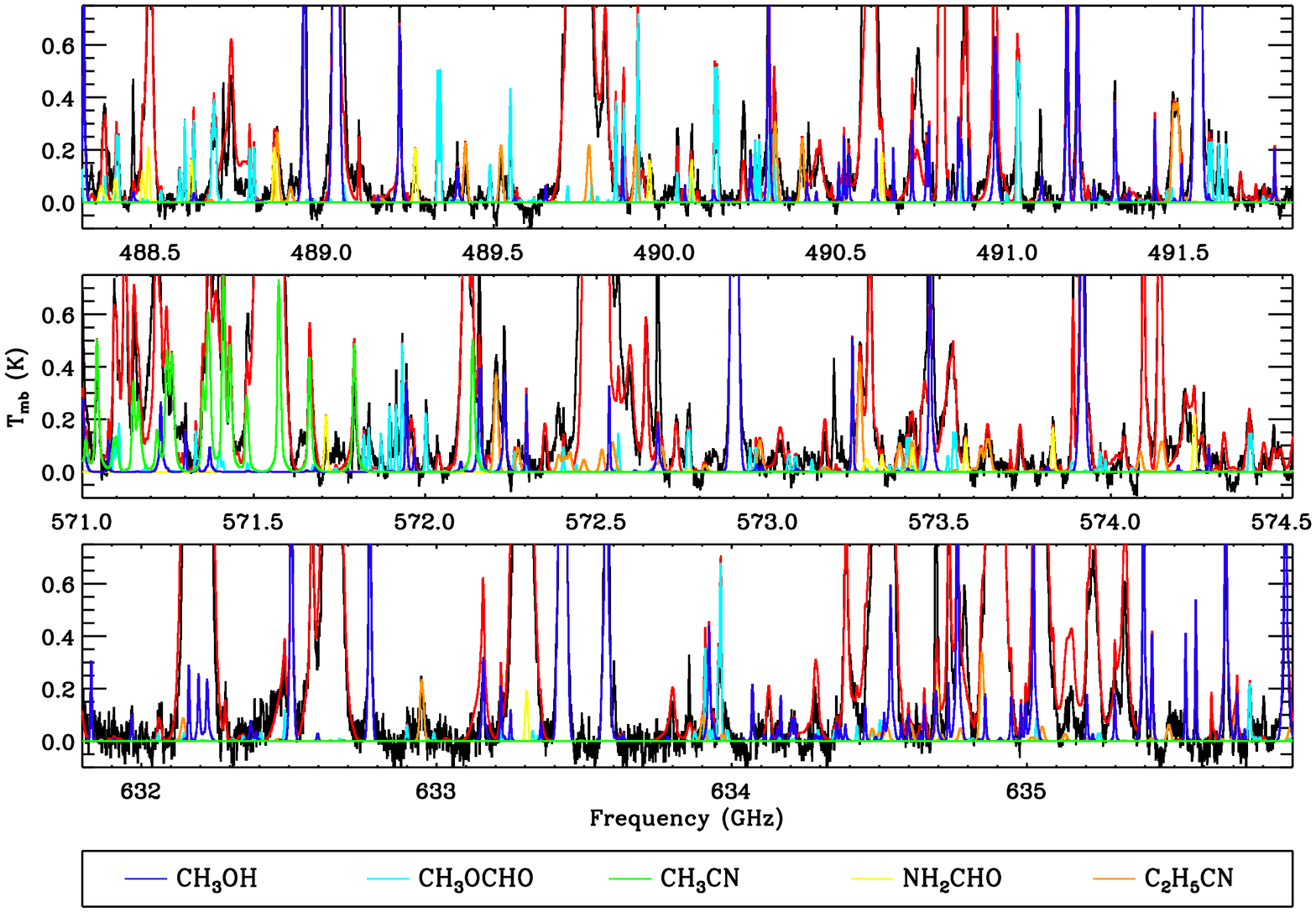}
\caption{Three selected spectral regions from the HIFI survey which highlight weak emission (\tmb~$<$~1~K) originating principally from complex organics. The same conventions as Fig.~\ref{p-sstrong} are followed here. \label{p-sweak}}
\end{figure}

\begin{figure}
\figurenum{17}
\epsscale{1.0}
\includegraphics[angle=90, scale=0.63]{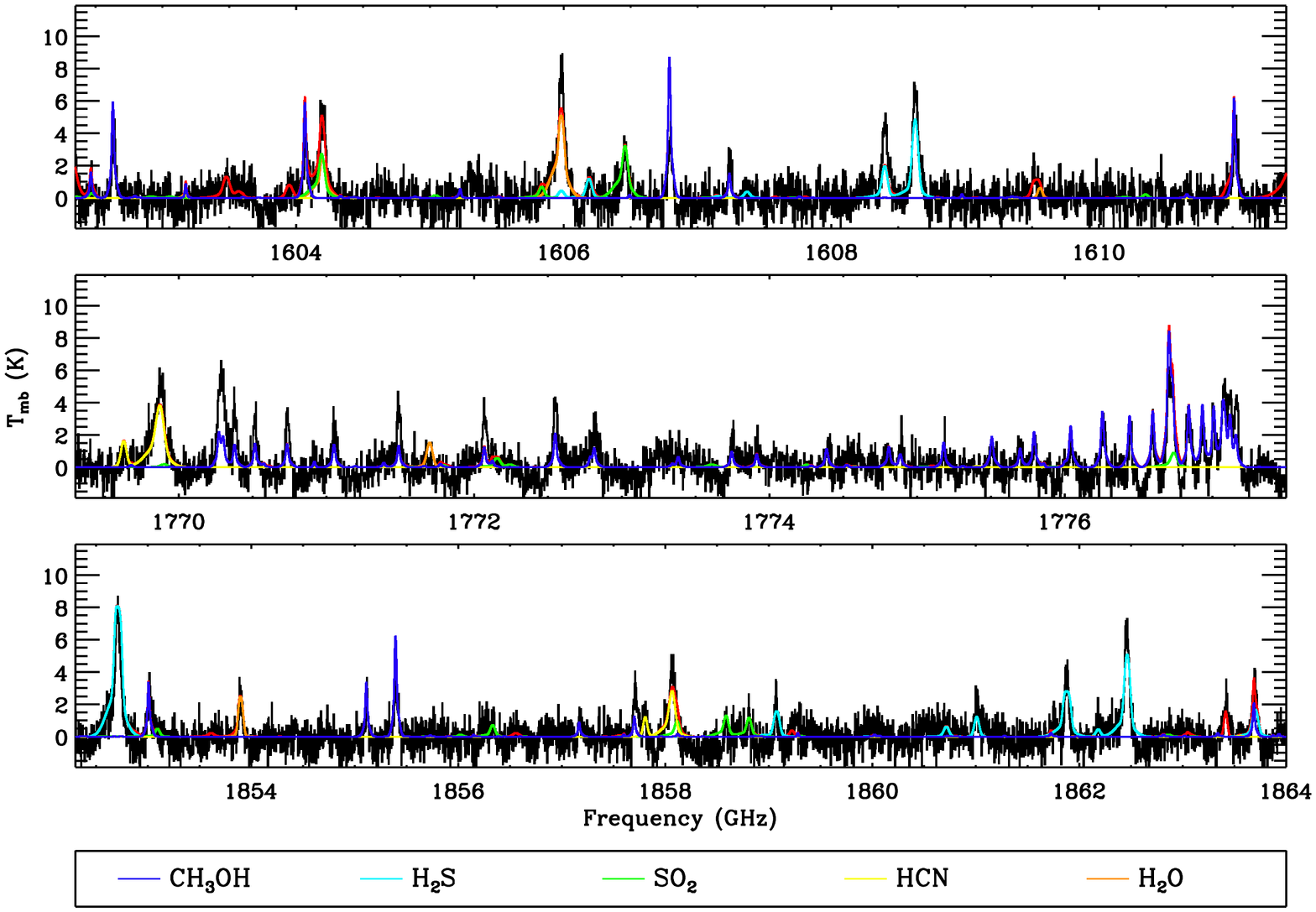}
\caption{Three selected spectral regions from the HIFI survey displaying emission from band 6 and 7, the highest frequency bands in the HIFI survey. The same conventions as Fig.~\ref{p-sstrong} are followed here. \label{p-shf}}
\end{figure}

In total, we detect \niso\ isotopologues of \nmols\ molecules. Emission from each species has been modeled simultaneously over the entire bandwidths of the HIFI and IRAM surveys. We, in general, find excellent agreement between the data and models. Summing the molecular fits, we obtain the full band model for Orion KL. Fig.~\ref{p-sstrong} plots three sections from the HIFI spectrum with the full band model overlaid as a solid red line. Individual models for the five most emissive molecules in these regions are also overlaid as different colors. Fig.~\ref{p-sstrong} focuses on frequencies less than 1280~GHz and intensity scales $\ge$~5~K. The plot thus gives a flavor for how well the full band model reproduces strong lines in bands 1a -- 5b. Fig.~\ref{p-sweak}, on the other hand, plots three regions at the low frequency end of the survey at \tmb\ levels less than 0.8~K, highlighting weak emission fit by the full band model largely from complex organics. A similar sample of three spectral regions in bands 6 and 7, the highest frequency bands in the HIFI survey, is plotted in Fig.~\ref{p-shf}. We see from this plot that the high frequency bands are dominated by emission from lighter species, the exception being CH$_{3}$OH, which is the only complex organic detected at these frequencies. 

In order to quantify how well each molecular fit reproduces the data, we compute a reduced chi squared, \redchisq, statistic for each model. The \redchisq\ calculations are described in the Appendix and reported in Table~\ref{t-chisq} in ascending order along with the database from which we obtained each spectroscopic catalog. Because we are mainly focused on the analysis of the HIFI spectrum in this study, the \redchisq\ statistic is computed only at HIFI frequencies. Our models make a number of simplifying assumptions. First, the molecular fits approximate temperature, and in the case of MADEX, density gradients in a simple way (i.e. adding multiple sub-components). Second, the XCLASS models assume LTE level populations. Third, both XCLASS and MADEX do not include radiative excitation effects which are likely important for some species, especially those with detected vibrational modes. And fourth, we assume the emitting source size does not change as a function of excitation energy, though we tried to mitigate this issue by changing the source size toward certain spatial/velocity components when temperature gradients are invoked. As a result, we do not expect all of our fits to have \redchisq~$\sim$~1. These calculations, however, do convey which models reproduce the data best. From Table~\ref{t-chisq}, we see a range of \redchisq\ values from \nxclass\ XCLASS models. The number of models is larger than \niso, the number detected isotopologues, because, in some instances we fit vibrationally excited emission (see Sec.~\ref{s-vib}) or different spin isomers with separate XCLASS models. We note that 3 XCLASS fits, which model weak or heavily blended emission ($^{13}$C$^{18}$O, OD, and HN$^{13}$C), do not have enough usable channels to compute reliable \redchisq\ values.  Table~\ref{t-chisq} shows that over half, 47 out of 80, of the XCLASS models have \redchisq~$\le$~1.5 indicating excellent agreement between the data and models. Another group of 18 have \redchisq~=~1.6 -- 3.0, which by eye fit quite well, but do not agree with the data as closely as those models with \redchisq~$\le$~1.5. Finally, 15 models have \redchisq~$>$~3.0, which do not reproduce the data as well as the former two categories. 

There are several factors that contribute to high \redchisq\ values in some of our XCLASS models.  First, we are unable to get excellent fits for some molecules partially because they are extremely optically thick. Species that fall into this category are: o-NH$_{3}$, H$_{2}$CO, SO$_{2}$, H$_{2}$S, CH$_{3}$OH-A, and CH$_{3}$OH-E. Second, radiative pumping likely plays a significant role in the excitation of several molecules with high \redchisq\ values producing deviations from our LTE models. H$_{2}$O, CH$_{3}$OH, H$_{2}$S, NH$_{3}$, and NH$_{2}$, most of which have detected vibrationally/torsionally excited lines in the HIFI band, may, along with their isotopologues, represent such species. We also include in this category the CH$_{3}$CN,$\nu_{8}=1$ model which has a \redchisq~=~4.0, but note that the ground vibrational mode models fit the data well. In addition to pumping, some of these molecules may not be in LTE possibly indicating they are tracing lower density gas. As a result, our LTE models may be ill suited to reproduce the observed emission. Finally, emission from OH, NH$_{3}$, and HCl contain absorption components which we do not fit in this study (see Sec.~\ref{s-indiv}), thus increasing our calculated \redchisq\ values for these molecules. Table~\ref{t-chisq} also reports \redchisq\ statistics for the MADEX models, most of which have values commensurate with their XCLASS counterparts, though they tend to be higher compared to XCLASS. This is, in part, a result of how we performed the MADEX modeling, which was less \emph{ad hoc} than our XCLASS approach. That is, using MADEX, we took a set of sub-components with given \tkin\ and \nhh\ and adjusted only the column densities to get the best fit, altering the kinetic temperatures and \hh\ densities only when \ntot\ adjustments didn't produce enough improvement in the fit. An additional reason for the higher MADEX \redchisq\ values, especially for SO$_{2}$ and its isotopologues, which have particularly discrepant \redchisq\ values, is that the MADEX models do not include reddening. Consequently, the largest deviations between the data and fits occur at high frequencies ($\nu$~$\gtrsim$~1~THz) where SO$_{2}$ emission is still strong and the models tends to over predict the observations. 

Using our XCLASS models as a template for the data, we compute, for each molecular species, the number of detected lines, \nf, and total integrated intensity, \intflux, within the HIFI survey. Integrated intensities are computed separately for each spatial/velocity component and are reported in Table~\ref{t-intflux} with molecules organized such that the total integrated intensity from all components (last column) is in descending order. For these calculations, we consider a ``line" to be any feature with a discernible peak given the resolution of HIFI. Hence, a detected ''line" which contains multiple unresolved hyperfine transitions, for example, is only counted once. In order for a line to be considered detected, it had to have a peak intensity $\ge$~3 times the local RMS, where we used a single RMS for each band. From Table~\ref{t-intflux}, we see that CH$_{3}$OH (sum of A and E) is the most emissive molecule, both in terms of total integrated intensity and number of detected lines. After methanol, the total integrated intensity is dominated by molecules with strong plateau components (e.g. CO, SO$_{2}$, and H$_{2}$O). Dimethyl ether is the second most emissive complex organic in terms of total integrated intensity, while methyl formate has the highest number of detected lines after methanol and its isotopologues. Summing \nf\ from all modeled species we obtain a total of $\sim$~\nlines\ identified lines in the Orion KL HIFI survey at or above 3~$\sigma$.

\subsection{Abundances}
\label{s-abund}

We compute molecular abundances by dividing the column densities in Tables~\ref{t-xclass} and \ref{t-madex} by \nhh\ values given in Table~\ref{t-standard}. We emphasize each complex organic is typically constrained by over 400 lines that, when combined with the mm data, encompasses an extremely broad range of excitation energies. The emission from lighter molecules (e.g. HCN, CS, etc.), with more widely spaced transitions, is also globally constrained in terms of \eup\ because of the wide bandwidth. For the rare isotopologue models, we obtained molecular abundances by multiplying \ntot\ by an assumed isotopic ratio. When more than one isotopologue is observed for a given molecule, they are averaged together. The same isotopologues are used to compute abundances from the XCLASS and MADEX models. That is, we do not use compact or extended ridge components which are invoked in some MADEX models but are not included in corresponding XCLASS fits (see Secs.~\ref{s-hcn}, \ref{s-hnc}, \ref{s-so}, and \ref{s-so2}). We assume the following isotopic ratios: $^{12}$C/$^{13}$C = 45, $^{32}$S/$^{33}$S = 75, $^{32}$S/$^{34}$S = 20, $^{16}$O/$^{18}$O = 250, $^{14}$N/$^{15}$N = 234, and $^{35}$Cl/$^{37}$Cl~=~3. The C, S, and O ratios are taken from \citet{tercero10}, who derive these values by modeling OCS and H$_{2}$CS emission present in the IRAM 30~m line survey using MADEX. The N isotopic ratio is taken from \citet{adande12}, who compute this value by analyzing CN hyperfine transitions, from which reliable optical depths can be measured. We assume a solar isotopic ratio for Cl \citep{asplund09}.

Molecular abundances, derived from our XCLASS models, are given in Table~\ref{t-abund}. For models with temperature gradients, we add the column densities from the individual sub-components together and take this sum as \ntot\ in our abundance calculations. Table~\ref{t-abund} also reports abundances derived from our more advanced MADEX models, which often employ temperature and density gradients. We follow the same convention when computing abundances based on MADEX, summing column densities from the individual sub-components. We estimate the uncertainty in our derived abundances to be approximately 40\%. This error estimate is described in the Appendix. Fig.~\ref{p-abund} is a bar chart which plots abundance as a function of molecule. Each spatial/velocity component is plotted in a different panel. Abundances computed by XCLASS and MADEX are plotted as solid black and dashed red lines, respectively. Hot core models with \vlsr~$\ge$~6.5~km/s are identified as originating from \hotcores. XCLASS abundances for these models are plotted in blue. The molecules are labeled along the x-axis and are roughly organized such that the cyanides are on the left side of the plot, the sulfur bearing species are in the middle, and the complex oxygen bearing organics are on the right. Comparing the overall abundance levels of the different spatial/velocity components, the richness of the hot core and compact ridge in complex organics is readily apparent. We also observe the well established chemical differentiation of O- and N-bearing organics between the compact ridge and hot core \citep{blake87}, and note that complex oxygen bearing organics observed toward the hot core have \vlsr\ values consistent with \hotcores, which is spatially closer to the compact ridge (see Sec.~\ref{s-hcs}). From Table~\ref{t-abund} and Fig.~\ref{p-abund}, we see that XCLASS and MADEX abundances typically agree quite well with one another (within a factor of $\sim$~2 -- 3). We, however, report HCN and SO$_{2}$ abundances toward the compact ridge that are offset by factors of 5.6 and 4.3, respectively, as well as an HNC hot core abundance that is offset by a factor of 6.3. Such discrepancies for HCN and SO$_{2}$ can be explained by the fact that the compact ridge component in the line profiles of these molecules is weak and that these species also have extended ridge components which blend with the compact ridge, making derived \ntot\ values less certain. The discrepant HNC hot core abundance is likely brought about by many of the HN$^{13}$C lines being blended with other species, leading to greater uncertainty in the derived column density.

\subsection{Vibrationally Excited Emission}
\label{s-vib}

We detect vibrationally excited emission from several molecules within the HIFI scan. These species are HCN ($\nu_{2}$=1 and $\nu_{2}$=2), H$^{13}$CN ($\nu_{2}$=1),  HNC ($\nu_{2}$=1), HC$_{3}$N ($\nu_{7}$=1), CH$_{3}$CN ($\nu_{8}$=1), NH$_{3}$ ($\nu_{2}$=1), SO$_{2}$ ($\nu_{2}$=1), and H$_{2}$O ($\nu_{2}$=1). Emission from these vibrationally excited modes are fit independently of the ground vibrational mode in XCLASS because separate JPL and CDMS catalogs exist for these states. We note that torsionally excited emission from CH$_{3}$OH and CH$_{3}$OCHO is also detected in the HIFI scan. For these molecules, however, the ground and vibrationally excited modes are included in the same catalog.

As with different isotopologues, we attempted to fit self consistent models between vibrationally excited and $\nu$~=~0 models, i.e. the same \trot\ and \ntot, though adjustments to \trot\ in the vibrationally excited models were sometimes employed to improve the fit. For most of the molecules with detected vibrationally excited modes, total column densities inferred from ground and vibrationally excited states agree to within a factor of 3. We, however, required significantly higher total column densities to fit vibrationally excited HCN (hot core) and CH$_{3}$CN (plateau only) emission relative to the ground vibrational state. The $\nu_{2}$=1 H$^{13}$CN and $\nu_{2}$=2 HCN models suggest total HCN column densities toward the hot core that are higher than that predicted by the HC$^{15}$N ground vibrational state model by factors of approximately 14 and 4, respectively, once \ntot\ values are scaled according to appropriate isotopic ratios (see Sec.~\ref{s-abund}). We note here that emission from HCN, $\nu_{2}$=1 and H$^{13}$CN, $\nu$=0 is likely optically thick. Similarly, the CH$_{3}$CN, $\nu_{8}$=1 plateau model requires a total column density that is a factor of 3.4 times higher than that predicted by $^{13}$CH$_{3}$CN and CH$_{3}$$^{13}$CN isotopologues once \ntot\ is properly scaled. Because we do not derive abundances for NH$_{3}$ and H$_{2}$O using XCLASS, we do not include them in the analysis below. 

The need for higher total column densities in the HCN and CH$_{3}$CN vibrationally excited models implies that these energetic modes are more heavily populated than predicted by our $\nu$~=~0 models, and is possibly the result of pumping by mid/far-IR photons. The ratio in column density between such modes yields a vibration temperature, \tvib, which may more closely reflect the local radiation temperature than the rotation temperature in our $\nu$~=~0 models. We estimate \tvib\ from our own LTE (XCLASS) models by calculating the total column density within individual vibrational modes, \ntot($\nu$). This is done by summing individual upper state columns of all rotation levels within a vibrational mode. That is,
\begin{equation}
N_{tot}(\nu) = \frac{N_{tot}}{Q(T)} \sum \limits_{i}  g_{i}e^{-\frac{-E_{i}}{kT_{rot}}}
\end{equation}
where the index $i$ corresponds to a rotation level, \ntot\ is the total molecular column density, $Q(T)$ is the partition function, $g_{i}$ is the statistical weight, and $E_{i}$ is the excitation energy. The vibration temperature can then be calculated by solving for temperature in the Boltzmann equation,
\begin{equation}
\label{e-boltz}
T_{vib} = - \frac{\Delta E}{k} ln \left[ \frac{g_{l}}{g_{u}} \frac{N_{tot}(\nu_{u})}{N_{tot}(\nu_{l})} \right]^{-1}.
\end{equation}
Here, $\Delta E$ is the excitation energy between vibrational modes in the ground rotation state, $k$ is the Boltzmann constant, $g_{l}$ and $g_{u}$ are the statistical weights of the lower and upper vibrationally excited modes, and \ntot($\nu_{l}$) and \ntot($\nu_{u}$) are the total column densities of the lower and upper vibrationally excited modes, respectively. When calculating \tvib\ for $\nu_{2}$=1~/~$\nu$=0 (HCN), $\nu_{2}$=2~/~$\nu_{2}$=1 (HCN), and $\nu_{8}$=1~/~$\nu$=0 (CH$_{3}$CN), we take g$_{l}$~/~g$_{u}$~=~1/2, 2/3, and 1/2, respectively, because $\ell$ splitting in the vibrationally excited modes break each rotation level into 2 ($\nu_{2}$=1 and $\nu_{8}$=1) or 3 ($\nu_{8}$=1) states. 

Vibration temperature estimates are reported in Table~\ref{t-vib}. \tvib, as measured by HCN ($\nu_{2}$=1~/~$\nu_{2}$=0), toward the hot core is markedly higher than our measured rotation temperature in the ground vibrational state (\trot~=~210~K). This suggests pumping plays a significant role in the excitation of the $\nu_{2}$=1 and $\nu_{2}$=2 vibrationally excited modes. If this temperature does indeed reflect the local continuum emitted by dust, it points to a radiation temperature $\gtrsim$~250~K. The $\nu_{2}$=2~/~$\nu_{2}$=1 HCN \tvib\ estimate is somewhat lower than the $\nu_{2}$=1~/~$\nu_{2}$=0 temperature, most likely because of the extremely high excitation energy of the $\nu_{2}$~=~2 mode ($\sim$2000~K). The local radiation field, therefore, may not be intense enough to populate this mode as efficiently as $\nu_{2}$~=~1. The vibration temperature for CH$_{3}$CN toward the plateau is also reported in Table~\ref{t-vib}. As with HCN, our derived \tvib\ is notably higher than the \trot\ reported for the ground vibrational state (\trot~=~130~K) indicating that CH$_{3}$CN is possibly being pumped by hot dust in the outflow. 

\subsection{Unidentified Lines}
\label{s-uline}

The residual between the observed spectrum and the full band model allows us to estimate the number of unidentified, ``U", lines. We identified channels emitted from U lines by first uniformly smoothing the data to a velocity resolution of $\sim$~1~km/s to increase the S/N. Residuals between the full band model and data were then computed by stepping across the spectrum in frequency intervals corresponding roughly to 1000~km/s. At each interval, we calculated the RMS noise level and subtracted a local baseline. A channel was then flagged as emitting from a U line if it was greater than 3 times the local RMS and 3 times the intensity level predicted by the full band model. Based on these criteria, we flag \nuchan\ channels in emission. If we assume a line width of 5~km/s, this corresponds to \nulines\ U lines. According to the full band model, we identify $\sim$~\nlines\ molecular lines at or above the 3~$\sigma$ noise level, leading to a U line fraction of $\sim$~\ulfrac. This calculation was checked by visual inspection to make sure most flagged channels do indeed appear to be originating from U lines. We, however, note that in some instances our methodology does produce ``false positives", which are mostly the result of complex line shapes not modeled particularly well especially in the wings, isolated regions with relatively high noise, weak spurs not removed during the data reduction process, and irregular baselines. This is especially true in bands 6 and 7 where more than half of the flagged channels may be ``false positives". We therefore emphasize that this calculation should be viewed as a rough estimate.

Fig.~\ref{p-uline} (upper panel) is a histogram that plots the number of U lines as a function of frequency. From the plot, we see that the number of U lines is highest at the low frequency end of the survey and quickly drops off at higher frequencies. The lower panel of Fig.~\ref{p-uline} plots the fraction of U line channels as a function of frequency. That is, the number of U line channels, x$_{\rm ul}$, divided by x$_{\rm ul}$ + x$_{\rm em}$, where x$_{\rm em}$ is the number of emission channels ($>$~3~$\sigma$) according to the full band model. The plot shows that this fraction stays between 6 and 15\% across the entire HIFI band, and suggests a possible trend in the U line fraction that increases with frequency, though given the uncertainties in U line flagging (i.e. ``false positives") especially at higher frequencies, this trend may not be real. We, thus, conclude that there is not a large over density of U lines relative to the total number of detected features at the low frequency end of the HIFI band ($\nu$~$\lesssim$~1200~GHz) where the line density is highest. Nor is the U line fraction drastically larger (within a factor of 2) at higher frequencies, with the exception of the highest frequency end of the HIFI survey ($\nu$~$\gtrsim$~1740~GHz) where the U line fraction is largest.

\section{Description of Individual Molecular Fits}
\label{s-indiv}

We present details of individual XCLASS and MADEX models in this section. Figure numbers within Figure Sets 13 and 14 corresponding to each XCLASS and MADEX model, respectively, are given below. Unless otherwise stated, we compute critical densities using collision rates obtained from the LAMDA database \citep{schoier05}. In the following discussion, the J quantum number corresponds to the total angular momentum of the molecule, i.e. 
{\bf J}~=~{\bf N }~+~{\bf S}~+~{\bf L} where {\bf N }, {\bf S}, and {\bf L} are the rotation, electron spin, and electron orbital angular momentum quantum numbers, respectively. For species without electronic angular momentum, {\bf J}~=~{\bf N} and rotation transitions are referenced by $\Delta$J. For molecules with electronic angular momentum, we label the rotation levels as N$_{J}$. Energy levels for symmetric rotors are labeled as J$_{K}$ where K corresponds to the angular momentum along the axis of symmetry. For asymmetric rotors, energy levels are labeled as J$_{Ka,Kc}$, where Ka and Kc represent the angular momentum along the axis of symmetry in the oblate and prolate symmetric top limits, respectively. In instances when we required a temperature gradient to fit the observed emission using XCLASS, we only report the rotation temperature of the sub-component which dominates the flux at HIFI frequencies. We focus our analysis on the molecular emission because it overwhelmingly dominates the line flux within the HIFI survey. We thus do not model atomic species, but note the detections of CI and CII and the non-detection of NII in Sec.~\ref{s-c}.

\subsection{CH$_{3}$CN}

We observe methyl cyanide (CH$_{3}$CN) toward the hot core, plateau, and compact ridge (Figs.~13.1 -- 13.5). In the HIFI band, we detect transitions over an approximate range in excitation energy of 330 -- 1800~K over K ladders spanning J = 27 up to 49. In the IRAM survey, we detect transitions from \eup~$\approx$~18~K to 660~K (J~=~5 to 13). We also detect emission from both $^{13}$C isotopologues as well as the $\nu_{8}$~=~1 vibrationally excited mode from levels with excitation energies $>$~2000~K. We derive \trot~=~260, 230, and 120~K for the hot core, compact ridge, and plateau, respectively. Only the hot core requires a temperature gradient in XCLASS.  We also model methyl cyanide along with its rarer isotopologues using MADEX (Figs.~14.1 -- 14.4). Collision rates are derived from the expressions and calculations of \citet{green86} for collisions with \hh. For these models, we invoke a temperature/density gradient for the hot core, and use single temperature/density models for the plateau and compact ridge. Total column densities for the $^{13}$C isotopologues derived via XCLASS and MADEX are equal to one another. Using a 2D line survey of Orion~KL taken with the IRAM 30~m telescope (Marcelino et al. 2014, in preparation), 
\citet{bell14} report several CH$_{3}$CN \ntot\ values toward IRc2, the infrared clump closest to the hot core, based on MADEX models of different K ladders most of which lie between 1\sn$^{16}$~\cms\ and 2\sn$^{16}$~\cms. Our hot core column density thus agrees to within a factor of $\sim$~2 with theirs. 

\subsection{C$_{2}$H$_{3}$CN}

We observe vinyl cyanide (C$_{2}$H$_{3}$CN) only toward the hot core (Fig.~13.6). In the HIFI and IRAM surveys we detect transitions over ranges in \eup\ of $\sim$~600 -- 750~K and 20 -- 600~K, respectively. The HIFI scan is thus only sensitive to the most highly excited lines. A temperature gradient is required to fit the observed emission. The cooler component, only visible in the IRAM survey, has a wider line width (\dv~=~12.5~km/s) than the hotter component (\dv~=~5.5~km/s) indicating the former is possibly probing more turbulent material. We derive \trot~=~215~K toward the hot core. 

\subsection{C$_{2}$H$_{5}$CN}

We observe ethyl cyanide (C$_{2}$H$_{5}$CN) exclusively toward the hot core (Fig.~13.7). In the HIFI scan, we detect transitions over a range in \eup\ of $\sim$~120 -- 550~K, while in the IRAM survey we detect transitions with excitation energies as low as $\sim$~15~K. The line profiles observed in the IRAM scan have line widths that are approximately a factor of 2 wider than observed in the HIFI data, even for transitions with similar excitation energies. Given its intricate structure, it is possible the IRAM data is pointed toward a more turbulent region within the hot core. As a result, our model fits the HIFI data well, but tends to over predict transitions observed in the IRAM survey because of the wider line widths. We derive \trot~=~136~K toward the hot core, but note that we require a compact 300~K component to fit the most highly excited emission. \citet{daly13} modeled the emission of ethyl cyanide in the IRAM survey using MADEX assuming LTE. They include three temperature sub-components in their hot core model (\trot~=~275~K, 110~K, and 65~K) and derive a total column density that is a factor of 1.5 times higher than our value.

\subsection{HC$_{3}$N}

Cyanoacetylene (HC$_{3}$N) is detected toward the hot core and the plateau (Figs.~13.8 -- 13.9). In the HIFI and IRAM scans, we observe transitions from J~=~53 -- 52 to approximately 77 -- 76 (\eup~$\approx$~620 -- 1300~K) and J~=~9 -- 8 to 30 -- 29 (\eup~$\approx$~20 -- 200~K), respectively. In addition, we observe the $\nu_{7}$~=~1 vibrationally excited mode toward both spatial/velocity components up to excitation energies of $\sim$~1320~K (J~$\approx$~67). We require a temperature gradient to fit the hot core component while the plateau is fit well by a single temperature model. We adopted a consistent set of rotation temperatures for both the $\nu$~=~0 and $\nu_{7}$~=~1 models. Our derived rotation temperatures for the hot core and plateau are 210~K and 115~K, respectively. 
\citet{esplugues13b} presents a more detailed non-LTE analysis of HC$_{3}$N emission in the HIFI and IRAM surveys. Using MADEX and assuming two temperature/density sub-components for both the hot core and plateau, they derive total column densities that agree well, (within a factor of 1.5 and 1.1 for the hot core and plateau, respectively), with the LTE \ntot\ values reported here. 

\subsection{HCN}
\label{s-hcn}

We observe hydrogen cyanide (HCN) toward all spatial velocity components (Figs.~13.10 -- 13.16). In addition to the main isotopologue, we also detect H$^{13}$CN, HC$^{15}$N, and DCN in the ground vibrational state. Transitions from J~=~6 -- 5 to 21 -- 20 (\eup~$\approx$~90 -- 980~K) are detected in the HIFI scan. The IRAM survey contains the 1 -- 0 and 2 -- 1 transitions. We also observe emission in the $\nu_{2}$~=~1 (HCN and H$^{13}$CN) and $\nu_{2}$~=~2 (HCN) vibrationally excited states only toward the hot core. The $\nu_{2}$~=~1 HCN line profiles contain two velocity components, one wide (\dv~=~15~km/s) and one more narrow (\dv~=~7~km/s), both at \vlsr~=~5.0~km/s, consistent with the hot core. In the $\nu_{2}$~=~2 state, we detect emission at excitation energies as high as $\sim$~2400~K. In XCLASS, we do not require temperature gradients to fit the compact or extended ridge. Additional cooler components, however, are needed to reproduce the observed IRAM line intensities in the hot core and plateau. Our derived \trot\ values are 210~K for the hot core, 120~K for the compact ridge, 130~K for the plateau, and 30~K for the extended ridge. We also model all HCN isotopologues in the ground vibrational state with MADEX (Figs.~14.5 -- 14.8), using temperature/density gradients for all spatial/velocity components except the extended ridge. The HC$^{15}$N MADEX model has a compact ridge component to fit emission at low excitation, which we do not include in the corresponding XCLASS model because including just an extended ridge component is sufficient to fit the line profiles with XCLASS. A more detailed analysis of emission from HCN and its isotopologues in both the HIFI and IRAM surveys using MADEX is presented in Marcelino et al. (2014, in preparation). Collision rates are taken from \citet{dumouchel10} for collisions with He scaled to \hh. 

\subsection{HNC}
\label{s-hnc}

Hydrogen isocyanide (HNC) is detected toward all spatial/velocity components in the main isotopologue (Figs.~13.17 -- 13.19). We also observe HN$^{13}$C toward the hot core and plateau. Between the HIFI and IRAM surveys, we detect the ground state line up to approximately the J~=~16 -- 15 transition (\eup~$\approx$~590~K). The $\nu_{2}$~=~1 vibrationally excited mode is observed toward the hot core in the main isotopologue up to excitation energies of $\sim$~1120 (J~$\sim$~14). In XCLASS, we require temperature gradients to fit both the hot core and plateau. The compact ridge and extended ridge are fit well by single temperature models. Our derived \trot\ values are 220~K for the hot core, 80~K for the compact ridge, 115~K for the plateau, and 20~K for the extended ridge. We fit the $\nu_{2}$~=~1 vibrationally excited emission using the same \trot\ as the ground vibrational state. HNC and HN$^{13}$C, in the ground vibrational state, are also fit with MADEX (Figs.~14.9 -- 14.10). As with HCN, we used temperature/density gradients to fit all spatial/velocity components except the extended ridge. The HN$^{13}$C MADEX model has a compact ridge component to fit emission at low excitation, which we do not include in the corresponding XCLASS model because including just an extended ridge component is sufficient to fit the line profiles with XCLASS. A more detailed analysis of emission from HNC and its isotopologues in both the HIFI and IRAM surveys using MADEX is presented in Marcelino et al. (2014 , in preparation). Collision rates are taken from \citet{dumouchel10} for collisions with He scaled to \hh.

\subsection{CN}

Cyanide radical (CN) emission is detected toward the plateau and extended ridge (Fig.~13.20). We do not observe hyperfine structure at HIFI frequencies, but do detect the doublets brought about by the unpaired electron. Combining the IRAM and HIFI scans, we observe from the ground rotation transition to the N = 8 -- 7 doublet (\eup~=~196~K). Both the plateau and extended ridge are well fit by single temperature XCLASS models, although the N = 1 -- 0 lines, (detected in the IRAM scan), are somewhat under predicted in the extended ridge by no more than a factor of $\sim$~1.5. Our derived rotation temperatures are 43~K and 21~K for the plateau and extended ridge, respectively, indicating CN is probing cool material toward these regions.

\subsection{HNCO}
\label{s-hnco}

We observe isocyanic acid (HNCO) along with its rarer isotopologue HN$^{13}$CO toward the hot core and plateau (Figs.~13.21 -- 13.22). In the HIFI scan, we detect transitions over excitation energies of $\sim$~40 -- 1050~K. In the IRAM survey, we detect somewhat lower energy lines over an \eup\ range of $\sim$~10 -- 600~K. Temperature gradients are required to fit both components. We derive \trot\ values of 209~K and 168~K for the hot core and plateau, respectively. Non-LTE excitation effects are likely important for this molecule because transitions between K$_{a}$ ladders are not collisionally excited. Rather, these states are pumped via the far-IR \citep{churchwell86}. Our derived values for \trot\ thus more closely couple to the dust temperature than the kinetic temperature of the gas. As a result, our LTE fits do deviate somewhat from the observations. In particular, our model tends to under predict A-type transitions and over predict B-type transitions. Nevertheless, our values for \trot\ and \ntot\ agree well with previous spectral survey studies who have also noted this effect \citep[see e.g.][]{comito05, schilke01, schilke97}. 

\subsection{HCO$^{+}$}
\label{s-hcop}

Formylium (HCO$^{+}$) and its rarer isotopologue H$^{13}$CO$^{+}$ are detected toward all spatial/velocity components (Figs.~13.23 -- 13.24). Between the IRAM and HIFI scans, we observe transitions from the ground state line up to J~=~17 -- 16 (\eup~=~655~K). Using XCLASS, we require a temperature gradient only for the plateau. The cooler sub-component, most prominent in the IRAM survey, is significantly wider (\dv~=~50~km/s) than the warmer sub-component (\dv~=~30~km/s) indicating the presence of higher velocity material. Our derived rotation temperatures are 190~K for the hot core, 80~K for the compact ridge, 135~K for the plateau, and 27~K for the extended ridge. We use a compact ridge source size of 25\arcsec\ in our XCLASS model because fits with \ssize~=~10\arcsec, our standard value for the compact ridge, under predict the observed line intensities for the main isotopologue. The reason for this is that the model emission becomes too optically thick before the observed peak intensity is reached. This may indicate that HCO$^{+}$ is more spatially extended than other compact ridge tracers, or that emission we are attributing to the compact ridge is indeed originating from hotter than typically observed gas in the extended ridge. HCO$^{+}$ maps presented by \citet{vogel84}, suggest the latter scenario is possible. They observe highly extended emission (\ssize~$\ge$~45\arcsec) toward two peaks to the NE and SW of IRc2 at \vlsr~$\sim$~11.4 and 7.4~km/s, respectively. Comparing the position of these peaks with a similar map of SO, they argue HCO$^{+}$ is tracing a region of the extended ridge that is impeding outflowing gas. Peak line intensities toward the 7.4 km/s peak suggest gas kinetic temperatures $\ge$~78~K, which is similar to the \trot\ we derive for our compact ridge component. We also modeled HCO$^{+}$ and H$^{13}$CO$^{+}$ using MADEX (Figs.~14.11 -- 14.12). Temperature/density gradients are invoked for all spatial/velocity components except the extended ridge. A more detailed analysis of emission from HCO$^{+}$ and its isotopologues in both the HIFI and IRAM surveys using MADEX is presented in Marcelino et al. (2014, in preparation). Collision rates for the MADEX models are taken from \citet{flower99} for collisions with \hh. 

\subsection{CCH}

We observe ethynyl (CCH) emission toward the hot core (s) and extended ridge (Fig.~13.25). In the HIFI survey, we detect transitions from N~=~6 -- 5 to 10 -- 9 (\eup~=~90 -- 230~K). In the IRAM scan, we detect the N~=~1 -- 0 and 2 -- 1 transitions (\eup~$\leq$~13~K). Hyperfine components are resolved in the IRAM data but not in the HIFI scan. The doublets, however, brought about by the unpaired electron are separated enough that they are resolved in both surveys. We did not need to invoke temperature gradients to fit either the \hotcores, \trot~=~53~K, or extended ridge, \trot~=~37~K. The hot core rotation temperature is very low compared to other molecules detected toward this region, indicating CCH is probing very cool gas relative to other hot core tracers.

\subsection{CS}

We observe carbon monosulfide (CS) along with the rarer $^{13}$CS, C$^{34}$S, and C$^{33}$S isotopologues toward all spatial/velocity components (Figs.~13.26 -- 13.29). In the HIFI and IRAM scans, we detect transitions from J~=~10 -- 9 to 26 -- 25 (\eup~$\approx$~130 -- 820~K) and 2 -- 1 to 5 -- 4 (\eup~$\approx$~10 -- 40~K), respectively. We require a temperature gradient only for the plateau. We adopted a consistent set of rotation temperatures for the rarer isotopologues. These \trot\ values are 100~K for the \hotcores, 225~K for the compact ridge, 155~K for the plateau, and 35~K for the extended ridge. Using non-LTE MADEX models, \citet{tercero10} also analyze emission from CS and its isotopologues within the IRAM survey. Their column densities, among the rarer isotopologues, agree with ours to within a factor of 3 toward all spatial/velocity components. 

\subsection{H$_{2}$S}

Hydrogen sulfide (H$_{2}$S) is observed toward the hot core, plateau, and extended ridge (Figs.~13.30 -- 13.32). In addition to the main isotopologue, we also detect H$_{2}$$^{34}$S and H$_{2}$$^{33}$S. In the HIFI scan, we observe emission over a range in excitation energies of $\sim$~50 -- 1230~K. The IRAM data contains the 1$_{1,0}$ -- 1$_{0,1}$ transition (\eup~=~28~K) for all three isotopologues, providing an additional constraint at low excitation. A detailed non-LTE study of H$_{2}$S emission toward the hot core is presented by 
\citet{crockett14} and we use the abundance derived in that study here. We use single temperature models for all but the plateau, which requires a gradient. A consistent set of rotation temperatures is used for the rarer isotopologues. Our derived \trot\ values are 145~K for the hot core, 115~K for the plateau, and 50~K for the extended ridge. For the hot core, we use a source size of 6\arcsec\ to be consistent with 
\citet{crockett14}.

\subsection{H$_{2}$CS}

We observe thioformaldehyde (H$_{2}$CS) emission toward the \hotcores\ and compact ridge (Fig.~13.33). In the HIFI scan, we detect emission over a range in \eup\ of $\sim$~200 -- 600~K. While in the IRAM survey, we detect transitions at lower excitation over an \eup\ range of $\sim$~10 -- 200~K. Temperature gradients are not required for either component in our XCLASS model. We derive rotation temperatures of 120~K and 100~K for the hot core and compact ridge, respectively. \citet{tercero10} present analysis of the H$_{2}$CS emission in the IRAM survey. Despite the fact that they assume a different compact ridge source size (15\arcsec) and a different temperature for the hot core (225~K) in their analysis, we derive column densities that agree to within a factor of 2.5. They also derive an ortho/para ratio of $\sim$~2.0~$\pm$~0.7 for both the compact ridge and hot core (small variations exist between isotopologues), which, given the uncertainty, is consistent with the equilibrium value of 3. We used a spectroscopic catalog that did not separate the ortho and para species (CDMS catalog) and found that a single ortho + para model fit the data well, which, for the temperatures considered in our models, points to an ortho/para ratio of 3.   

\subsection{OCS}

We observe carbonyl sulfide (OCS) emission toward all spatial/velocity components (Fig.~13.34). In the HIFI and IRAM scans, we detect transitions from J~=~40 -- 39 to approximately 61 -- 60 (\eup~$\approx$~480 -- 1100~K) and 7 -- 6 to 23 -- 22 (\eup~$\approx$~20 -- 160~K), respectively.  We do not detect vibrationally excited emission in the HIFI dataset, but do in the IRAM scan ($\nu_{2}$~=~1). Temperature gradients are not needed to fit any component using XCLASS. Our derived rotation temperatures are 190~K for the hot core, 165~K for the compact ridge, 110~K for the plateau, and 10~K for the extended ridge. The OCS emission within the IRAM survey, including the vibrationally excited emission, is analyzed in detail by \citet{tercero10}. As a result, we do not include a vibrationally excited OCS model here. Our derived column densities agree to within a factor of 2.5 with that derived in the Tercero study. 

\subsection{SO}
\label{s-so}

Sulfur monoxide (SO) is detected toward the hot core, plateau, and extended ridge (Figs.~13.35 -- 13.37). We also observe the rarer isotopologues $^{33}$SO and $^{34}$SO toward the hot core and plateau. In the HIFI scan, we detect transitions over a range in \eup\ of $\sim$~10 -- 1200~K. While in the IRAM survey, we detect additional low energy rotation transitions with \eup~$\lesssim$~100~K. Using XCLASS, we require temperature gradients to fit both the hot core and plateau. A consistent set of rotation temperatures is adopted for the rarer species. We derive \trot\ values of 258~K for the hot core, 163~K for the plateau, and 60~K for the extended ridge. Line profiles are dominated by the plateau especially at low excitation. Hence, our reported extended ridge model is not very well constrained because it manifests as a small bump atop the strong plateau component. Our model reproduces the observed line intensities of the $\Delta$N~=~1, $\Delta$J~=~1 transitions well, which are quite strong in the HIFI scan (\tpeak\ values can be as high as 25~K). We, however, note that our model systematically under predicts the less emissive $\Delta$N~=~1, $\Delta$J~=~0 rotation transitions (observed \tpeak~$\lesssim$~3~K) by approximately a factor of 5. We even detect several $\Delta$N~=~3 lines at the 1~K level, which our model also under predicts. We also fit SO and its isotopologues using MADEX (Figs.~14.13 -- 14.15).  The plateau is modeled with a temperature/density gradient while the hot core and extended ridge are fit with single components. Our MADEX models also under predict the $\Delta$N~=~1, $\Delta$J~=~0 and $\Delta$N~=~3 transitions. The $^{34}$SO MADEX model includes an extended ridge component to fit extremely weak emission in the IRAM survey, which we do not include in the corresponding XCLASS model. Collision rates are taken from \citet{lique06}, who report rates for He, scaled to \hh. 
\citet{esplugues13a} modeled the emission of SO and its isotopologues within the IRAM survey using MADEX in non-LTE mode. Their column densities, for the rarer isotopologues, agree to within a factor of 3 with our LTE \ntot\ values. 

\subsection{SO$_{2}$}
\label{s-so2}

The main isotopologue of sulfur dioxide (SO$_{2}$) is observed toward all spatial velocity components (Figs.~13.38 -- 13.41). The rarer isotopologues $^{33}$SO$_{2}$ and $^{34}$SO$_{2}$ are also detected toward the hot core and plateau. Additionally, we observe the $\nu_{2}$ = 1 vibrationally excited mode toward the hot core. In the HIFI scan, we detect transitions with \eup\ from $\sim$~70~K to $\sim$~1100~K, the most highly excited lines originating from the $\nu_{2}$ = 1 state. The IRAM survey probes excitation energies as low as $\sim$10~K. Modeling the data with XCLASS, we adopt a single set of rotation temperatures for all isotopologues. We derive \trot\ values of 240~K for the hot core, 100~K for the compact ridge, 150~K for the plateau, and 50~K for the extended ridge. Temperature gradients are required to fit the hot core and compact ridge. We fit the $\nu_{2}$ = 1 emission using the same \trot\ as the ground vibrational state. We also modeled SO$_{2}$ and its isotopologues in the ground vibrational state with MADEX (Figs.~14.16 -- 14.18). Unfortunately, collision rates do not exist at high enough excitation to produce reliable non-LTE models at HIFI frequencies. Attempts to use estimated collision rates, moreover, produced suspicious level populations. As a result, the MADEX SO$_{2}$ models presented here are computed assuming LTE. Just as with the MADEX SO models, the plateau is fit with a temperature gradient. The hot core, extended ridge, and compact ridge are fit with single temperature models. Our $^{33}$SO$_{2}$ and $^{34}$SO$_{2}$ MADEX fits include compact ridge and extended ridge components to fit extremely weak emission in the IRAM survey, which we do not include in the corresponding XCLASS models. 
\citet{esplugues13a} modeled the emission of SO$_{2}$ and its isotopologues within the IRAM survey using MADEX assuming LTE. They derive column densities for the rarer isotopologues that agree to within a factor of 3 with ours.  

\subsection{HCS$^{+}$}

We detect the main isotopologue of thioformylium (HCS$^{+}$) exclusively toward the compact ridge (Fig.~13.42). In the HIFI scan, we observe transitions from J = 12 -- 11 (\eup~=~160~K) to approximately 20 -- 19 (\eup~=~430~K). The IRAM data contains the  J = 2 -- 1, 4 -- 3, and 5 -- 4 transitions (\eup~=~6 -- 31~K). All observed HCS$^{+}$ lines in both the IRAM and HIFI scans are weak (\tpeak~$\lesssim$~1~K). A single temperature model with \trot~=~105~K fits the data well. As with the other sulfur carbon molecules detected in this survey, the HCS$^{+}$ emission in the IRAM scan was analyzed using MADEX by \citet{tercero10}. Our derived column density (\ntot~=~1.8\sn$^{14}$~\cms) is a factor of $\sim$~2 times higher than that reported by the Tercero study for the compact ridge. Our results are thus in close agreement with theirs despite our assumption of LTE and the fact that \citet{tercero10} assume \ssize~=~15\arcsec\ for the compact ridge, which is likely the biggest reason our HCS$^{+}$ column density is somewhat higher than theirs.

\subsection{SiS}

We detect the main isotopologue of silicon monosulfide (SiS) toward the plateau (Fig.~13.43). We observe transitions from J = 27 -- 26 to approximately 41 -- 40 (\eup~$\approx$ 330 -- 750~K) in the HIFI scan, while the IRAM survey contains transitions from J = 5 -- 4 to 15 -- 14 (\eup~$\approx$~10 -- 110~K). The observed emission is fit well by a single temperature XCLASS model with \trot~=~145~K. \citet{tercero11} analyze the SiS emission within the IRAM dataset. In that study, they model the SiS lines by including contributions from the hot core, plateau, and the so called 15.5~km/s component. At HIFI frequencies, we only observe SiS emission from the plateau. As a result, we do not include models for the other two components. The Tercero study derives a column density of 3.5\sn$^{14}$~\cms\ toward the plateau using MADEX, not assuming LTE. Our derived column density of 4.3\sn$^{14}$~\cms\ thus agrees well with the Tercero result, despite our assumption of LTE. 

\subsection{SiO}
\label{s-sio}

Silicon monoxide (SiO) emission is detected toward the plateau (Figs.~13.44 -- 13.46). In addition to the main isotopologue, we also detect emission from $^{29}$SiO and $^{30}$SiO. The HIFI survey contains transitions from J = 12 -- 11 to 29 -- 28 (\eup = 163 -- 905~K) and the IRAM scan contains transitions from J = 2 -- 1 to 6 -- 5 (\eup = 6 -- 44~K). We fit the line emission in XCLASS by splitting the model into high velocity plateau (\vlsr~=~10.0~km/s, \dv~=~35~km/s) and low velocity plateau (\vlsr~=~7 -- 8.0~km/s, \dv~=~15~km/s) components. These two components do not necessarily correspond to the LVF and HVF as defined in Sec.~\ref{s-orion}. Both components require temperature gradients in order to fit both the HIFI and IRAM observations. We were able to adopt a consistent set of rotation temperatures for all isotopologues. We derive \trot\ values of 120~K and 270~K for the high and low velocity plateau components, respectively. \citet{tercero11} modeled the SiO emission in the IRAM survey using MADEX in a similar way. In addition to low and high velocity plateau components, they include hot core, compact ridge, and extended ridge components to fit the line profiles. We do not detect compact or extended ridge components at HIFI frequencies and thus do not include models for those components here. It is, however, possible that the component we attribute to the low velocity plateau may include emission from the hot core. Nevertheless, we find a two component model reproduces the data well. Furthermore, our $^{29}$SiO and $^{30}$SiO column densities (the sum of all temperature sub-components) are commensurate with those derived in the Tercero study.

\subsection{HCl}
\label{s-hcl}

We observe the J = 1 -- 0, 2 -- 1, and 3 -- 2 transitions of hydrochloric acid (HCl) and H$^{37}$Cl in the HIFI scan (Figs.~13.47 -- 13.48). The line profiles are wide (\dv~=~25~km/s), indicating an origin in the plateau. Because of the wide line profiles, we do not resolve the hyperfine components. The J = 2 -- 1 and 3 -- 2 lines show blue shifted absorption in both the main and $^{37}$Cl isotopologues, while the ground state lines do not. The absorption component appears also to be wide (\dv~$\gtrsim$~15~km/s), though it is difficult to constrain because of the blending between absorption and emission. The wide line width suggests the absorbing layer also originates from the plateau. We do not fit the absorption component in this work. A temperature gradient is not required to fit the emission and we derive \trot~=~73~K for both isotopologues. In order to reproduce the observed peak line intensities, we needed to use a 50\arcsec\ source size as opposed to 30\arcsec, indicating HCl is quite extended toward the plateau. It is also possible that emission from other components, (e.g. the hot core), may be contributing to the line profiles. However, given the dominance of the plateau and the presence of blue shifted absorption in the excited lines, adding such components would not be meaningfully constrained. 

\subsection{NS}

We observe nitric sulfide (NS) emission toward the hot core and compact ridge (Fig.~13.49). In the HIFI scan, we detect NS transitions from J = 21/2 -- 19/2 to approximately J = 47/2 -- 45/2 (\eup = 134 -- 960~K). In the IRAM survey, we observe emission from the J = 5/2 -- 3/2 and J = 7/2 -- 5/2 transitions (\eup~$\approx$~10 -- 340~K), the hyperfine structure being visible in the IRAM scan but not at HIFI frequencies. We, however, do observe the splitting from the unpaired electron in the HIFI spectrum. We did not require temperature gradients for either component in our XCLASS model and derive \trot~=~105 and 200~K for the hot core and compact ridge, respectively. We note that our model systematically predicts line centers $\sim$~10 -- 20~MHz higher than observed for many transitions $\gtrsim$~1~THz. Such differences are commensurate with the quoted uncertainties in the line catalog (JPL database) used to produce the model, which begin to noticeably increase as a function of frequency for $\nu$~$\gtrsim$~700~GHz.

\subsection{NO}

We observe nitric oxide (NO) emission toward all spatial/velocity components (Fig.~13.50). In the HIFI scan, we detect transitions from J~=~11/2 -- 9/2 to approximately J~=~29/2 -- 27/2 (\eup~$\approx$~80 -- 540~K). In the IRAM survey, we detect hyperfine lines from the ground state rotation transition J~=~3/2 -- 1/2. As with the other nitrogen bearing radicals in this survey, we do not detect hyperfine structure in the HIFI scan, but do observe the doublets produced by the spin angular momentum of the unpaired electron. We do not need temperature gradients for any spatial/velocity component and derive \trot\ values of 180~K for the hot core, 70~K for the compact ridge, 155~K for the plateau, and 60~K for the extended ridge. 

\subsection{H$_{2}$CO}

We observe formaldehyde (H$_{2}$CO) emission toward the \hotcores, compact ridge, and plateau (Figs.~13.51 -- 13.53). We also detect the rarer isotopologue H$_{2}$$^{13}$CO toward the \hotcores\ and compact ridge, as well as the singly deuterated species, HDCO, only toward the compact ridge. In the HIFI and IRAM surveys, we detect transitions over ranges in excitation energy of $\sim$~100 -- 1000~K and $\sim$~10 -- 90~K, respectively. Single temperature models are used for all spatial/velocity components except the plateau. For H$_{2}$CO and H$_{2}$$^{13}$CO, we use a source size of 15\arcsec\ for the compact ridge because models with \ssize~=~10\arcsec, our standard source size for the compact ridge, predict line intensities that are too weak. We derive \trot\ values of 135~K for the \hotcores, 50~K for the compact ridge, and 88~K for the plateau. A more detailed analysis of formaldehyde emission in the HIFI scan is presented by 
\citet{neill13b}, who use the RADEX non-LTE code \citep{vandertak07} to model the formaldehyde emission toward the compact ridge in order to obtain a robust D/H ratio. As discussed in the 
\citet{neill13b} study, the HDCO line profiles differ somewhat from other compact ridge tracers in that they contain two narrow components (both with \dv~=~2.2~km/s): one at \vlsr~=~7.7~km/s, near the canonical velocity of the compact ridge, and the other at \vlsr~=~10.0~km/s. The detection of the higher velocity component points to the presence of a highly deuterated clump, the origin of which remains unclear. As a result, our HDCO model contains two velocity components at  7.7 and 10.0~km/s. Both of which assume a source size 12.5\arcsec\ to remain consistent with 
\citet{neill13b}.

\subsection{H$_{2}$CCO}

We detect ketene (H$_{2}$CCO) emission only toward the compact ridge (Fig.~13.54). In the HIFI scan, the observed line intensities are quite weak (\tpeak~$<$~0.5~K). Nevertheless, we detect transitions in the HIFI scan over an approximate range in \eup\ of 300 -- 500~K. While in the IRAM survey we observe transitions with excitation energies from  $\sim$~20~K to 150~K. The line profiles in the IRAM scan contain a wider component (\vlsr~$\approx$~6~km/s, \dv~$\approx$~15~km/s) that we attribute to the hot core. Because this component is only visible in the IRAM data, we do not include a hot core model in this work. We derive \trot~=~100~K with no temperature gradient. 

\subsection{H$_{2}$O}

We detect water (H$_{2}$O) emission toward the \hotcores, compact ridge, and plateau (Figs.~13.55 -- 13.61). In addition to the main isotopologue, we also observe H$_{2}$$^{18}$O, H$_{2}$$^{17}$O, HDO, HD$^{18}$O, and D$_{2}$O. Emission from the $\nu_{2}$~=~1 vibrationally excited mode is also detected from the main isotopologue. In the HIFI survey, we detect from the ground state transition up to excitation energies $\gtrsim$~2500~K, the most highly excited lines originating from $\nu_{2}$~=~1. Absorption is also observed in low energy lines (\eup~$\lesssim$~200~K). We derive \trot\ values of 270, 150, and 150~K for the hot core, compact ridge, and plateau, respectively. More detailed non-LTE analyses of the water emission in the HIFI scan are presented by \citet{melnick10} and, more recently, by 
\citet{neill13a}, in which water abundances as well as D/H ratios toward the different spatial/velocity components are derived. We adopt the abundances derived in 
\citet{neill13a} here. 

\subsection{CH$_{2}$NH}

Methanimine (CH$_{2}$NH) is detected toward the hot core and extended ridge (Fig.~13.62). In the HIFI and IRAM scans, we detect transitions over ranges in \eup\ of $\sim$~15 -- 370~K and 5 -- 50~K, respectively. Temperature gradients are not required for either component. We derive \trot~=~130 and 40~K toward the hot core and extended ridge, respectively. 

\subsection{NH$_{2}$CHO}

We observe formamide (NH$_{2}$CHO) emission only toward the compact ridge (Fig.~13.63). The HIFI and IRAM scans contain transitions over ranges in excitation energy of approximately 280 -- 660~K and 10 -- 150~K, respectively. We detect only the more emissive a-type transitions, the b-types having over a factor of 4 lower dipole moment than the a-types. Even at low excitation, the lines are weak (\tpeak~$<$~0.5~K) in the HIFI survey. There are hints of a wider component in the IRAM scan, possibly from the hot core \citep[see][]{motiyenko12}, but we do not include a model for that component here. A single temperature model with \trot~=~190~K fits the observed emission well. \citet{motiyenko12} model formamide emission within the IRAM survey. They obtain a compact ridge column density that agrees closely (within 10\%) with our value. Our derived \trot\ agrees well with that reported in \citet{sutton95}, which we used as a starting point for our own fit. Our derived \ntot\ is roughly a factor of $\sim$~3 higher than that reported in the Sutton study, most likely because \citet{sutton95} reports a beam averaged column density for formamide. 

\subsection{C$_{2}$H$_{5}$OH}

Ethanol (C$_{2}$H$_{5}$OH) emission is observed toward the compact ridge (Fig.~13.64). In the HIFI and IRAM surveys, we detect transitions over approximate ranges in \eup\ of 100 -- 470~K and 30 -- 210~K, respectively. We do not need a temperature gradient to fit the observed emission and derive \trot~=~110~K. As with several other complex organics detected in this survey, the observed line profiles are weak (\tpeak~$<$~1~K).

\subsection{CH$_{3}$OCH$_{3}$}

Dimethyl ether (CH$_{3}$OCH$_{3}$) is detected toward the \hotcores\ and compact ridge (Fig.~13.65). In the HIFI and IRAM scans, we observe transitions over ranges in excitation energy of $\sim$~90 -- 620~K and 5 -- 510~K, respectively.  Single temperature models are sufficient to fit both components. We derive \trot~=~100 and 110~K for the \hotcores\ and compact ridge, respectively. 

\subsection{CH$_{3}$OCHO}

We observe methyl formate (CH$_{3}$OCHO) only toward the compact ridge (Fig.~13.66). Transitions in the HIFI and IRAM surveys are detected over \eup\ ranges of $\sim$~160 -- 740~K and 10 -- 300~K, respectively. A fraction of the detected lines are from the v$_{t}$~=~1 torsionally excited state. \citet{favre11}  observe multiple transitions of CH$_{3}$OCHO using the Plateau de Bure Interferometer (PdBI). In that study, they fit the observed line profiles using two velocity components, one at \vlsr~=~7.5~km/s and the other at 9.2~km/s, both of which have \dv~=~1.7~km/s. We find that a single compact ridge model with \vlsr~=~8.0~km/s, \dv~=~2.8~km/s, and \trot~=~110~K reproduces the HIFI observations well, though some line profiles hint at the possibility for two components. When both velocity components are combined, our derived \trot\ agrees well with theirs toward MF1 (\trot~=~100~K), the largest methyl formate clump observed toward the compact ridge. Our derived column density is also commensurate with that derived by \citet{favre11} toward MF1.

\subsection{CH$_{3}$OH}
\label{s-ch3oh}

We observe methanol (CH$_{3}$OH) emission toward the \hotcores\ and compact ridge (Figs.~13.67 -- 13.72). In addition to the main isotopologue, we also detect $^{13}$CH$_{3}$OH as well as the singly deuterated CH$_{3}$OD and CH$_{2}$DOH isotopologues toward these same spatial/velocity components. In both the HIFI and IRAM surveys, we detect transitions over a range in excitation energy of approximately 20 -- 1200~K, the most highly excited transitions originating from the torsionally excited $\nu_{t}$~=~1 mode.  We derive \trot~=~128 and 140~K for the \hotcores\ and compact ridge, respectively. More detailed analysis of methanol emission within the HIFI scan is presented by \citet{wang11} and 
\citet{neill13b}. 

\subsection{HF}
The J = 1 -- 0 transition of hydrogen fluoride (HF) is detected in the HIFI scan (Fig.~13.73), but is heavily blended with transitions of SO$_{2}$ and CH$_{3}$OH. After subtracting emission from contaminating lines, \citet{phillips10} finds the HF 1 -- 0 transition displays a strong blue shifted absorption component, which extends over a \vlsr\ range of $\sim$~$-$9 -- $-$45~km/s,  and a weaker red shifted emission component which extends  over a \vlsr\ range of $\sim$~12 -- 50~km/s. The broad line profile indicates an origin in the plateau. \citet{phillips10} estimates an HF column density in the absorption component, which they attribute to the LVF, of 2.9\sn$^{13}$~\cms. Because we only observe the ground state and the line profile is dominated by absorption, we do not provide an HF model. 

\subsection{CO}

We observe CO, $^{13}$CO, C$^{18}$O, and C$^{17}$O toward all spatial/velocity components and $^{13}$C$^{18}$O toward the extended ridge and compact ridge (Figs.~13.74 -- 13.78). In the HIFI scan, we detect transitions from 5 -- 4 up to 17 -- 16 (\eup~=~80 -- 810~K), while the IRAM survey contains the 1 -- 0 ground state transition. Our reported \trot\ values are the same as those used in \citet{plume12}, who analyze the C$^{18}$O and C$^{17}$O emission in the HIFI scan to obtain robust \hh\ column density measurements toward the different spatial/velocity components within Orion KL, except for the plateau, toward which we use a higher \trot. These \trot\ values are 150~K for the hot core, 125~K for the compact ridge, 130~K for the plateau, and 40~K for the extended ridge. As described in Sec.~\ref{s-xclass}, we adopt these \nhh\ estimates in this study modulo small adjustments for the plateau and compact ridge. With this said, we note that our reported plateau C$^{18}$O column density is higher by a factor of 1.8 than the \nhh\ value given in Table~\ref{t-standard} multiplied by 2.0\sn$^{-7}$, our assumed C$^{18}$O abundance. The reason for this discrepancy is that a somewhat wider plateau line width (\dv~=~25~km/s) than used in the Plume study (\dv~=~20~km/s) appears to produce better agreement between the model and data. As a result, we were forced to increase the C$^{18}$O column density by a factor of 1.8 to properly fit the data. 

\subsection{NH$_{2}$}

Only the ortho species of amidogen (NH$_{2}$) is detected toward the hot core and extended ridge (Fig.~13.79). In the HIFI scan, we detect from the ground state up to excitation energies of $\sim$~375~K. No transitions are detected in the IRAM survey. We derive \trot~=~130 and 50~K toward the hot core and extended ridge, respectively. 

\subsection{NH$_{3}$}
\label{s-nh3}

We observe ammonia (NH$_{3}$) toward the hot core, plateau, and extended ridge (Figs.~13.80 -- 13.84). We also detect $^{15}$NH$_{3}$ and NH$_{2}$D as well as the $\nu_{2}$~=~1 vibrationally excited mode from the main isotopologue. Emission from the $\nu_{2}$~=~1 mode and NH$_{2}$D are only detected toward the hot core. The main and $^{15}$N isotopologues are observed via low energy rotation transitions which include the ground state ortho and para lines up to the 3$_{0}$ -- 2$_{0}$ ortho transition (\eup~$\approx$~170~K). Because the deuterium atom breaks the symmetry of the ammonia molecule, we  observe NH$_{2}$D transitions within and across K ladders up to excitation energies of $\sim$~600~K. Vibrationally excited emission is detected from energy levels as high as $\sim$~2000~K. A detailed analysis of the NH$_{2}$D emission in the HIFI scan is presented by 
\citet{neill13b} for the purpose of obtaining a robust ammonia D/H ratio. The observed NH$_{3}$ and $^{15}$NH$_{3}$ transitions do not cover a large range in excitation energy. The lines from the main isotopologue, moreover, have a blue shifted absorption component, which we do not fit here. Consequently, we derive a \trot~=~250~K toward the hot core based on the $\nu_{2}$~=~1 and NH$_{2}$D emission. We also derive \trot~=~23 and 35~K for the extended ridge and plateau, respectively, though these values are not as well constrained because of the limited coverage in excitation energy. Because of the limited coverage in \eup, we do not report NH$_{3}$ abundances in this work. 

\subsection{OH}
\label{s-oh}

We observe the hydroxyl radical (OH) toward the plateau and the deuterated isotopologue (OD) toward the compact ridge (Figs.~13.85 -- 13.86). For both species, we detect only the $^{2}\Pi_{1/2}$ J~=~3/2 -- 1/2 doublet in the HIFI scan. We used \trot~=125~K, characteristic of the compact ridge, to fit the OD lines. We, however, note that this rotation temperature is not well constrained because only one \eup\ is sampled.  

\subsection{CH$^{+}$}
We detect the J~=~2 -- 1 transition of methyliumylidene (CH$^{+}$) in absorption (Fig.~13.87). The line profile has a \vlsr~$\approx$~8~km/s and \dv~$\approx$~5~km/s indicating a possible origin in the extended ridge, though the measured \dv\ is slightly wider than typical for this spatial/velocity component. We possibly detect the J~=~1 -- 0 transition with both emission and absorption components, although it is not clear because the baseline is irregular at this frequency. Because of the small \eup\ coverage and the fact that this species is detected in absorption, we do not include a model for CH$^{+}$. 

\subsection{H$_{2}$O$^{+}$}
We detect the ortho ground state rotation transition (1$_{1,1}$ -- 0$_{0,0}$) of the water cation (H$_{2}$O$^{+}$) in absorption (Fig.~13.88). Both fine structure components, J~=~3/2 -- 1/2 and J~=~1/2 -- 1/2, are detected but the latter is heavily blended with emission lines from other species. As discussed by \citet{gupta10}, who analyze the H$_{2}$O$^{+}$ emission in detail, the J~=~3/2 -- 1/2 transition contains a narrow absorption component at \vlsr~$\sim$~9~km/s consistent with the extended ridge. The J~=~3/2 -- 1/2 transition also displays a broad blue-shifted absorption component that extends to \vlsr~$\sim$~$-$45~km/s, similar to the J~=~1--0 transition of HF \citep[see][]{phillips10}, which likely originates from the LVF. \citet{gupta10} derive column densities of 7~$\pm$~2\sn$^{12}$~\cms\ and 1.0~$\pm$~0.3\sn$^{13}$~\cms\ for the narrow and broad components, respectively. Because of the small \eup\ coverage and the fact that this molecule is only detected in absorption, we do not provide an H$_{2}$O$^{+}$ model. 

\subsection{OH$^{+}$}
We detect the ground state rotation transition of the hydroxyl cation (OH$^{+}$) in absorption (Fig.~13.89). The J~=~1 -- 1 and 2 -- 1 fine structure components are clearly detected, while the J~=~0 -- 1 component is heavily blended. \citet{gupta10} analyze the OH$^{+}$ emission in detail. They find that the OH$^{+}$ lines are similar to those of H$_{2}$O$^{+}$ and HF, and contain both a narrow component at \vlsr~$\sim$~9~km/s, consistent with the extended ridge, and a wider blue-shifted component which originates in the LVF.  \citet{gupta10} derive column densities of 9~$\pm$~3\sn$^{12}$~\cms\ and 1.9~$\pm$~0.7\sn$^{13}$~\cms\ for the narrow and broad components, respectively. Because of the small \eup\ coverage and the fact that this molecule is only detected in absorption, we do not provide an OH$^{+}$ model. 

\subsection{Atomic Species}
\label{s-c}

We detect the $^{3}P_{1}$ -- $^{3}P_{0}$ and $^{3}P_{2}$ -- $^{3}P_{1}$ transitions of neutral atomic carbon (CI, Fig.~13.90) and the $^{2}P_{3/2}$ -- $^{2}P_{1/2}$ transition of singly ionized carbon (CII, Fig.~13.91). The CI and CII line profiles contain emission and absorption components. There are at least two emission components in the CI profiles: one wide (\vlsr~$\sim$~8.5~km/s,  \dv~$\sim$~13~km/s), suggesting an origin in the plateau or \hotcores, and one more narrow (\vlsr~$\sim$~7.5~km/s, \dv~$\sim$~3~km/s), indicating emission from the compact ridge. In the CI lines,  the absorption component has \vlsr~$\sim$~11.5~km/s and \dv~$\sim$~2~km/s, suggesting absorption from the extended ridge. The CII line also contains at least two emission components possibly originating from the plateau (\vlsr~$\sim$~7.5~km/s,  \dv~$\sim$~17~km/s) and compact ridge (\vlsr~$\sim$~8.0~km/s,  \dv~$\sim$~4~km/s). The absorption component in the CII line is more prominent than in the CI transitions, and is wider (\vlsr~$\sim$~8.0~km/s,  \dv~$\sim$~6~km/s) than what is typical for the extended ridge. Atomic emission toward Orion~KL is more extended than the DBS mode 3\arcmin\ chopper throw \citep[see e.g.][]{werner84, genzel89, herrmann97, ikeda99, ikeda02}. Emission in the off beam, therefore, likely contributes to some of the perceived absorption observed in the CI and CII lines. Lastly, we do not detect the $^{3}P_{1}$ -- $^{3}P_{0}$ transition of NII at 1461.1~GHz. Because of the off beam contamination and the fact that we are focused on molecular emission in this study, we do not include models for CI and CII. 

\section{Discussion}
\label{s-disc}

The \redchisq\ values reported in Table~\ref{t-chisq}, in general, indicate excellent agreement between the data and models. Most of the complex organics, in particular, have low \redchisq\ values ($\lesssim$~1.5), indicating LTE is a good approximation for these species and suggests an origin in high density gas ($>$~10$^{6}$~\cmc). As a result, abundances of complex molecules derived from our LTE fits should be robust estimates toward Orion~KL. The fact that CH$_{3}$CN abundances derived from our XCLASS and more advanced non-LTE models agree closely with one another supports this assertion. LTE models, therefore, may be sufficient to derive robust abundances for complex organics toward other massive hot cores, where gas densities are high. The exceptions to this are some isotopologues (including the main isotopologue) of methanol and the $\nu_{8}$~=~1 CH$_{3}$CN model which have \redchisq~$\ge$~3.5. These high \redchisq\ values are likely due to a combination of issues brought on by high optical depth, non-LTE excitation, and radiative pumping (see Sec.~\ref{s-lstats}). 

In addition to molecular abundances, our XCLASS models provide rotation temperatures for each species, which, under the assumption of LTE, estimates the local gas kinetic temperature. The distribution of rotation temperatures toward each spatial/velocity component therefore yields insight into the thermal structure within each region. Fig.~\ref{p-trot} shows histograms of rotation temperature in four panels, each representing a different spatial/velocity component. For molecules fit with temperature gradients, we only include the \trot\ which corresponds to the sub-component that dominates the emission in the HIFI survey. Species with multiple detected isotopologues are only counted once. We exclude rotation temperatures derived for NH$_{3}$ toward the plateau and extended ridge, as well as OD toward the compact ridge because they were fit using transitions over a small range in \eup\ (see Secs.~\ref{s-nh3} and \ref{s-oh}). Within each panel, the black solid curve corresponds to the \trot\ distribution of all molecules detected toward that component. The solid green, dashed red, and dashed blue lines represent cyanides, sulfur bearing molecules, and complex oxygen bearing organics, respectively. More discussion on the chemical implications of our results will be given in a future work (Crockett et al. 2014, in preparation).

Examining Fig~\ref{p-trot}, we see the spread in rotation temperature is much larger toward the hot core relative to the other spatial/velocity components, suggesting the hot core has the most heterogeneous thermal structure toward Orion~KL. In particular, there are a significant number of hot core molecules with \trot~$>$~175~K, which is where the other distributions, (save a few species in the compact ridge), terminate. Interferometric studies of molecular emission toward the hot core have certainly demonstrated that this region is both chemically and physically intricate, having structure on scales $\lesssim$~1-2\arcsec\ \citep{wang10, goddi11}. Examining which molecules have the highest \trot\ toward the hot core, we see that  cyanides (green histogram) account for 5 of the 11 molecules with \trot~$>$~200~K. We note the cyanide in the 125 -- 150~K bin, C$_{2}$H$_{5}$CN with \trot~=~136~K, requires an additional compact 300~K sub-component to fit lines at high excitation. In general, sulfur bearing molecules do not have particularly high rotation temperatures toward the hot core, although S-bearing species which also contain oxygen (i.e. OCS, SO, and SO$_{2}$) all have rotation temperatures $>$~175~K. 

It is possible that the hot core \trot\ distribution is influenced by radiative pumping from the strong IR continuum characteristic of this region \citep{genzel89, debuizer12}.  Detailed non-LTE studies of emission from H$_{2}$O 
\citep{neill13a} and H$_{2}$S 
\citep{crockett14} within the Orion KL HIFI scan require an enhanced far-IR continuum ($\lambda$~$<$~100~\um) compared to what is observed to explain the emission of both these light hydrides. Because much of the far-IR is optically thick, this suggests a possible origin near a hidden self-luminous source or sources. Other studies have also needed to invoke strong IR radiation fields to explain NH$_{3}$ \citep{hermsen88} and HDO \citep{jacq90} emission toward the hot core. We note that all hot core tracers with detected vibrational modes, except CH$_{3}$OH, have rotation temperatures $>$~175 K. Most of these vibrationally excited modes are likely populated via pumping from the mid/far-IR. Thus it is possible that the local radiation field is redistributing the population levels of these molecules to reflect a higher \trot\ than the actual kinetic temperature. 

The compact ridge \trot\ distribution is much narrower than that of the hot core with a little less than half of the distribution falling in the 100 -- 125~K bin, indicating a more quiescent, uniform environment. Most of the complex O-bearing molecules have rotation temperatures near the peak of this distribution, with the exception of NH$_{2}$CHO which has \trot~=~190~K. There are, however, several high temperature outliers. Just as in the hot core, CH$_{3}$CN and OCS are on the high end of the \trot\ distribution. Unlike the hot core, HCN and HNC do not appear to be particularly warm (\trot~=~120~K and 80~K, respectively). The other species populating the high temperature part of the compact ridge distribution are NS and CS, both of which do not have particularly high \trot\ values toward the hot core (\trot~$\approx$~100~K). 

To determine if these higher \trot\ molecules correspond to a spatially distinct region within the compact ridge, we once again employ the ALMA-SV line survey. Emission from two molecules which have high \trot\ measurements toward the compact ridge, OCS (\trot~=~165~K) and CS (\trot~=~225~K), are present in the ALMA scan. Fig.~\ref{p-almacr} plots integrated emission maps of the OCS 20 -- 19 and CS 5 -- 4 in the left and right panels as blue contours, respectively. The emission is integrated between \vlsr~=~7.6 -- 8.3~km/s to avoid contributions from the hot core and plateau. We also overlay emission from CH$_{3}$OCHO 18$_{7,11}$ -- 17$_{7,10}$, in both panels as yellow contours. Methyl formate is a prominent compact ridge tracer with a derived \trot~=~110~K, close to the peak of the compact ridge \trot\ distribution. The CH$_{3}$OCHO contours surround a cavity that is likely carved out by the LVF \citep{irvine87, plambeck09, favre11}. The two southern methyl formate peaks, furthermore, correspond to MF1 and MF3 using the labeling conventions of \citet{favre11}. The continuum is also overlaid on both panels as a gray scale, which mainly traces IRc2, the infrared source closest to the hot core. From the plot, we see that both the OCS and CS peaks are offset from MF1 and MF3, occurring midway between the two and closer to the edge of the CH$_{3}$OCHO cavity. The higher rotation temperatures may therefore be a result of the interactions between the LVF and the compact ridge. 

The plateau has a narrower \trot\ distribution than the hot core. Although not as centrally peaked as the compact ridge, the plateau does not have a conspicuous high temperature ($>$~175~K) tail in its distribution. Unlike the hot core, cyanides are not among the hottest molecules detected toward this component. Although we note torsionally excited CH$_{3}$CN emission is observed toward the plateau, from which we compute a vibration temperature of 171~K (see Sec.~\ref{s-vib}). This indicates methyl cyanide may be probing hotter material toward the plateau than the rotation temperature (\trot~=~120~K) would suggest. As with the hot core, sulfur bearing species are spread throughout the distribution with the hottest being the O/S-bearing species SO and SO$_{2}$. OCS, on the other hand, traces somewhat cooler gas. We note that the highest \trot\ molecule detected toward the plateau is HNCO (\trot~=~168~K), which also displays one of the highest rotation temperatures toward the hot core (\trot~=~209~K). This is likely the result of pumping from the far-IR, which \citet{churchwell86} has shown is significant for this molecule (see Sec.~\ref{s-hnco}).

Finally, the extended ridge displays the narrowest of all four temperature distributions centered at $\sim$~40~K, exemplifying the cool, quiescent nature of this region. Although there is not much variation in \trot, sulfur bearing molecules tend to be warmer than average.  Cyanides, on the other hand, are on the cooler side of the distribution. 

\section{Conclusions}
\label{s-con}

We have presented a full band analysis of the HIFI spectral survey toward Orion KL, which spans a frequency range from 480 to 1907~GHz. In all, we detect emission from \nmols\ molecules (\niso\ isotopologues). Combining this dataset with a ground based mm line survey, we modeled the molecular emission using the XCLASS program, which assumes LTE level populations. A subset of the molecules detected in this survey was also modeled using the MADEX non-LTE code. A \redchisq\ analysis demonstrates that there is excellent agreement between the data and models for most molecules. In particular, complex organics are well fit by our LTE models, which indicates these species are emitting from high density gas (\nht~$>$~10$^{6}$~\cmc). According to our LTE fits, we detect $\sim$~\nlines\ lines in the HIFI scan and estimate a U line fraction of \ulfrac. Methanol is the most emissive molecule in the HIFI survey both in terms of total integrated intensity and number of detected lines.  We detect vibrationally/torsionally excited emission from 10 species and find evidence for radiative pumping of HCN in the hot core and CH$_{3}$CN in the plateau. We also isolate a spatial/velocity component that has line parameters ``in between" that expected from the hot core and compact ridge (\dv~$\sim$~5 -- 10~km/s, \vlsr~$\sim$~6.5 -- 8~km/s), which likely peaks $\sim$~1\arcsec\ south of the hot core sub-mm continuum peak consistent with the high water column density clump discovered by 
\citet{neill13a}.

Using column densities derived from our models, we computed molecular abundances toward the hot core, compact ridge, plateau, and extended ridge. Abundances derived from our XCLASS and more advanced MADEX models typically agree to within a factor of 3, indicating our LTE abundances are reliable. Plotting the distribution of rotation temperatures toward each spatial/velocity component reveals the HIFI scan probes a diversity of environments toward Orion KL. The \trot\ distribution for the hot core is significantly wider than the others, suggesting the hot core harbors a more complex thermal structure than the other spatial/velocity components. Although the compact ridge \trot\ distribution is quite narrow, it does have a high temperature tail. Interferometric maps of OCS and CS, two molecules at the high temperature end of the compact ridge distribution, show these molecules peak closer to the LVF than CH$_{3}$OCHO, a species with a rotation temperature near the peak of the compact ridge \trot\ distribution. The higher temperature tail thus may be the result of interactions between the LVF and compact ridge. The relative and absolute abundances derived in this study, along with the publicly available data and molecular fits, which extend from below 100~GHz to beyond 1.9~THz, represent a legacy for comparison to other sources and chemical models.

\acknowledgments
HIFI has been designed and built by a consortium of institutes and university departments from across Europe, Canada and the United States under the leadership of SRON Netherlands Institute for Space Research, Groningen, The Netherlands and with major contributions from Germany, France and the US. Consortium members are: Canada: CSA, U.Waterloo; France: CESR, LAB, LERMA, IRAM; Germany: KOSMA, MPIfR, MPS; Ireland, NUI Maynooth; Italy: ASI, IFSI-INAF, Osservatorio Astrofisico di Arcetri-INAF; Netherlands: SRON, TUD; Poland: CAMK, CBK; Spain: Observatorio Astron\'{o}mico Nacional (IGN), Centro de Astrobiolog\'{i}a (CSIC-INTA). Sweden: Chalmers University of Technology - MC2, RSS \& GARD; Onsala Space Observatory; Swedish National Space Board, Stockholm University - Stockholm Observatory; Switzerland: ETH Zurich, FHNW; USA: Caltech, JPL, NHSC. HIPE is a joint development by the Herschel Science Ground Segment Consortium, consisting of ESA, the NASA Herschel Science Center, and the HIFI, PACS and SPIRE consortia. 

The National Radio Astronomy Observatory is a facility of the National Science Foundation operated under cooperative agreement by Associated Universities, Inc. 

This paper makes use of the following ALMA data: ADS/JAO.ALMA\#2011.0.00009.SV. ALMA is a partnership of ESO (representing its member states), NSF (USA) and NINS (Japan), together with NRC (Canada) and NSC and ASIAA (Taiwan), in cooperation with the Republic of Chile. The Joint ALMA Observatory is operated by ESO, AUI/NRAO and NAOJ.

Support for this work was provided by NASA through an award issued by JPL/Caltech.

\appendix
\section{Reduced \chisq\ Calculations and Uncertainty Estimates}
\label{s-apx}

We calculate reduced chi squared, \redchisq, statistics for each model which are reported in Table~\ref{t-chisq}. This calculation is performed by first uniformly smoothing the data to a velocity resolution of $\sim$~0.5~km/s.  Residuals between the molecular fits and data are then computed by stepping across the spectrum in frequency intervals corresponding roughly to 1000~km/s. At each interval, we calculate the RMS noise level and subtract a local baseline. For a channel to be included in the \redchisq\ calculation of a given molecular fit, the model emission has to be $>$~3 times the local RMS and responsible for 95\% of the total flux when compared to the entire full band model. That is, the intensity ratio between the fit being considered and the full band model has to be $\ge$~0.95 in order to avoid channels that contain emission from more than one molecule (i.e. blended lines).  The uncertainty in each channel, \sigtot, is computed by adding in quadrature the contributions from several different sources of error,
\begin{equation}
\sigma_{tot} = \sqrt{\sigma_{bl}^{2} + \sigma_{rms}^{2} + \sigma_{cal}^{2} + \sigma_{pt}^{2} +  \sigma_{bf}^{2} + \sigma_{\tau_{d}}^{2}}.
\label{e-sigtot}
\end{equation}
Here, $\sigma_{bl}$ is a baseline offset uncertainty,  $\sigma_{rms}$ is the local RMS noise level,  $\sigma_{cal}$ is the calibration uncertainty,  $\sigma_{pt}$ is the pointing error, and $\sigma_{bf}$ is the error in beam filling factor brought on by the source size uncertainty, and $\sigma_{\tau_{d}}$ is the error due to our uncertainty in the dust optical depth. The method used to compute each source of error is given below. All uncertainties are propagated using the ``error propagation equation" of \citet[][their Eq.~3.13]{bevington03},

$\bm{\sigma_{bl}}$ -- This represents offsets in the baseline that still exist despite having subtracted a baseline within each 1000~km/s interval. These offsets exist at approximately the 0.05~K level, we thus take $\sigma_{bl}$~=~0.05~K. 

$\bm{\sigma_{rms}}$ -- This is the RMS noise level in the baseline. Emission line channels are rejected by excluding channels predicted to be in emission by the full band model. We also use a sigma clip algorithm to remove any additional channels in emission from U lines. 

$\bm{\sigma_{cal}}$ -- We assume a 10\% calibration uncertainty for each channel \citep{roelfsema12}. 

$\bm{\sigma_{pt}}$ -- The pointing uncertainty is computed assuming a Gaussian beam and the nominal \herschel\ absolute pointing error (APE) of 2\arcsec\ \citep{pilbratt10}. The percent error in the intensity of each channel brought on by the APE, $\delta_{p}$, is thus,
\begin{equation}
\delta_{p} = 1-exp\left[-~\frac{APE^{2}}{2(\theta_{b}/2.355)^{2}} \right],
\end{equation}
where $\theta_{b}$ is the beam size. 

$\bm{\sigma_{bf}}$ -- Because we do not vary the source size in our molecular fits, we include a beam filling factor uncertainty. We take the error in source size to be 20\%. Assuming the shape of the telescope beam and source are Gaussian, the beam filling factor, $\eta_{bf}$, is,
\begin{equation}
\eta_{bf} = \frac{\theta_{s}^{2}}{\theta_{s}^{2} + \theta_{b}^{2}}.
\label{e-bff}
\end{equation}
The beam filling factor is directly proportional to the observed intensity. Propagating the uncertainty, we derive an expression for the percent error in the intensity of each channel due to our assumed \ssize\ uncertainty, $\delta_{ss}$, 
\begin{equation}
\delta_{ss} = 2 \ \eta_{bf}(1-\eta_{bf}) \frac{\sigma_{s}}{\theta_{s}}
\end{equation}
where $\sigma_{s}$ is the error in the source size and $\sigma_{s}$~/~$\theta_{s}$ is the percent error in the source size. In instances when more than one spatial/velocity component is detected for a given species, an intermediate ``effective" source size, $\theta_{eff}$, is used for the \ssize\ in Eq.~\ref{e-bff}. Adopted $\theta_{eff}$ values are given in Table~\ref{t-chisq}.

$\bm{\sigma_{\tau_{d}}}$ -- We find that extinction due to dust significantly reduces the observed line profiles particularly at high frequencies ($\gtrsim$~800~GHz). We account for the uncertainty in \taud\ by assuming a 1~$\sigma$ error of 30\% in \nhh\ and propagating this uncertainty using Eq.~\ref{e-taud}. This is the same error assumed by 
\citet{crockett14}, and encompasses \nhh\ values reported in various studies \citep{plume12, favre11, mundy86, genzel89} at the 3~$\sigma$ level when we set \nhh~=~2.5\sn$^{24}$~\cms\ (see Sec.~\ref{s-xclass}). The observed main beam temperature goes as \tmb~$\propto$~$e^{-\tau_{d}}$, the percent error due to \taud\ is thus equal to $\sigma_{\tau_{d}}$. 

Our derived column densities are based on optically thin emission, which, in this limit, is directly proportional to the observed intensity. The uncertainty in \tmb\ is given by Eq.~\ref{e-sigtot}. We note that, in our XCLASS models, line intensity is also a function of rotation temperature. However, for most molecules, we observe transitions over a comprehensive range in excitation energy. As a result, we are able to reasonably decouple \ntot\ and \trot\ from one another.  We estimate an \ntot\ uncertainty of 25\% for XCLASS, which we compute using Eq.~\ref{e-sigtot} at 1~THz assuming the sources of error given above. This value is consistent with \ntot\ uncertainties estimated in previous studies which use MADEX to model the Orion~KL IRAM survey 
\citep[Marcelino et al. 2014, in preparation;][]{tercero10, tercero11, esplugues13a, esplugues13b}.
Because the modeling methodology is similar here, we adopt the same \ntot\ 25\% error for MADEX. Applying the same calculation (using Eq.~\ref{e-sigtot} at 1~THz) to Table~\ref{t-intflux}, we estimate a 25\% error in the integrated intensities reported for individual spatial/velocity components. We note that this uncertainty may be an underestimate for molecular fits with high \redchisq\ values (i.e. \redchisq~$>$~3.0, see Sec.~\ref{s-lstats}).

We estimate a 10\% uncertainty for \trot\ in our XCLASS models. This uncertainty is commensurate with the \trot\ percent error derived by 
\citet{crockett14} for \hts, who report \trot~=~141~$\pm$~12~K, yielding a percent error of 9\%. We take this uncertainty as representative of all molecular species and note that changes of 10\% in \trot\ produce noticeable deviations by eye in models that fit the data well. 

We adopt errors in the \vlsr\ values reported in Tables~\ref{t-xclass} and \ref{t-madex} of $\pm$~1~km/s for the hot core, compact ridge, and extended ridge, and $\pm$~2~km/s for the plateau. We also estimate \dv\ uncertainties of $\pm$~0.5~km/s, 1.5~km/s, and 5.0~km/s, for the compact/extended ridge, hot core, and plateau, respectively. These uncertainties are derived by visual comparison of the models and data. Changing the \vlsr\ or \dv\ by increments equal to or greater than these uncertainties produce noticeable differences in model line profiles. The errors are smaller for spatial/velocity components with narrower line profiles because variations in \dv\ and \vlsr\ are more easily detected in these ``sharper" components. We note that our uncertainty estimates for \dv\ and \vlsr\ toward each spatial/velocity component are also commensurate with the standard deviation of these quantities for all molecules detected toward each component. 

We estimate a 40\% error in molecular abundance. We arrive at this value by assuming a 25\% error in column density, a 30\% error in \nhh, and propagating the uncertainty, which amounts to adding these quantities in quadrature with one another. 

\tvib\ errors reported in Table~\ref{t-vib} are computed assuming a 25\% error in the column density of each vibrational mode and using Eq.~\ref{e-boltz} to propagate the uncertainties.

\clearpage

\bibliographystyle{apj}
\bibliography{ms.bbl}

\clearpage

\clearpage

\clearpage

\clearpage

\clearpage

\clearpage

\clearpage

\clearpage

\clearpage

\clearpage

\clearpage

\clearpage

\clearpage

\figsetstart
\figsetnum{13}
\figsettitle{Individual XCLASS models}

\figsetgrpstart
\figsetgrpnum{13.13}
\figsetgrptitle{H$^{13}$CN}
\figsetplot{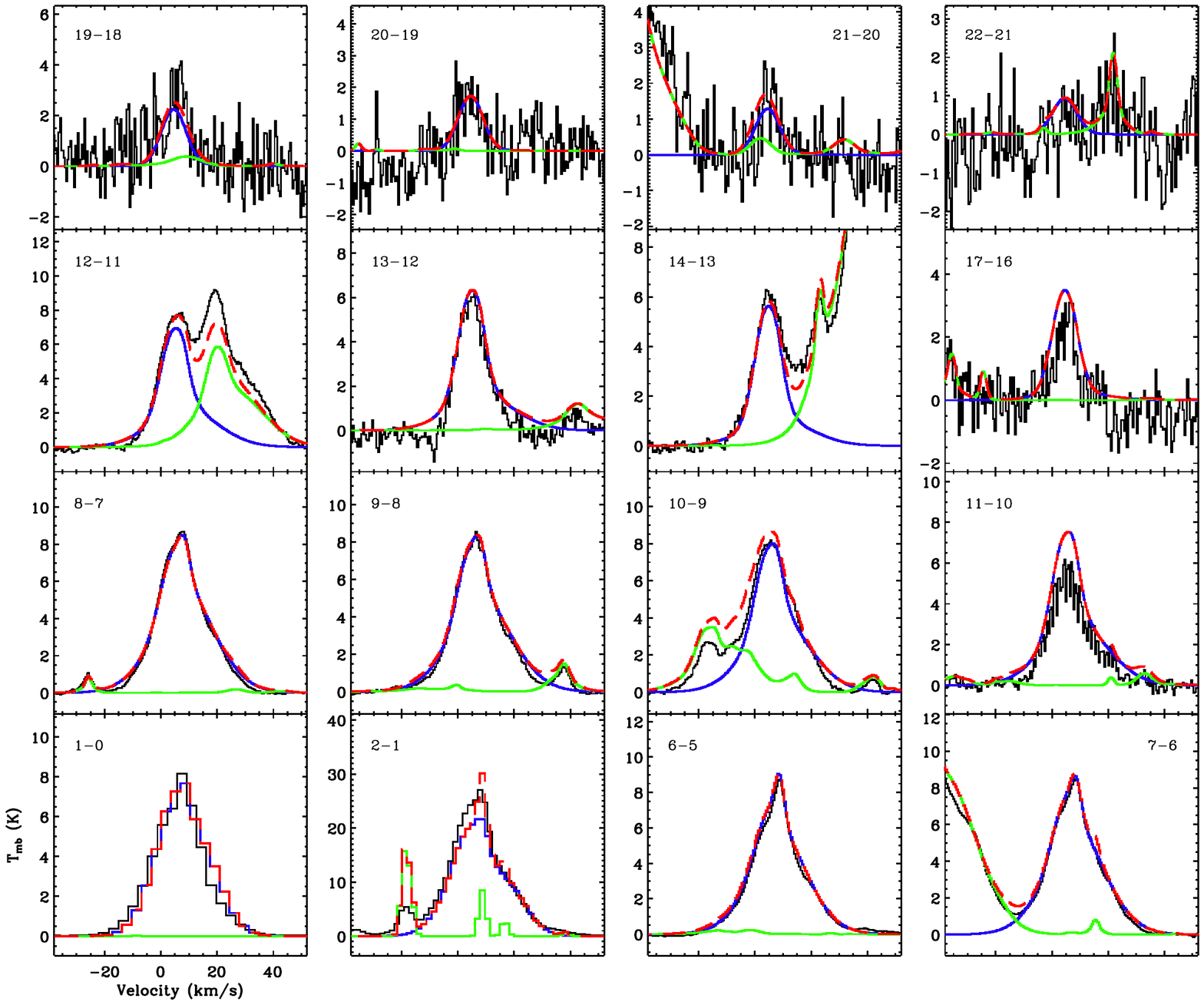}
\figsetgrpnote{Observed transitions in the HIFI and IRAM surveys (black) with the XCLASS 
model for the molecule being considered overlaid as a blue solid line. Modeled 
emission from all other species (solid green) along with the total fit (dashed 
red) are also overlaid. Quantum numbers labeling each transition are given in 
each panel and excitation energy increases from the lower left panel to the 
upper right. See Sec.~5 for a description of the quantum numbers. 
}
\figsetgrpend

\figsetgrpstart
\figsetgrpnum{13.33}
\figsetgrptitle{H$_{2}$CS}
\figsetplot{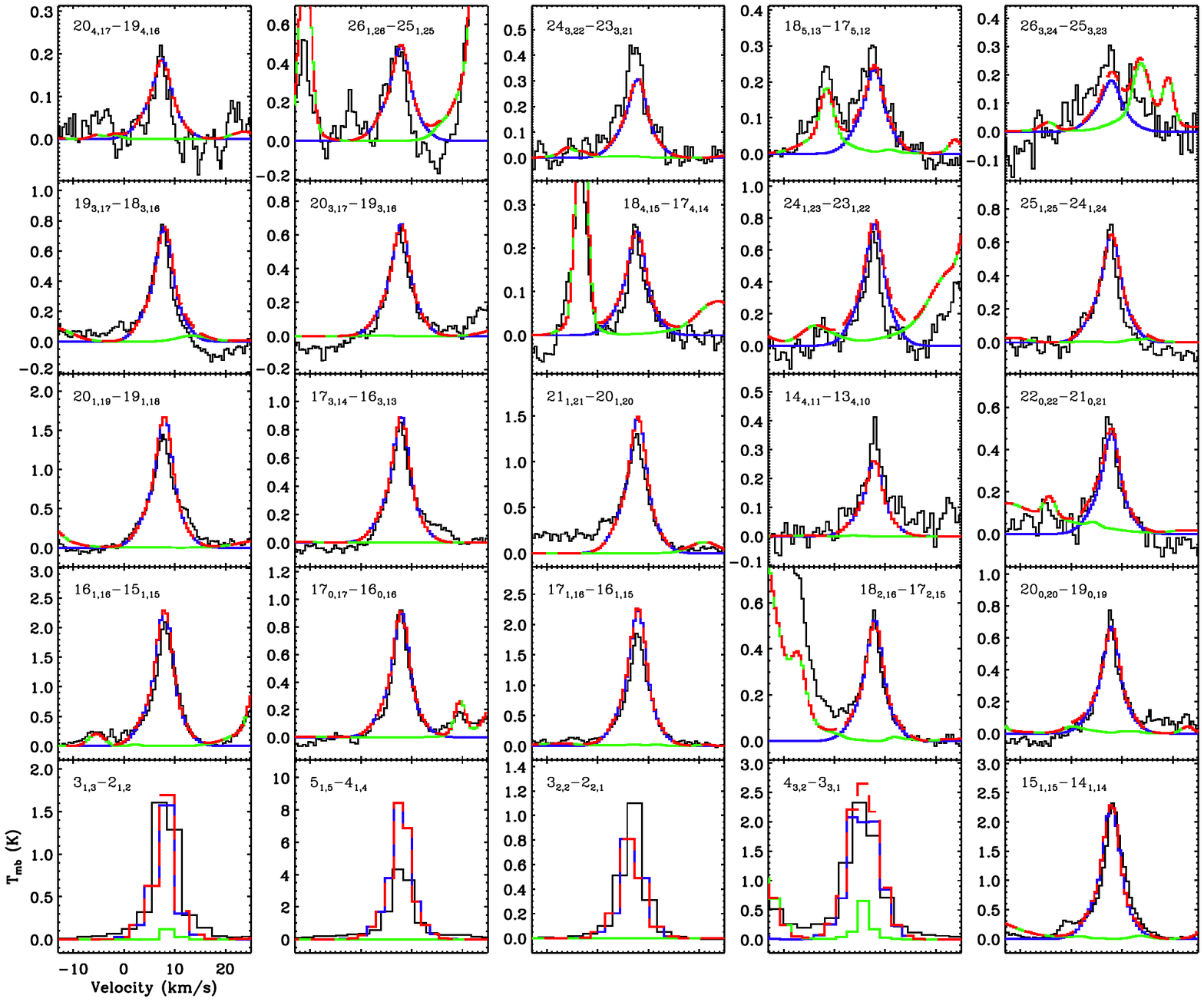}
\figsetgrpnote{Observed transitions in the HIFI and IRAM surveys (black) with the XCLASS 
model for the molecule being considered overlaid as a blue solid line. Modeled 
emission from all other species (solid green) along with the total fit (dashed 
red) are also overlaid. Quantum numbers labeling each transition are given in 
each panel and excitation energy increases from the lower left panel to the 
upper right. See Sec.~5 for a description of the quantum numbers. 
}
\figsetgrpend

\figsetgrpstart
\figsetgrpnum{13.66}
\figsetgrptitle{CH$_{3}$OCHO}
\figsetplot{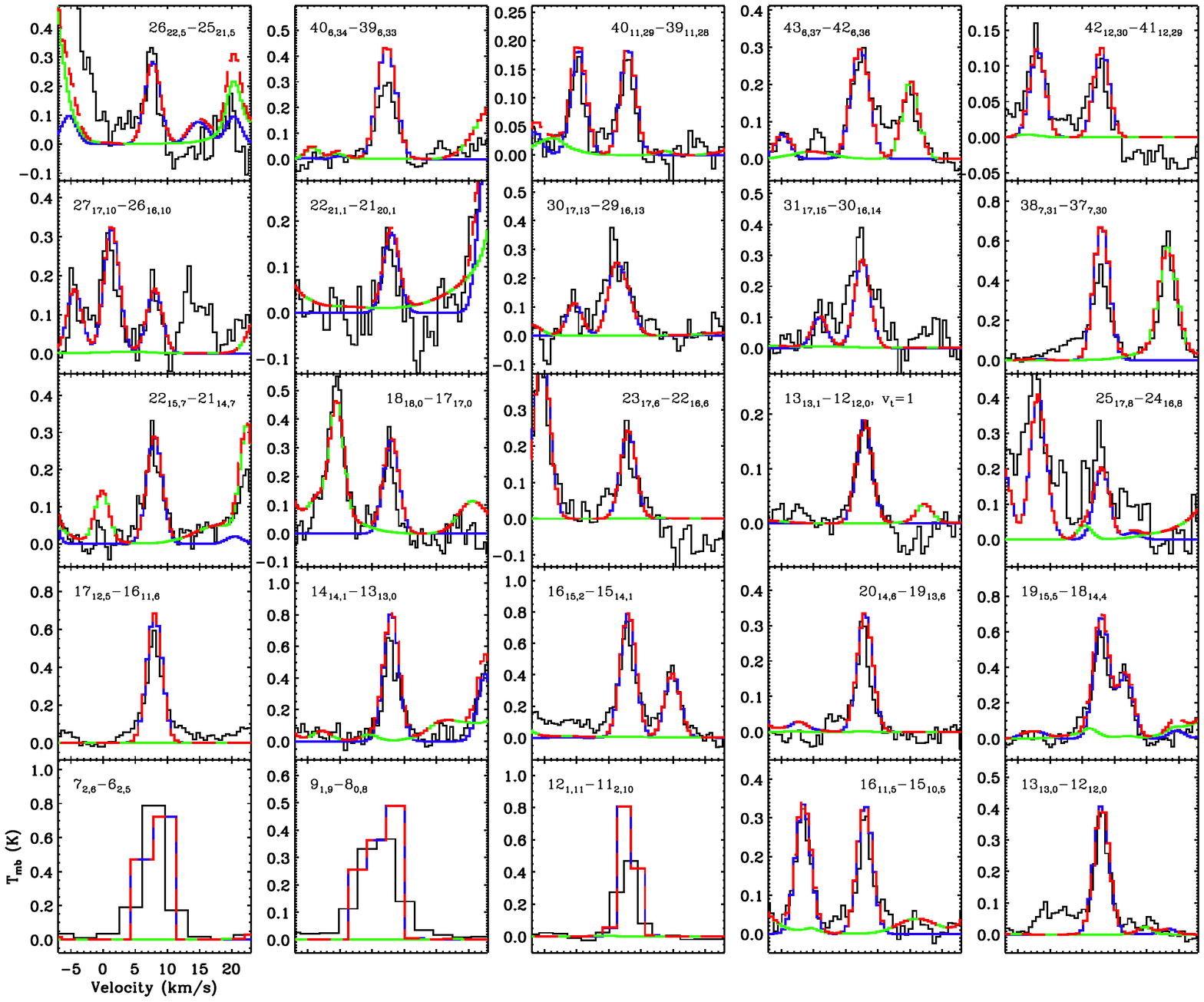}
\figsetgrpnote{Observed transitions in the HIFI and IRAM surveys (black) with the XCLASS 
model for the molecule being considered overlaid as a blue solid line. Modeled 
emission from all other species (solid green) along with the total fit (dashed 
red) are also overlaid. Quantum numbers labeling each transition are given in 
each panel and excitation energy increases from the lower left panel to the 
upper right. See Sec.~5 for a description of the quantum numbers. 
}
\figsetgrpend

\figsetend

\clearpage

\clearpage

\clearpage

\figsetstart
\figsetnum{14}
\figsettitle{Individual MADEX models}

\figsetgrpstart
\figsetgrpnum{14.1}
\figsetgrptitle{CH$_{3}$CN-A}
\figsetplot{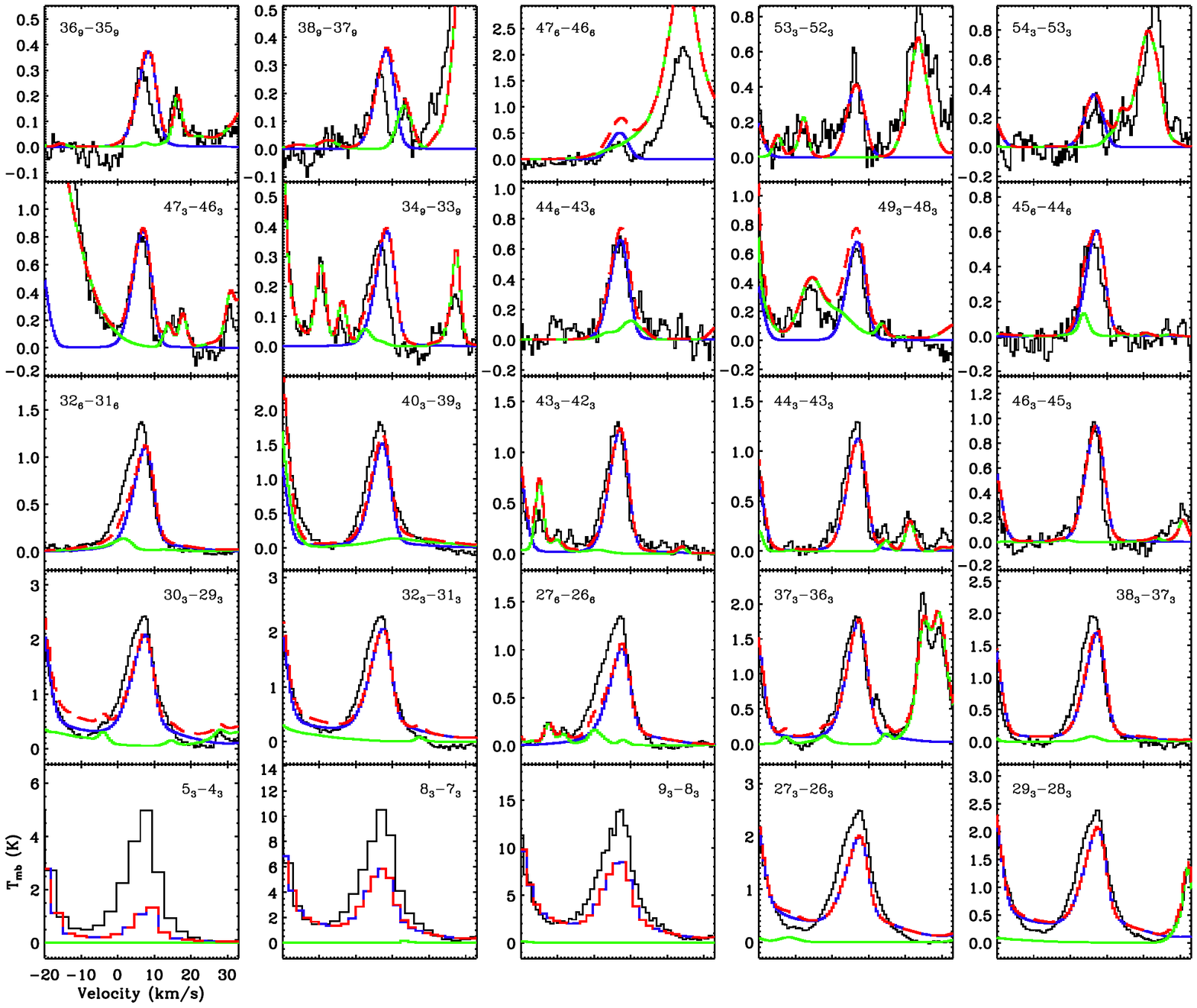}
\figsetgrpnote{Observed transitions in the HIFI and IRAM surveys (black) with the MADEX 
model for the molecule being considered overlaid as a blue solid line. Modeled 
emission from all other species (solid green) along with the total fit (dashed 
red) are also overlaid. Quantum numbers labeling each transition are given in 
each panel and excitation energy increases from the lower left panel to the 
upper right. See Sec.~5 for a description of the quantum numbers. 
}
\figsetgrpend

\figsetgrpstart
\figsetgrpnum{14.14}
\figsetgrptitle{$^{34}$SO}
\figsetplot{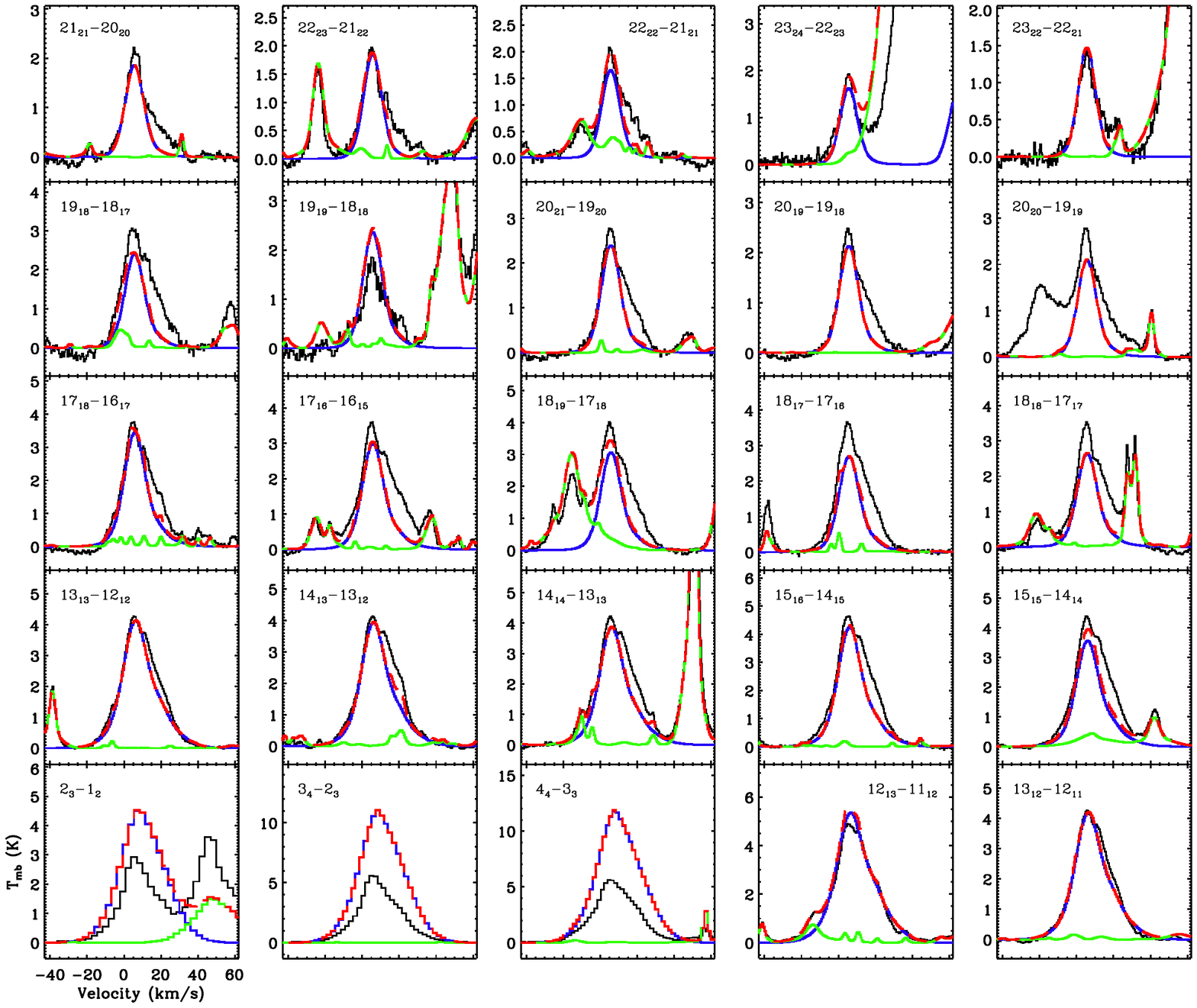}
\figsetgrpnote{Observed transitions in the HIFI and IRAM surveys (black) with the MADEX 
model for the molecule being considered overlaid as a blue solid line. Modeled 
emission from all other species (solid green) along with the total fit (dashed 
red) are also overlaid. Quantum numbers labeling each transition are given in 
each panel and excitation energy increases from the lower left panel to the 
upper right. See Sec.~5 for a description of the quantum numbers. 
}
\figsetgrpend

\figsetend

\clearpage

\clearpage

\clearpage

\clearpage

\clearpage

\begin{figure}
\figurenum{18}
\epsscale{1.0}
\includegraphics[angle=90, scale=0.63]{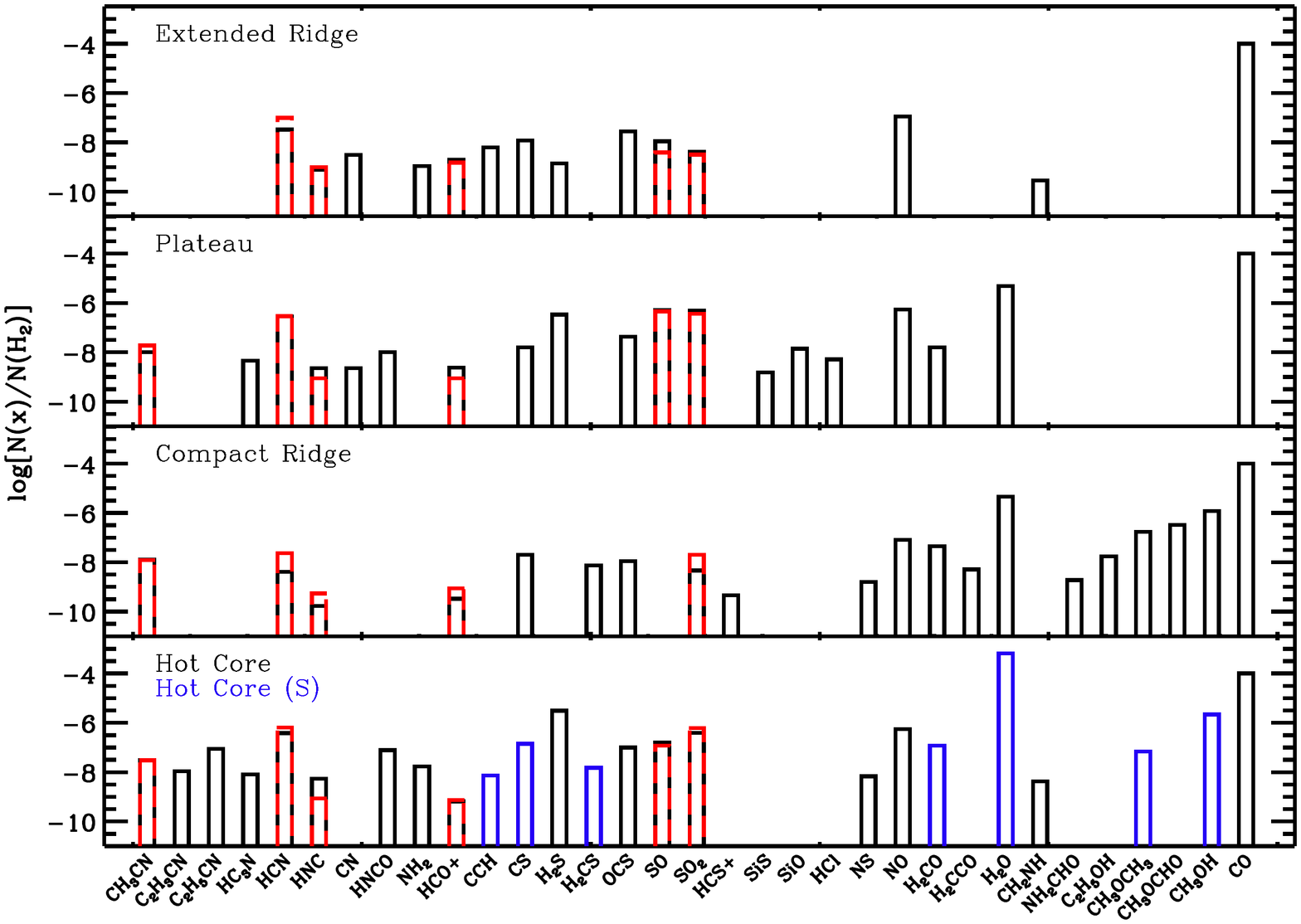}
\caption{Derived abundances plotted as a function of molecule. Each panel corresponds to a different spatial/velocity component which is labeled in the upper left hand corner of each panel. Abundances derived using XCLASS and MADEX are plotted as solid black and dashed red lines, respectively. XCLASS abundances plotted in blue indicate an origin from hot core (s).  \label{p-abund}}
\end{figure}

\clearpage

\begin{figure}
\figurenum{19}
\epsscale{0.83}
\plotone{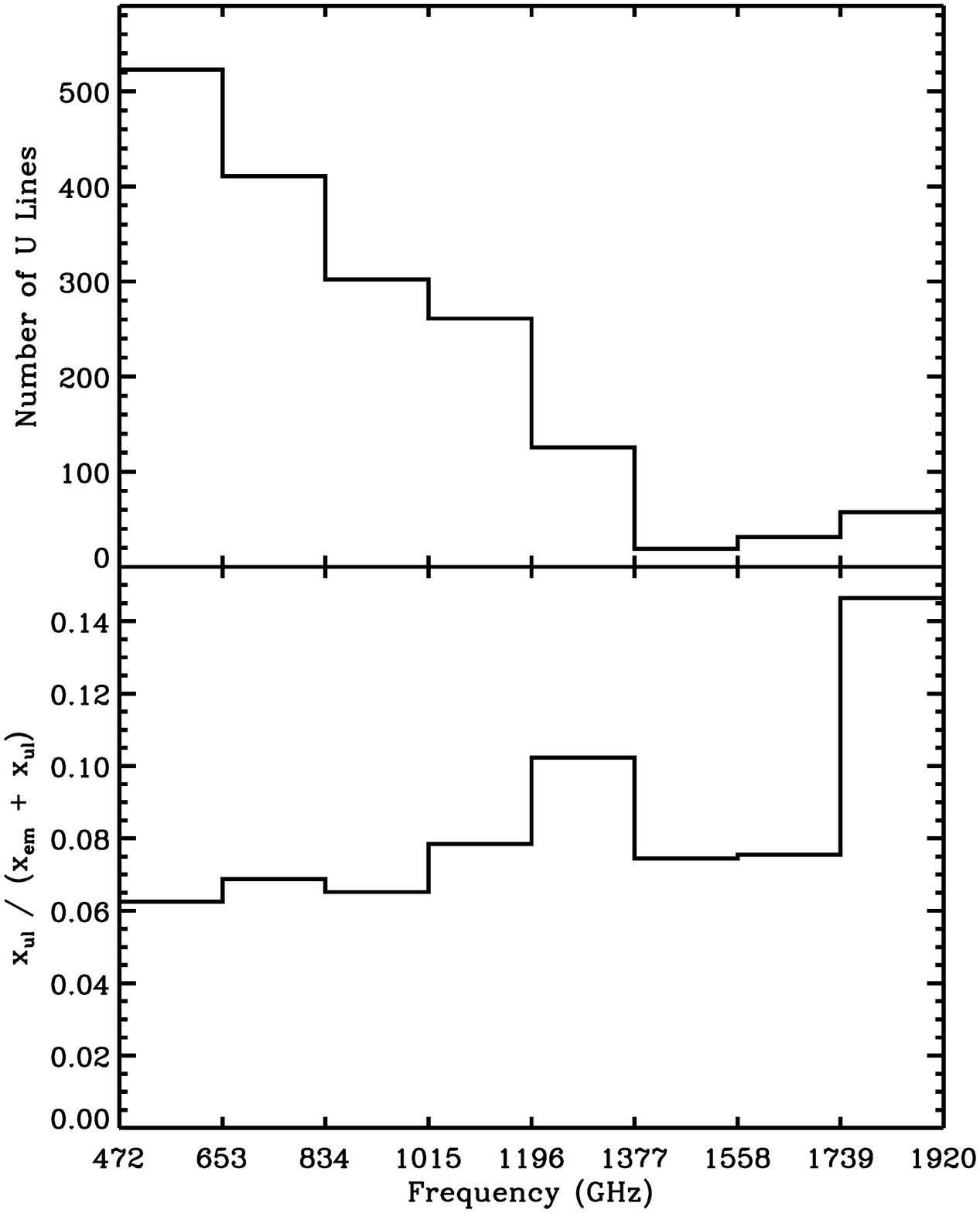}
\caption{Histogram of the number of U lines as a function of frequency (top) and the fraction of U line channels as a function of frequency (bottom). Both histograms have a bin width of 181~GHz. \label{p-uline}}
\end{figure}

\clearpage

\begin{figure}
\figurenum{20}
\epsscale{1.0}
\includegraphics[angle=90, scale=0.63]{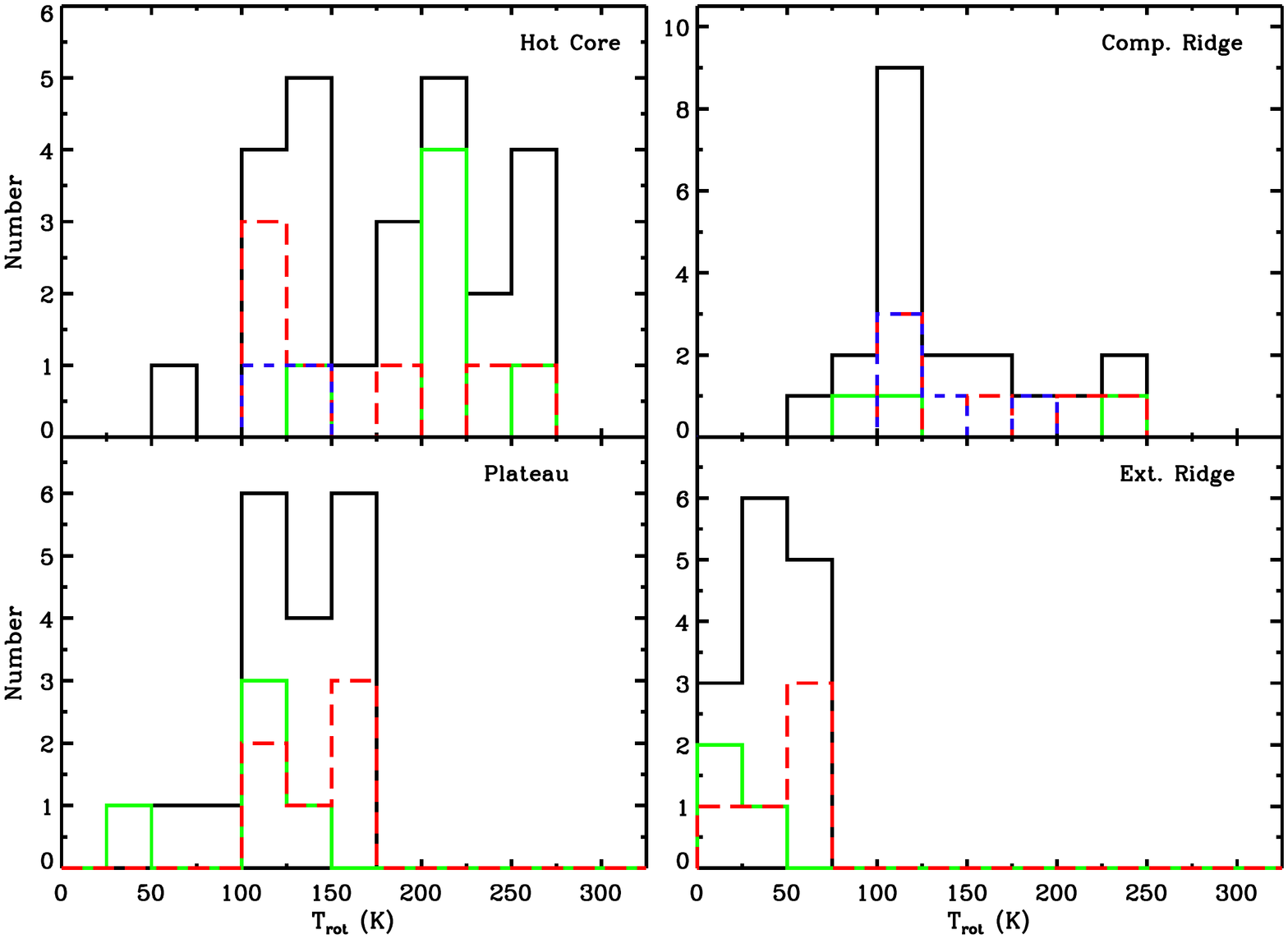}
\caption{Rotation temperature histograms originating from the individual spatial/velocity components. Rotation temperatures from cyanides, sulfur bearing molecules, and complex oxygen bearing organics are plotted as solid green, dashed red, and dashed blue lines, respectively. \trot\ values from all molecules are plotted as a black solid line. \label{p-trot}}
\end{figure}

\clearpage

\begin{figure}
\figurenum{21}
\epsscale{0.95}
\plotone{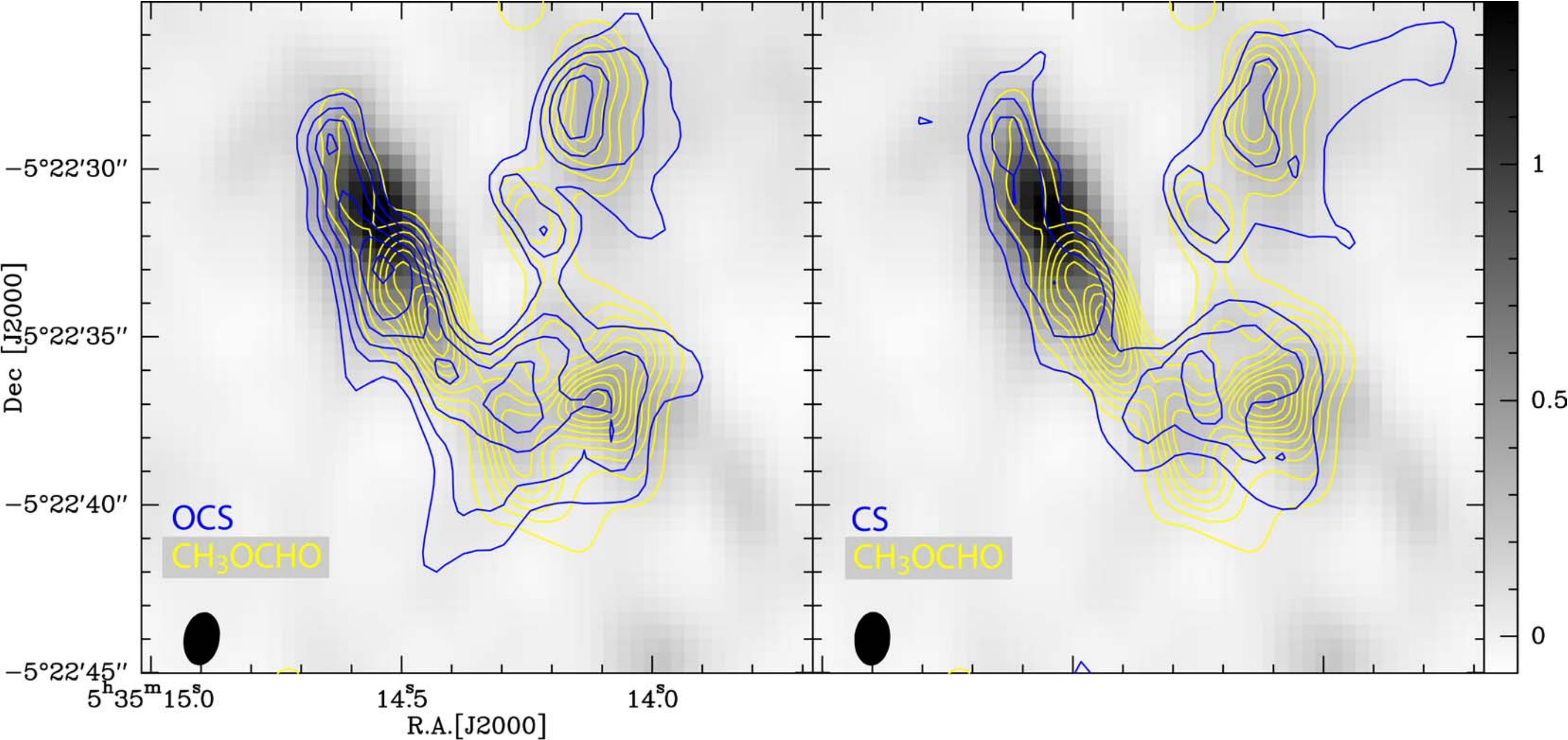}
\caption{Integrated intensity maps of OCS 20 -- 19 and CS 5 -- 4 between \vlsr~=~7.6 -- 8.3~km/s (blue contours) from the ALMA-SV survey. The first contour corresponds to 1.32 Jy beam$^{-1}$ km s$^{-1}$ and higher contours increase in steps of 1.28 Jy beam$^{-1}$ km s$^{-1}$. The yellow contours in both panels correspond to a channel map of CH$_{3}$OCHO 18$_{7,11}$ -- 17$_{7,10}$ at \vlsr~=~7.8~km/s. The first contour is 337 mJy beam$^{-1}$, which is also contour step. The continuum at 230.9~GHz in Jy beam$^{-1}$ is overlaid as a greyscale. 
\label{p-almacr}}
\end{figure}

\clearpage

\begin{deluxetable}{llcccc}
\tablecaption{HIFI Observations \label{t-obs}}
\tablecolumns{6}
\tablewidth{0pt}
\tablehead{\colhead{Band\tablenotemark{a}} & \colhead{Observation Date}  & \colhead{OD} & \colhead{OBSID} & \colhead{Frequency Range} & \colhead{RMS}  \\ 
\colhead{} & \colhead{}  & \colhead{} & \colhead{} & \colhead{(GHz)} & \colhead{(mK)}}
\startdata
1a		&	1 March 2010		 &  291	&	1342191504	&	479.5 -- 561.5	    &  20 \\
1b		&	2 March 2010		 &  292	&	1342191592	&	554.5 -- 636.5	    &  20 \\
2a		&	12 April 2010  		&  333	&	1342194540	&	626.0 -- 726.0	    &  20 \\
2b		&	7 March 2010 		&  297 	&	1342191755	&	714.1 -- 801.3	    &  30 \\
3a		&	29 September 2010 & 504  	&	1342205334 	&	799.0 -- 860.0	    &  40 \\
3b		&	19 March 2010 	& 309	&	1342192329	&	858.1 -- 961.0	    &  40 \\
4a		&	2 March 2010 		& 293	&	1342191601	&	949.1 -- 1061.0    &  50 \\
4b		&	3 March 2010 		& 294	&	1342191649	&	1049.9 -- 1122.0  &  80 \\
5a		&	6 March 2010 		& 296   	&	1342191725	&	1108.1 -- 1242.8  &  100 \\
5b		&	19 March 2011 	& 674	&	1342216387 	&	1227.1 -- 	1280.0 &  100 \\
6a (HC)	&	6 March 2010 		& 296	&	1342191727	&	1425.9 -- 1535.0  &  700 \\
6b (HC)	&	22 March 2010 	& 312   	&	1342192562	&	1573.3 -- 1702.8  &  800 \\
7a (HC)	&	15 April 2010 		& 336  	&	1342194732	&	1697.2 -- 1797.8  &  800 \\
7b (HC)	&	23 March 2010 	& 313  	&	1342192673	&	1788.4 -- 1906.5  &  800 \\
6a (CR)	&	6 March 2010 		& 296	&	1342191728	&	1425.9 -- 1535.0  &  700 \\
6b (CR)	&	22 March 2010 	& 312  	&	1342192563	&	1573.3 -- 1702.8  &  800 \\
7a (CR)	&	15 April 2010 		& 336	&	1342194733	&	1697.2 -- 1797.8  &  800 \\
7b (CR)	&	23 March 2010 	& 313   	&	1342192674	&	1788.4 -- 1906.5  &  800 \\
\enddata
\tablenotetext{a}{For bands 6 and 7, ``HC" and ``CR" indicate hot core and compact ridge pointings, respectively.}
\end{deluxetable}

\clearpage

\begin{deluxetable}{lcc}
\tablecaption{Adopted Values for Source Size and \hh\ Column Density \label{t-standard}}
\tablecolumns{5}
\tablewidth{0pt}
\tablehead{\colhead{Component} & \colhead{\ssize\ (\arcsec)} & \colhead{\nhh\ (\cms)}}
\startdata
Hot Core            & 10   & 3.1~$\times$~10$^{23}$ \\
Compact Ridge  & 10   & 3.9~$\times$~10$^{23}$ \\
Plateau              & 30   & 2.8~$\times$~10$^{23}$ \\
Extended Ridge & 180 & 7.1~$\times$~10$^{22}$ \\
\enddata
\tablecomments{Values for \ssize\ and \nhh, including references, are explained in Sec.~\ref{s-xclass}}
\end{deluxetable}

\clearpage

\begin{deluxetable}{lcccc}
\tablecaption{Optically Thick Molecules \label{t-thick}}
\tablecolumns{5}
\tablewidth{0pt}
\tablehead{\colhead{Molecule} & \colhead{Hot Core} & \colhead{Compact Ridge} & \colhead{Plateau} & \colhead{Extended Ridge}}
\startdata
CH$_{3}$CN         &  X  &  X  &  X  &  ...   \\
CH$_{3}$OH        &  X  &  X  &  ...  &  ...   \\
$^{13}$CO            &  X  &  X  &  X   &  X   \\
CS                         &  X  &  X  &  X   &  X   \\
H$_{2}$CO               &  X  &  X  &    &  ...   \\
H$_{2}$$^{18}$O     &  X  &  X  &  X  &  ...   \\
H$_{2}$S                 &  X  &  ...  &  X  &    \\
HCN                        &  X  &   X    &  X  & X   \\
HCN,$\nu_{2}$=1    &  X  &  ...  &  ...  & ...   \\
H$^{13}$CN            &  X  &    &  X  & X   \\
HCO$^{+}$              &  X  & X   &  X  & X   \\
HNC                        &  X  &    &  X  &  X    \\
HNCO                      &  X  &  ...  &  X  &  ...   \\
NH$_{3}$                &  X  &  ...  &  X  & X   \\
OH                           &  ...  &  ...  &  X  & ...   \\
SiO                         &  ...  &  ...  &  X  & ...   \\
SO                         &  X  &  ...  &  X  &    \\
SO$_{2}$              &  X  &    &  X  &    \\
\enddata
\tablecomments{``X" indicates an optically thick model, ``..." corresponds to a  non-detection, and no mark indicates an optically thin model, which will have an entry in Table~\ref{t-xclass} and possibly Table~\ref{t-madex}.}
\end{deluxetable}

\clearpage

\begin{deluxetable}{lccccc}
\tablecaption{XCLASS Model Parameters  \label{t-xclass}}
\tablecolumns{5}
\tablewidth{0pt}
\tablehead{\colhead{Molecule} & \colhead{\ssize\ (\arcsec)} & \colhead{\trot\ (K)} & \colhead{\ntot\ (\cms)} & \colhead{\vlsr\ (km/s)}  & \colhead{\dv\ (km/s)}}
\startdata
\cutinhead{Hot Core}
C$_{2}$H$_{3}$CN           &     20  &    130  &    7.5$\times 10^{14}$     &    3.5  &   12.5 \\
C$_{2}$H$_{3}$CN           &     10  &    215  &    2.7$\times 10^{15}$     &    5.0  &    5.5 \\
C$_{2}$H$_{5}$CN           &     10  &    136  &    2.1$\times 10^{16}$     &    5.0  &    7.0 \\
C$_{2}$H$_{5}$CN           &      5  &    300  &    6.5$\times 10^{15}$     &    5.0  &    7.0 \\
CH$_{2}$NH                 &     10  &    130  &    1.3$\times 10^{15}$     &    6.0  &    8.0 \\
CH$_{3}$CN,$\nu_{8}$=1     &     10  &    260  &    3.7$\times 10^{15}$     &    6.0  &    6.0 \\
$^{13}$CH$_{3}$CN          &     20  &     80  &    8.0$\times 10^{13}$     &    5.5  &    8.0 \\
$^{13}$CH$_{3}$CN          &     10  &    260  &    1.3$\times 10^{14}$     &    5.5  &    8.0 \\
\enddata
\tablecomments{We estimate the uncertainties of \trot\ and \ntot\ in our XCLASS models to be $\sim$~10\% and  25\%, respectively. Errors in \vlsr\ are $\pm$~1~km/s for the hot core, compact ridge, and extended ridge, and $\pm$~2~km/s for the plateau. Errors in \dv\ are $\pm$~0.5~km/s, 1.5~km/s, and 5.0~km/s for the compact/extended ridge, hot core, and plateau, respectively. See the Appendix for more details concerning error estimates.}
\end{deluxetable}

\clearpage

\begin{deluxetable}{lcccccc}
\tablecaption{MADEX Model Parameters  \label{t-madex}}
\tablecolumns{7}
\tablewidth{0pt}
\tablehead{\colhead{Molecule} &  \colhead{\ssize\ (\arcsec)} & \colhead{\tkin\ (K)}  & \colhead{\nht\ (\cmc)}  & \colhead{\ntot\ (\cms)} & \colhead{\vlsr\ (km/s)}  & \colhead{\dv\ (km/s)}}
\startdata
\cutinhead{Hot Core}
$^{13}$CH$_{3}$CN-A      &   10.0      &   250.0     &     1.5 $\times$ $10^{7}$   &     5.0 $\times$ $10^{13}$  &   5.5       &   9.0       \\  
$^{13}$CH$_{3}$CN-A      &   7.0       &   300.0     &     5.0 $\times$ $10^{7}$   &     2.5 $\times$ $10^{13}$  &   5.5       &   7.0       \\  
$^{13}$CH$_{3}$CN-A      &   5.0       &   400.0     &     1.0 $\times$ $10^{8}$   &     3.0 $\times$ $10^{13}$  &   6.0       &   5.0       \\  
$^{13}$CH$_{3}$CN-E      &   10.0      &   250.0     &     1.5 $\times$ $10^{7}$   &     5.0 $\times$ $10^{13}$  &   5.5       &   9.0       \\  
$^{13}$CH$_{3}$CN-E      &   7.0       &   300.0     &     5.0 $\times$ $10^{7}$   &     2.5 $\times$ $10^{13}$  &   5.5       &   7.0       \\  
$^{13}$CH$_{3}$CN-E      &   5.0       &   400.0     &     1.0 $\times$ $10^{8}$   &     3.0 $\times$ $10^{13}$  &   6.0       &   5.0       \\  
CH$_{3}$$^{13}$CN-A      &   10.0      &   250.0     &     1.5 $\times$ $10^{7}$   &     5.0 $\times$ $10^{13}$  &   5.5       &   9.0       \\  
CH$_{3}$$^{13}$CN-A      &   7.0       &   300.0     &     5.0 $\times$ $10^{7}$   &     2.5 $\times$ $10^{13}$  &   5.5       &   7.0       \\
\enddata
\tablecomments{We estimate \ntot\ uncertainties in our MADEX models to be $\sim$~25\%. Errors in \vlsr\ are $\pm$~1~km/s for the hot core, compact ridge, and extended ridge, and $\pm$~2~km/s for the plateau. Errors in \dv\ are $\pm$~0.5~km/s, 1.5~km/s, and 5.0~km/s for the compact/extended ridge, hot core, and plateau, respectively.} See the Appendix for more details concerning error estimates. There are no \hh\ density entries for SO$_{2}$ and its isotopologues because we assume LTE for this molecule within MADEX (see Sec.~\ref{s-so2}).
\end{deluxetable}

\clearpage

\begin{deluxetable}{lcccc}
\tablecaption{Reduced \chisq\ Calculations \label{t-chisq}}
\tablecolumns{5}
\tablewidth{0pt}
\tablehead{\colhead{Molecule} & \colhead{\sseff} & \colhead{\nchan} & \colhead{\redchisq} & \colhead{Catalog}}
\startdata
\cutinhead{XCLASS}
$^{13}$C$^{18}$O            &      100   &          1   &        ...   &     CDMS   \\
OD                          &       10   &          2   &        ...   &      JPL   \\
HN$^{13}$C                  &       20   &         22   &        ...   &     CDMS   \\
p-NH$_{3}$                  &       20   &        227   &        0.3   &      JPL   \\
H$^{37}$Cl                  &       50   &        109   &        0.3   &      JPL   \\
C$^{17}$O                   &       20   &         98   &        0.4   &     CDMS   \\
NO                          &       20   &       1388   &        0.4   &      JPL   \\
C$^{18}$O                   &       20   &        286   &        0.4   &     CDMS   \\
\enddata
\tablecomments{A ``..." indicates there were not enough channels to compute a robust \redchisq\ value.}
\tablenotetext{a}{Both $^{13}$C isotopologues of methyl cyanide were fit in the same model.}
\end{deluxetable}

\clearpage

\begin{deluxetable}{lrrrrrr}
\tablecaption{Integrated Intensity and Number of Lines Based on XCLASS Models \label{t-intflux}}
\tablecolumns{7}
\tablewidth{0pt}
\tablehead{ \colhead{Molecule} &  \colhead{\nf} &  \colhead{} &  \colhead{} &  \colhead{\intflux\ (K km s$^{-1}$)} &  \colhead{} &  \colhead{} \\
\cline{3-7}
\colhead{} & \colhead{} & \colhead{Hot Core} & \colhead{Compact Ridge} & \colhead{Plateau} & \colhead{Extended Ridge}  & \colhead{Total\tablenotemark{a}}}
\startdata
CO                            &    11             &    0.0            &    0.0            &    61790.8        &    0.0            &    61790.8        \\   
SO$_{2}$                      &    704            &    10379.1        &    217.3          &    46389.5        &    205.1          &    57191.0        \\   
CH$_{3}$OH-E                  &    1664           &    23313.7        &    16885.8        &    0.0            &    0.0            &    40199.5        \\   
H$_{2}$O                      &    35             &    0.0            &    0.0            &    38797.5        &    0.0            &    38797.5        \\   
CH$_{3}$OH-A                  &    1248           &    22328.1        &    15812.1        &    0.0            &    0.0            &    38140.2        \\   
SO                            &    66             &    1395.9         &    0.0            &    25307.2        &    244.6          &    26947.6        \\   
H$_{2}$CO                     &    129            &    3520.4         &    2258.8         &    4311.1         &    0.0            &    10090.3        \\   
HCN                           &    13             &    587.9          &    297.9          &    7025.5         &    107.6          &    8018.9         \\
\enddata
\tablecomments{We estimate a 25\% error in the integrated intensity of individual spatial/velocity components. Details concerning this uncertainty calculation are given in the Appendix.}
\tablenotetext{a}{Total integrated intensity from all spatial/velocity components.}
\end{deluxetable}

\clearpage

\begin{deluxetable}{lcccc}
\tablecaption{Abundances: \ntot(X)/\nhh \label{t-abund}}
\tablecolumns{5}
\tablewidth{0pt}
\tablehead{\colhead{Molecule} & \colhead{Hot Core} & \colhead{Compact Ridge} & \colhead{Plateau} & \colhead{Extended Ridge}}
\startdata
\cutinhead{XCLASS}
CH$_{3}$CN               &   3.0~$\times$~10$^{-8}$        &   1.3~$\times$~10$^{-8}$        &   1.0~$\times$~10$^{-8}$        &   ...                           \\ 
C$_{2}$H$_{3}$CN         &   1.1~$\times$~10$^{-8}$        &   ...                           &   ...                           &   ...                           \\ 
C$_{2}$H$_{5}$CN         &   8.9~$\times$~10$^{-8}$        &   ...                           &   ...                           &   ...                           \\ 
HC$_{3}$N                &   8.1~$\times$~10$^{-9}$        &   ...                           &   4.6~$\times$~10$^{-9}$        &   ...                           \\ 
HCN                      &   3.8~$\times$~10$^{-7}$        &   4.1~$\times$~10$^{-9}$        &   2.8~$\times$~10$^{-7}$        &   3.3~$\times$~10$^{-8}$        \\ 
HNC                      &   5.5~$\times$~10$^{-9}$        &   1.7~$\times$~10$^{-10}$       &   2.3~$\times$~10$^{-9}$        &   7.6~$\times$~10$^{-10}$       \\ 
CN                       &   ...                           &   ...                           &   2.3~$\times$~10$^{-9}$        &   3.1~$\times$~10$^{-9}$        \\ 
HNCO                     &   7.8~$\times$~10$^{-8}$        &   ...                           &   1.0~$\times$~10$^{-8}$        &   ...                           \\
\enddata
\tablecomments{We estimate abundance uncertainties derived from both XCLASS and MADEX to be $\sim$~40\%. See the Appendix for more details concerning error estimates.}
\tablenotetext{a}{Hot core abundance taken from \citet{crockett14}}  
\tablenotetext{b}{All abundance taken from \citet{neill13a}} 
\tablenotetext{c}{Assumed abundance}
\tablenotetext{d}{Abundance computed by averaging the column densities from the low and high velocity plateau components (see Sec.~\ref{s-sio}).}
\end{deluxetable}

\clearpage

\begin{deluxetable}{lccc}
\tablecaption{Computed Vibrational Temperatures \label{t-vib}}
\tablecolumns{4}
\tablewidth{0pt}
\tablehead{\colhead{Molecule} & \colhead{Modes} &  \multicolumn{2}{c}{\tvib\ (K)} \\
\cline{3-4} 
\colhead{} & \colhead{} &  \colhead{Hot Core} & \colhead{Plateau}}
\startdata
HCN               &  $\nu_{2}$=1 / $\nu$=0        &  270 $\pm$ 20 & ... \\
HCN               &  $\nu_{2}$=2 / $\nu_{2}$=1  &  240 $\pm$ 20 & ... \\
CH$_{3}$CN  &  $\nu_{8}$=1 / $\nu$=0        &  260\tablenotemark{a} & 170 $\pm$ 20 \\
\enddata
\tablecomments{The method by which we compute uncertainties for \tvib\ is described in the Appendix.}
\tablenotetext{a}{Because the hot core $\nu$~=~0 model is consistent with $\nu_{8}$~=~1, we report \trot\ here.}
\end{deluxetable}

\clearpage


\end{document}